%% file: manuscript.tex
\documentclass[preprint,12pt]{elsarticle}

\usepackage[utf8]{inputenc}
\usepackage[T1]{fontenc}
\usepackage{amsmath,amssymb}
\usepackage{graphicx}
\usepackage{booktabs}
\usepackage{multirow}
\usepackage{makecell}
\usepackage{tabularx}
\usepackage{threeparttable}
\usepackage{rotating}
\usepackage[table]{xcolor}
\usepackage{subcaption}
\usepackage{eurosym}
\usepackage{lineno}
\usepackage[left=2cm,right=2cm,top=2.2cm,bottom=2.2cm]{geometry}
\usepackage[hidelinks]{hyperref}

\journal{One Health}

\definecolor{sig90}{RGB}{220,232,245}
\definecolor{sig95}{RGB}{176,196,222}

\graphicspath{{figures/}}

\begin{document}

\begin{frontmatter}

\title{A spatiotemporal Bayesian hierarchical model of heat-related mortality in Catalonia, Spain (2012--2022): The role of environmental and socioeconomic modifiers}

\author[udg,ciber]{David Solano}
\author[udg,ciber]{Marta Solans}
\author[dipsalut,udg]{Xavier Perafita}
\author[idiap,ics,uvic]{Anna Ruiz-Comellas}
\author[udg,ciber]{Marc Saez\corref{cor1}}
\ead{marc.saez@udg.edu}
\author[udg,ciber]{Maria A. Barcel\'o}

\cortext[cor1]{Corresponding author: Prof.\ Dr.\ Marc Saez, PhD, CStat. Research Group on Statistics, Econometrics and Health (GRECS), University of Girona, Carrer de la Universitat de Girona 10, Campus de Montilivi, 17003 Girona, Spain. Tel.: +34-972-418338. E-mail: \texttt{marc.saez@udg.edu}}

\address[udg]{Research Group on Statistics, Econometrics and Health (GRECS), University of Girona, Spain}
\address[ciber]{Centro de Investigaci\'on Biom\'edica en Red de Epidemiolog\'ia y Salud P\'ublica (CIBERESP), Instituto de Salud Carlos III, Madrid, Spain}
\address[dipsalut]{Observatori---Organisme Aut\`onom de Salut P\'ublica de la Diputaci\'o de Girona (Dipsalut), Girona, Spain}
\address[idiap]{Unitat de Suport a la Recerca de la Catalunya Central, Fundaci\'o Institut Universitari per a la recerca a l'Atenci\'o Prim\`aria de Salut Jordi Gol i Gurina, Manresa, Spain}
\address[ics]{Health Promotion in Rural Areas Research Group, Ger\`encia d'Atenci\'o Prim\`aria i a la Comunitat Catalunya Central, Institut Catal\`a de la Salut, Manresa, Spain}
\address[uvic]{Department of Medicine, Faculty of Medicine, Universitat de Vic -- Central de Catalunya (UVic-UCC), Vic, Spain}

\begin{abstract}
Extreme heat represents a growing threat to public health as climate change intensifies. While its mortality impacts are well recognised, the interplay between heat, air pollution, and socioeconomic vulnerability remains insufficiently understood.

We quantified the associations of extreme maximum temperatures and heatwaves with daily mortality in Catalonia (2012--2022) and evaluated how air pollutants---particularly ozone, nitrogen dioxide and particulate matter---and socioeconomic factors confound or modify this relationship.

We conducted a time-series ecological analysis across 379 Basic Health Areas during summer months, linking mortality with meteorological, air pollution and socioeconomic indicators. Hierarchical Bayesian zero-inflated negative binomial models were fitted with nested spatiotemporal random effects (Mat\'ern SPDE field evolving over years; month nested within year; unstructured ABS frailty).

Among 210{,}860 summer deaths in the complete environmental sample, extreme heat and heatwaves were associated with higher mortality after multipollutant and socioeconomic adjustment (e.g.\ extreme heat at lag~2: RR~$1.105$, 95\% CrI $1.086$--$1.125$; heatwave at lag~3: RR~$1.142$, 95\% CrI $1.116$--$1.167$; posterior Prob~$\approx 1$). Ozone concentrations $\geq 100\,\mu$g/m$^3$ were positively associated with mortality, and extreme heat$\times$high ozone interactions further elevated risk. Areas with higher income inequality and older population structures also showed elevated mortality.

These findings indicate that heat indicators raise summer mortality in Catalonia even after environmental and socioeconomic adjustment, while compound ozone exposure and contextual vulnerability further shape risk. Adaptation strategies should therefore combine heat-health warnings with air quality management and equity-oriented interventions.
\end{abstract}


\end{frontmatter}

\section*{Highlights}
\begin{itemize}
\item Extreme heat and heatwaves raise mortality after multipollutant and SES adjustment
\item Ozone $\geq 100\,\mu$g/m$^3$ and heat$\times$ozone interactions amplify risk
\item Marginal heat effects and interaction contrasts are reported separately
\item Nested Bayesian spatiotemporal models capture residual space--time dependence
\item Inequality and ageing remain important contextual vulnerability markers
\end{itemize}

\section*{Graphical abstract}
\begin{center}
\includegraphics[width=\textwidth]{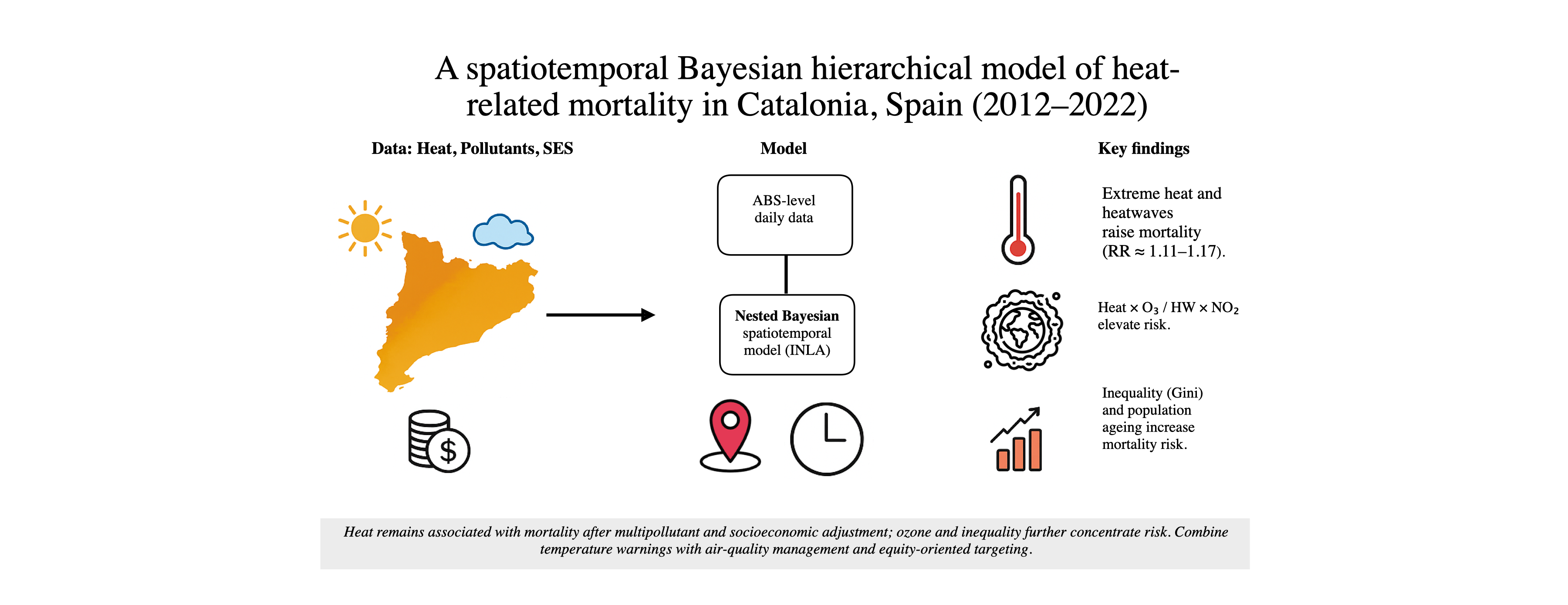}
\end{center}

\section{Introduction}
\label{sec:intro}

As climate change accelerates, heatwaves have emerged as a major public health concern among extreme weather events. During the summer of 2022, extreme heat (including but not limited to heatwaves) was associated with an estimated 61,672 deaths in Europe \citep{ballester2023heat}, with southern areas being particularly affected \citep{ballester2023heat,beck2024mortality}. A substantial proportion of these deaths has been attributed to anthropogenic climate change \citep{beck2024mortality}. Even before 2022, the Global Burden of Disease (GBD) study highlighted a sustained increase in exposure to extreme heat and its growing contribution to climate change-related mortality \citep{murray2020global}. Future projections indicate that the frequency, intensity, and duration of extreme heat events will continue to increase \citep{barriopedro2023heat,epa2025climate}, particularly in southern and central Europe \citep{batibeniz2025rapid,beck2024mortality}. In this context, L\"uthi et al.\ estimated that extreme heat-related mortality events, which historically occurred once every 100 years, may now recur every 10--20 years, or even more frequently, under global warming scenarios of $1.5^\circ$C or $2^\circ$C \citep{luthi2023rapid}.

Air pollution remains one of the leading environmental health risks worldwide. According to the GBD study, ambient air pollution, particularly fine particulate matter (PM$_{2.5}$) and ozone (O$_3$), is among the top contributors to global disease burden \citep{murray2020global,brauer2024global}. In 2019 alone, air pollution was responsible for more than 4.5 million deaths worldwide and accounted for approximately 8\% of all deaths \citep{murray2020global}. The associated burden is largely driven by cardiovascular and chronic respiratory diseases, including COPD \citep{murray2020global}. Specifically, fine particulate matter exposure has been linked to millions of premature deaths globally, with ozone contributing additional mortality \citep{murray2020global,fuller2022pollution}.

Several studies have evidenced the synergistic effect of high temperatures and air pollution, especially ozone and fine particulate matter, on mortality risk. In China, this interaction has been observed in various provinces (cardiovascular mortality and ozone \citep{xu2023ozone,qi2023modification}; myocardial infarction mortality and ozone \citep{xu2023extreme}; and stroke mortality and fine particulate matter \citep{deng2024short}) and counties (all-cause, nonaccidental, circulatory, cardiovascular, cerebrovascular, and respiratory disease mortality and ozone \citep{du2024exposure}). Similar associations have been reported in French cities, where ozone has been linked to all-cause mortality \citep{dear2005effects} as well as nonaccidental, cardiovascular, and respiratory disease mortality \citep{alari2023role}. In California, fine particulate matter has been associated with elevated risks of all-cause, cardiovascular, and respiratory mortality \citep{rahman2022effects}. Across nine European cities, coarse particulate matter (PM$_{10}$) was linked to all-cause mortality \citep{analitis2014effects}, while in the Greater London area ozone was associated with all-cause, respiratory and cardiovascular mortality \citep{gao2025synergistic}. These combined effects pose an increasing public health threat in densely populated urban areas, with particularly severe consequences for vulnerable populations including the elderly, children, individuals with chronic illnesses, and low-income communities \citep{fuller2022pollution,alari2023role}.

Furthermore, certain air pollutants may act as modifying \citep{qi2023modification,dear2005effects,rahman2022effects,analitis2014effects}, mediating \citep{alari2023role,gao2025synergistic,coenders2025moderation}, or confounding \citep{analitis2014effects} factors in the heat--mortality relationship. Tropospheric ozone is particularly relevant, as its formation is enhanced by strong sunlight and high temperatures. Once formed, however, ozone concentrations tend to be lower in urban traffic areas, due to the rapid oxidation of nitrogen oxide (NO) to nitrogen dioxide (NO$_2$), whereas higher ozone levels occur in less polluted regions with no local nitrogen oxide to consume \citep{miteco2025ozone}. Studies conducted in France \citep{dear2005effects,alari2023role}, European cities \citep{rahman2022effects} and NUTS level 3 regions \citep{chen2025trends} suggest that ozone may partly explain increased cardiovascular and respiratory mortality during heatwaves, whereas pollutants like nitrogen dioxide might act as confounders in this relationship.

Population susceptibility to heatwaves varies considerably according to geographic and socioeconomic factors. Key determinants, including age, gender, education, availability of green spaces, access to cooling systems, and housing conditions, play critical roles in shaping vulnerability during heatwaves \citep{luthi2023rapid,achebak2023drivers,feron2023compound,zhao2024global}. Variations in the levels and composition of air pollutants across these regions and populations further modulate their vulnerability. Surprisingly, among studies examining the interactions between heat extremes and exposure to air pollution \citep{xu2023ozone,qi2023modification,xu2023extreme,deng2024short,du2024exposure,dear2005effects,alari2023role,rahman2022effects,analitis2014effects,gao2025synergistic,coenders2025moderation,chen2025trends,achebak2023drivers,feron2023compound}, there are but a few that incorporate socioeconomic variables beyond meteorological and air pollutant variables; notably Qi et al.\ \citep{qi2023modification} and Feron et al.\ \citep{feron2023compound}, who adjusted their models by occupation (i.e.\ farmer or non-farmer) and contextual socioeconomic status, respectively. On the other hand, only Coenders et al.\ \citep{coenders2025moderation} accounted for potential spatial or temporal dependencies. That is to say, not only do the synergistic effects of heat extremes and exposure to air pollutants present geographical and/or temporal variability (i.e., individual heterogeneity), but this variability may also have a systematic structure. For example, after controlling for confounders, areas located closer together in space may share similar risk levels relative to those further apart.

Catalonia, a densely populated Mediterranean region experiencing growing exposure to intense heat extremes, constitutes a critical setting for investigating climate vulnerability. Indeed, in a previous longitudinal ecological study covering 6.3 million people (81.6\% of Catalonia's population), we found that during the summer months, heatwaves and maximum temperature extremes increased the risk of death \citep{barcelo2025excess}. Notably, these effects were delayed, occurring one week after maximum temperature extremes and three days following heatwaves. In addition, we found that several effect modifiers increased the risk of dying on days with extreme heat, namely: being 65 years or older, high relative humidity, extreme minimum temperature, and living in a basic health area (ABS, for its acronym in Catalan, \textit{\`Area B\`asica de Salut}) with a low income. However, our prior analysis did not explore the interaction between extreme heat and exposure to air pollutants and excluded 24\% of the Catalan ABSs (representing 18.24\% of Catalonia's population). In fact, some of those omitted ABSs corresponded to rural territories which exhibit higher summer temperatures. Given that ozone concentrations are typically higher in peri-urban and rural areas than in urban centres, this omission may underestimate relevant exposures.

This study aims to address these gaps by analysing the effects of extreme heat, including heatwaves, on daily mortality across all the Catalan ABSs during the summer months of 2012--2022, while also exploring their interactions with exposure to air pollutants. As in our previous study \citep{barcelo2025excess}, we: (i) control for confounders, including both observed (contextual socioeconomic variables) and unobserved factors (individual heterogeneity and spatial and temporal dependencies), and (ii) explicitly account for spatiotemporal variability using hierarchical Bayesian spatiotemporal models.

\section{Methods}
\label{sec:methods}

\subsection{Design}

We used a time-series ecological design for the summer months (June to September, inclusive) from 2012 to 2022. All the variables were analysed at the contextual level of the 379 ABSs into which Catalonia is divided.

Consequently, all inferences are limited to the contextual (small area) level. This ecological design does not allow for the estimation of the exposure--response relationship at the individual level, and the associations observed at the ABS level should not be interpreted as reflecting risk relationships at the individual level, in order to avoid the ecological fallacy.

\subsection{Variables and data sources}

Daily all-cause deaths were obtained from the Spanish National Statistics Institute (INE) at the census tract level, and aggregated at the ecological level for each ABS. We analysed all-cause deaths for all ages, individuals aged $\geq 65$ years, and individuals aged $\geq 85$ years.

Semi-hourly levels of maximum and minimum temperatures and relative humidity for 2012--2022 were retrieved from the 189 active automatic meteorological stations in the Network of Automatic Meteorological Stations (XEMA, by its Catalan acronym), operated by the Meteorological Service of Catalonia (METEOCAT) (open data) \citep{xema2025data}. Twelve meteorological stations located at altitudes $\geq 1{,}500$ metres were excluded. Daily data were obtained by averaging the semi-hourly values.

Hourly concentrations of air pollutants---airborne particulate matter with a diameter of $\leq 10\,\mu$m (PM$_{10}$), nitrogen dioxide (NO$_2$) and ozone (O$_3$)---were obtained from the 95 automatic monitoring stations of the Catalan Network for Pollution Control and Prevention (XVPCA) (open data) \citep{xema2025airquality}. As with meteorological data, hourly values were averaged to obtain daily data; in the case of O$_3$ the 8-hour moving average was used.

Not all ABSs had meteorological stations or air pollution monitoring stations within their territory. When there is spatial variation in the study region (i.e., between the different stations within the ABS), using the average can lead to bias and the underestimation of the health effect of interest \citep{wannemuehler2009conditional}. To address this, we used a hierarchical Bayesian spatiotemporal model \citep{saez2022spatial} that explicitly accounted for spatial variability, providing daily predictions of meteorological variables and air pollutants for each ABS. This approach yielded unbiased estimators with correct variances (details can be found in Saez and Barcel\'o \citep{saez2022spatial}).

Air pollutant concentrations were categorized based on the 2021 World Health Organization (WHO) Air Quality Guidelines (AQGs) and estimated reference levels \citep{who2021airquality}. For each pollutant, the reference category was set according to the WHO AQG for the annual average (except for ozone, which was based on the season average), while subsequent categories were defined using the corresponding 1-day guideline thresholds. Specifically: PM$_{10}$: $<15\,\mu$g/m$^3$ (reference), $15$--$44.9\,\mu$g/m$^3$, and $\geq 45\,\mu$g/m$^3$; NO$_2$: $<10\,\mu$g/m$^3$ (reference), $10$--$24.9\,\mu$g/m$^3$, and $\geq 25\,\mu$g/m$^3$; and O$_3$: $<60\,\mu$g/m$^3$ (reference), $60$--$99.9\,\mu$g/m$^3$, and $\geq 100\,\mu$g/m$^3$. The reference value of $60\,\mu$g/m$^3$ corresponds to the AQG for peak season average concentrations, which is particularly appropriate given that the study focuses on the summer months. The $\geq 100\,\mu$g/m$^3$ category collapses the WHO daily 8-hour guideline threshold with higher exposures; finer splits at $\geq 120\,\mu$g/m$^3$ yielded sparse heat$\times$ozone cells and unstable interaction estimates, and were therefore not retained in the primary models.

We also considered annual ABS-level data on socioeconomic and demographic variables for the 2012--2022 period. Socioeconomic variables were the average income per person (in euros) and the Gini index (in percentage) \citep{ine2022household}, while the demographic variable was the percentage of population aged 65 and over \citep{ine2022household}. All indicators were observed at the census tract level. We computed ABS-level values as population-weighted averages, using the census tract population data by age and sex provided by the INE \citep{ine2025continuous}. All covariates were categorized into quartiles, with the first quartile used as the reference category in all cases.

Continuous environmental and socioeconomic variables were categorised primarily to enhance interpretability and policy relevance. Air pollutant categories were defined according to the 2021 WHO Air Quality Guidelines, enabling comparison across recognised health-based thresholds and allowing potential non-linear associations to emerge without imposing linearity assumptions. Socioeconomic indicators were categorised into quartiles to reflect gradients of contextual vulnerability at the small-area level. We acknowledge that alternative continuous specifications could be considered; however, the categorical approach was retained to prioritise interpretability and threshold-based contrasts.

Finally, in the models we included the total population of each ABS as offset \citep{ine2025continuous}.

\subsection{Extreme heat temperatures and heatwaves}

For each ABS and month, we calculated the trigger temperatures as the 95th percentile of the frequency distribution of the daily maximum temperatures predicted by the spatiotemporal model \citep{saez2022spatial} for the period 2012--2021 (excluding the likely outlier of 2015). The year 2015 was excluded from the reference distribution because it was characterised by exceptionally high summer temperatures and excess mortality in Catalonia and Spain, representing an extreme climatic event within the study period \citep{barcelo2025excess}. Including 2015 in the percentile calculation would have inflated the trigger threshold and potentially reduced sensitivity for identifying extreme temperature days in subsequent years. The selection of the 95th percentile was based on the definition from the Spanish Ministry of Health's National Plan for Preventive Actions Against the Effects of Excess Temperatures on Health \citep{barcelo2025excess,ministeriosanidad2023plan,tobias2023heat}.

A heatwave indicator was defined as 1 when maximum temperature extremes occurred during three or more consecutive days in at least 10\% of the meteorological stations across Catalonia \citep{saez2022spatial}, in accordance with the official criteria established by the Spanish Ministry of Health \citep{ministeriosanidad2023plan}. Although this definition is based on a regional threshold, it was combined with ABS-specific trigger temperatures and daily predicted exposures derived from the hierarchical spatiotemporal model, allowing the analysis to retain small-area specificity while remaining aligned with national surveillance definitions.

\subsection{Data analysis}

We conceptually distinguish between confounding, effect modification, and mediation. Confounding refers to distortion of the association between heat exposure and mortality due to shared relationships with air pollutants or socioeconomic characteristics. Effect modification (interaction) refers to variation in the magnitude of the heat--mortality association across levels of a third variable (e.g., inequality or pollutant concentrations). Mediation, in contrast, would imply that heat influences mortality through an intermediate mechanism (e.g., increased pollutant formation). Our analytical approach was designed to evaluate confounding and effect modification; formal mediation analysis was not performed, and therefore no causal mediation pathways are inferred.

We estimated generalised linear mixed models (GLMMs) for extreme maximum temperature and for heatwaves, stratified by age (all ages, $\geq 65$, and $\geq 85$ years). For each exposure we report (i)~\emph{main-effects} models, which identify the marginal association of the heat indicator after adjustment for pollutants and socioeconomic covariates, and (ii)~\emph{interaction} models that add heat$\times$modifier terms. Letting $\mu_{it}$ denote the conditional mean mortality rate in ABS $i$ on day $t$, the interaction specification is
\begin{equation}
\begin{aligned}
\log(\mu_{it})
&= \alpha
+ \beta_H H_{i,t-\ell}
+ \boldsymbol{\beta}_P^{\!\top}\mathbf{P}_{it}
+ \boldsymbol{\beta}_S^{\!\top}\mathbf{S}_{it}
+ H_{i,t-\ell}\,
  \boldsymbol{\delta}^{\!\top}\mathbf{W}_{it}\\
&\quad
+ \eta_i
+ S_{i,y(t)}
+ v_{m(t),y(t)}
+ \log(N_i).
\end{aligned}
\label{eq:model}
\end{equation}
Here $H_{i,t-\ell}$ is the lagged heat indicator (extreme maximum temperature with $\ell=2$, or heatwave with $\ell=3$); $\mathbf{P}_{it}$ stacks categorical indicators for lagged mean PM$_{10}$, NO$_2$ and O$_3$ (WHO AQG strata; reference = lowest category); and $\mathbf{S}_{it}$ stacks quartile indicators for average net income, Gini index and the share of residents aged $\geq 65$ years (reference = first quartile). The vector $\mathbf{W}_{it}=(\mathbf{P}_{it}^{\!\top},\,\mathrm{RH}^{(Q4)}_{i,t-\ell},\,\mathbf{S}_{it}^{\!\top})^{\!\top}$ collects candidate effect modifiers---pollutants, high relative humidity (fourth quartile, same lag as $H$), and socioeconomic strata---so that $\boldsymbol{\delta}$ encodes the corresponding heat$\times$modifier interactions. Main-effects models omit the $H_{i,t-\ell}\,\boldsymbol{\delta}^{\!\top}\mathbf{W}_{it}$ term; in that case $\beta_H$ is the adjusted marginal heat association. In interaction models, $\beta_H$ is the heat association in the \emph{reference covariate stratum} only. We additionally report \emph{stratum-specific combined} heat effects $\exp(\beta_H)$ and $\exp(\beta_H+\delta_j)$ (holding other modifiers at reference), obtained from joint posterior draws of the fixed effects. The remaining terms are an unstructured ABS frailty $\eta_i$, a year-specific Mat\'ern spatial field $S_{i,y(t)}$, a month effect nested within year $v_{m(t),y(t)}$, and an offset $\log(N_i)$ for ABS population.

Lags $\ell$ were selected by minimising the Watanabe--Akaike information criterion (WAIC) \citep{watanabe2010asymptotic} over lags $1$--$7$ in main-effects screening models with a common sample size across lags, using the same nested spatiotemporal random-effect structure and zero-inflated negative binomial likelihood as the final models (outcome: all-age summer deaths). Extreme heat selected lag~$2$ and heatwave lag~$3$ (the latter coinciding with our previous Catalan analysis \citep{barcelo2025excess}). Pollutant exposures were the mean of daily concentrations over $\{t-\ell,\ldots,t-1\}$, not single-day values. Categorical strata for heat, humidity and pollutants were retained to prioritise threshold-based interpretability \citep{who2021airquality}.

Daily mortality at the small-area level displayed a high frequency of zero counts together with dispersion exceeding the Poisson mean--variance equality. For this reason, we specified a mixture count model within a hierarchical Bayesian framework. The outcome distribution was modelled using a negative binomial likelihood to accommodate extra-Poisson variability, combined with a zero-inflation component to account for structural zeros at the ABS-day level.

Alternative likelihood specifications available in the Integrated Nested Laplace Approximation (INLA) \citep{r-inla2025} were compared using the WAIC \citep{watanabe2010asymptotic}, and the zero-inflated negative binomial formulation provided the most favourable fit,
\begin{equation}
\mathrm{Prob}(y_{it}\mid \ldots)
= w_{it}\,\mathbf{1}[y_{it}=0]
+ (1-w_{it})\times \mathrm{NegBinom}(y_{it}),
\label{eq:zinb}
\end{equation}
where $y_{it}$ denotes the observed number of deaths in ABS $i$ on day $t$, and $w_{it}$ represents the probability mass assigned to structural zero counts.

\subsection{Random effects}

The modelling framework builds upon our previously developed hierarchical spatiotemporal approach \citep{saez2022spatial}, extended here to nested space--time dependence so that residual spatial structure is allowed to evolve over years rather than being assumed independent of the temporal trend.

The hierarchical structure incorporated multiple latent components. An area-specific exchangeable random effect ($\eta_i$) captured time-invariant ABS-level residual variation. Temporal seasonality within years was modelled as a first-order random walk over summer months nested in year ($v_{m,y}$; iid across years). Long-range residual spatial dependence was represented by a Gaussian Mat\'ern field implemented via the SPDE approach \citep{lindgren2011explicit} within INLA, with the field grouped by year and linked across years by an RW1 group model ($S_{i,y}$). Calendar year indices were coded as contiguous integers $1,\ldots,n_{\mathrm{years}}$ for the INLA grouping. Nesting the spatial field within year allows residual spatial structure to evolve over time, which is appropriate when exposure effects are not separable in space and time.
\begin{equation}
\mathrm{Cov}\bigl(S(x_i),S(x_{i'})\bigr)
= \frac{\sigma^2}{2^{\nu-1}\Gamma(\nu)}
\bigl(\kappa\|x_i-x_{i'}\|\bigr)^{\nu}
K_{\nu}\bigl(\kappa\|x_i-x_{i'}\|\bigr).
\label{eq:matern}
\end{equation}
$K_{\nu}$ is the modified Bessel function of the second type and order $\nu>0$. $\nu$ is a smoothness parameter, $\sigma^2$ is the variance, and $\kappa>0$ is related to the range $\rho=\sqrt{8\nu}/\kappa$, the distance at which the spatial correlation is close to 0.1 \citep{lindgren2011explicit}.

\subsection{Assessment of covariate adjustment}

Primary inference is based on fully adjusted models including heat indicators, air pollutants, humidity and socioeconomic quartiles, together with the spatiotemporal random effects described above. The hierarchical structure allows for partial grouping between areas and years, accounts for spatial and temporal autocorrelation, and improves the stability of estimates for small areas. While this approach does not eliminate ecological bias, it strengthens contextual inference compared to conventional ecological correlation analyses.

\subsection{Inference}

In all cases, the inferences were made following a Bayesian perspective using the INLA approach \citep{rue2009approximate,rue2017bayesian}, applied in its experimental mode \citep{van2023new}. We used priors that penalize complexity (PC priors). These priors are robust in that they do not impact results \citep{simpson2017penalising}.

\section{Results}
\label{sec:results}

\subsection{Descriptive analyses}

\subsubsection{Meteorological characteristics}

In 2022, a year marked by record-breaking heat and unusually high excess mortality in Spain, the summer period was characterised by a higher share of ABS-days under extreme-heat and heatwave conditions than in 2012--2021 (Tables~\ref{tab:1a} and~\ref{tab:1d}).

\begin{figure}[htbp]
\centering
\begin{subfigure}{0.48\textwidth}
\includegraphics[width=\linewidth]{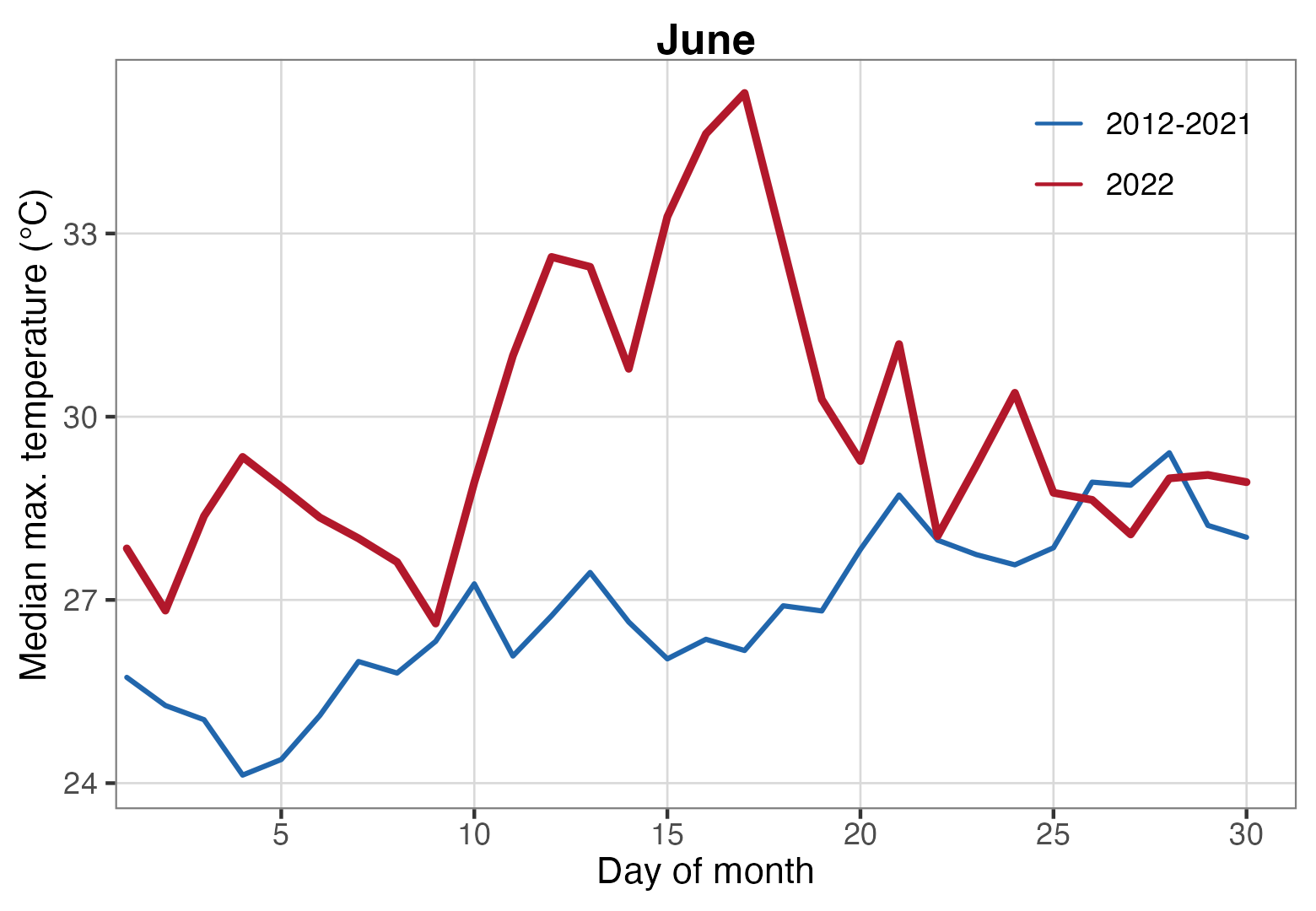}
\end{subfigure}\hfill
\begin{subfigure}{0.48\textwidth}
\includegraphics[width=\linewidth]{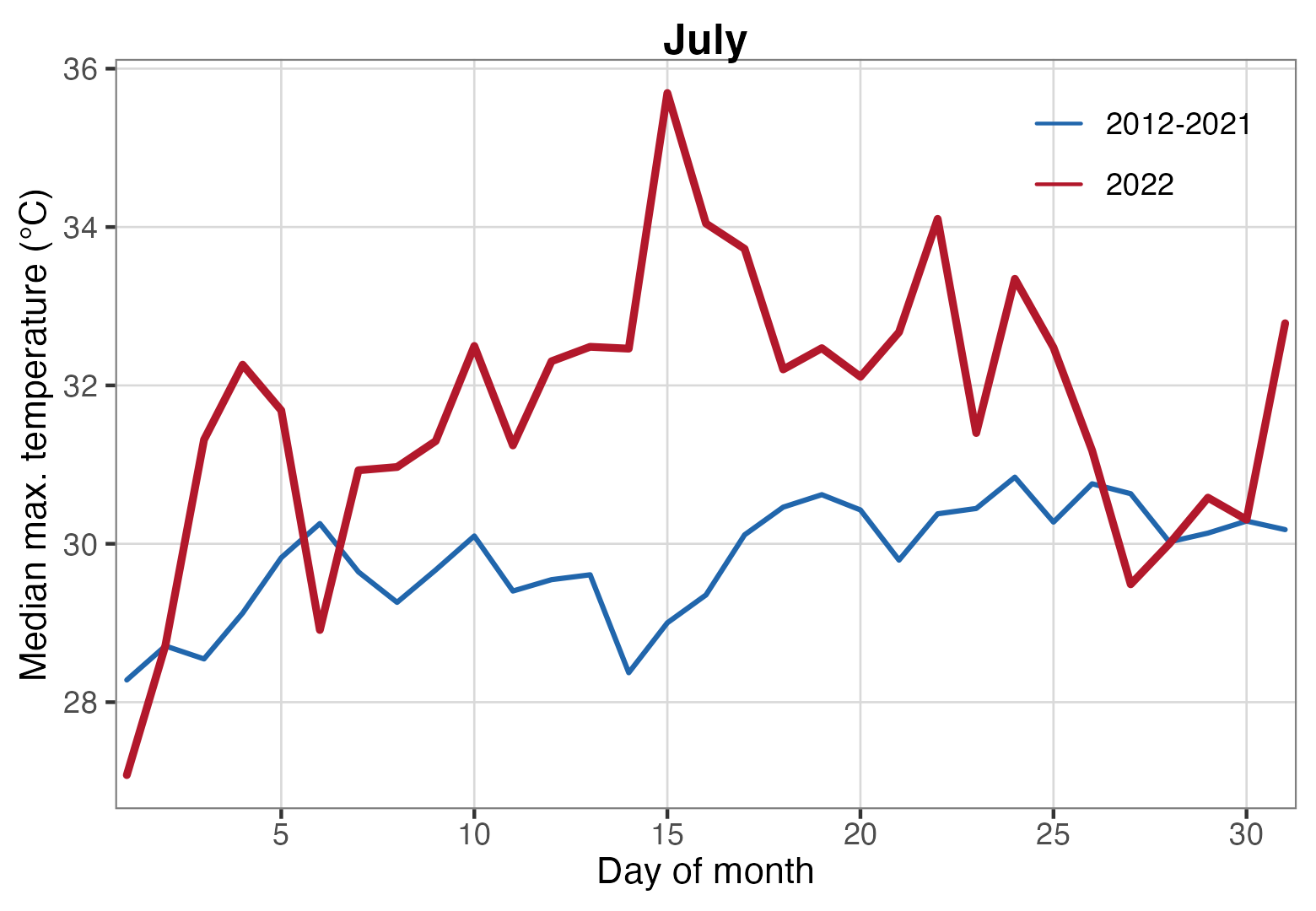}
\end{subfigure}\\[0.6em]
\begin{subfigure}{0.48\textwidth}
\includegraphics[width=\linewidth]{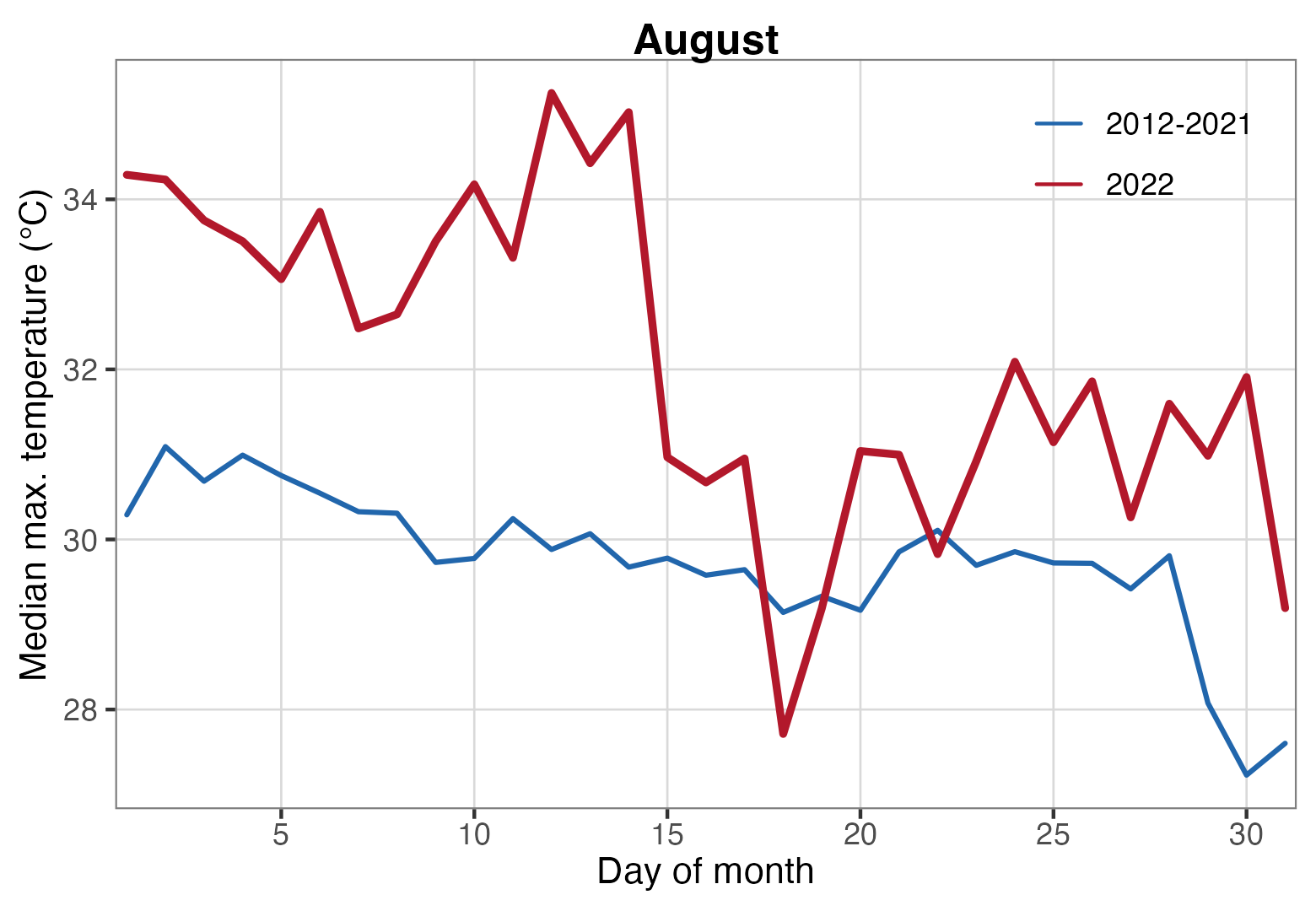}
\end{subfigure}\hfill
\begin{subfigure}{0.48\textwidth}
\includegraphics[width=\linewidth]{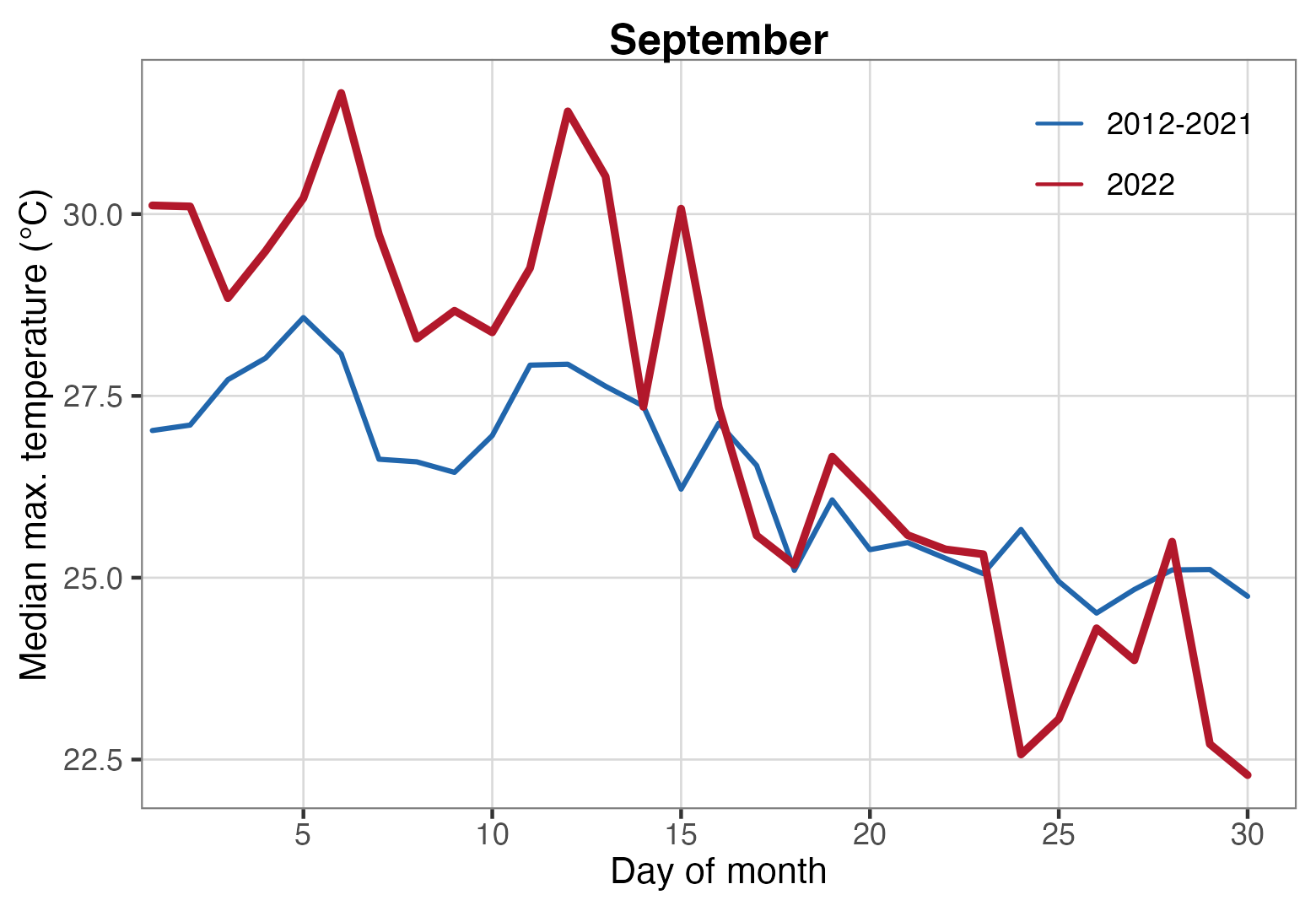}
\end{subfigure}
\caption{Temporal evolution of the median maximum temperature from 2012--2021 and in 2022.}
\label{fig:temp-ts}
\end{figure}

\begin{figure}[htbp]
\centering
\includegraphics[width=\textwidth]{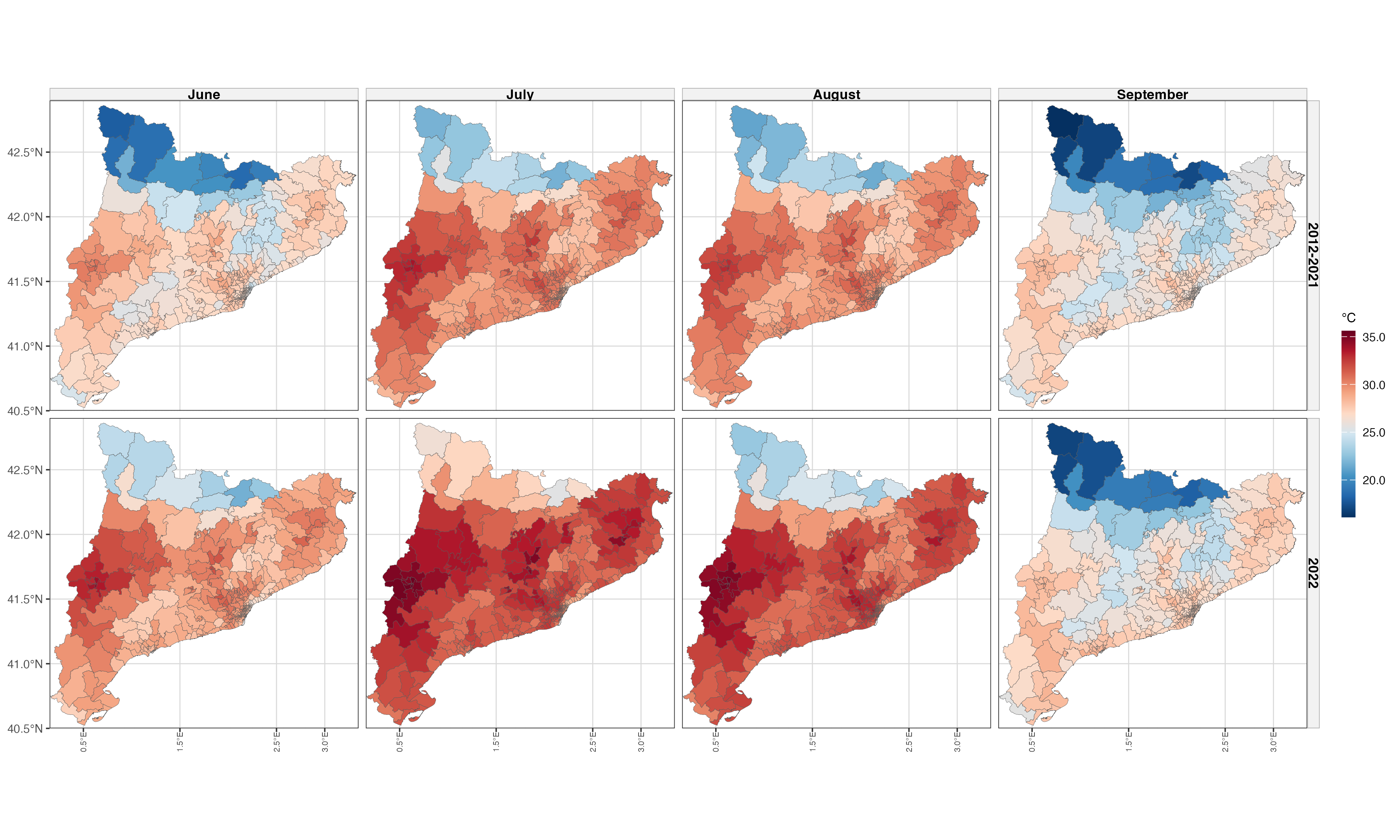}
\caption{Spatial distribution of median daily maximum temperature by ABS during summer months (June--September): 2012--2021 (top row) and 2022 (bottom row). Spatiotemporal predictions \citep{saez2022spatial}.}
\label{fig:temp-maps}
\end{figure}

Maximum temperatures during summer 2022 were consistently higher than the 2012--2021 average, particularly in June and August, when median maximum temperatures were $2.39^\circ$C and $2.16^\circ$C higher, respectively (Table~\ref{tab:1a}, Figure~\ref{fig:temp-ts}). Extreme-heat prevalence rose from $3.2\%$ to $12.7\%$ of June ABS-days and from $9.6\%$ to $37.2\%$ of August ABS-days; heatwave prevalence in August rose from $6.4\%$ to $31.5\%$ (Table~\ref{tab:1d}). The territorial extent of areas experiencing the highest maximum temperatures was substantially greater in 2022, especially in June (Figure~\ref{fig:temp-maps}).

Median minimum and mean temperatures followed a similar pattern, with higher values across all months---particularly in June and August (Table~\ref{tab:1a} and Figure~S1 in Supplemental information). Median relative humidity showed broadly similar values between 2022 and 2012--2021, with August 2022 slightly drier (Table~\ref{tab:1b} and Figure~S2 in Supplemental information).

\subsubsection{Air pollution}

Median nitrogen dioxide concentrations during summer 2022 were lower than in 2012--2021 in all summer months (Table~\ref{tab:1b} and Figure~S4 in Supplemental information). Median ozone concentrations were markedly higher in July and August 2022 ($+6.73$ and $+6.22\,\mu$g/m$^3$ in medians, respectively) than in the preceding decade (Figure~\ref{fig:o3-ts}). Median PM$_{10}$ was higher in June--July 2022 than in 2012--2021 (Figure~S3 in Supplemental information).

\begin{figure}[htbp]
\centering
\includegraphics[width=\textwidth]{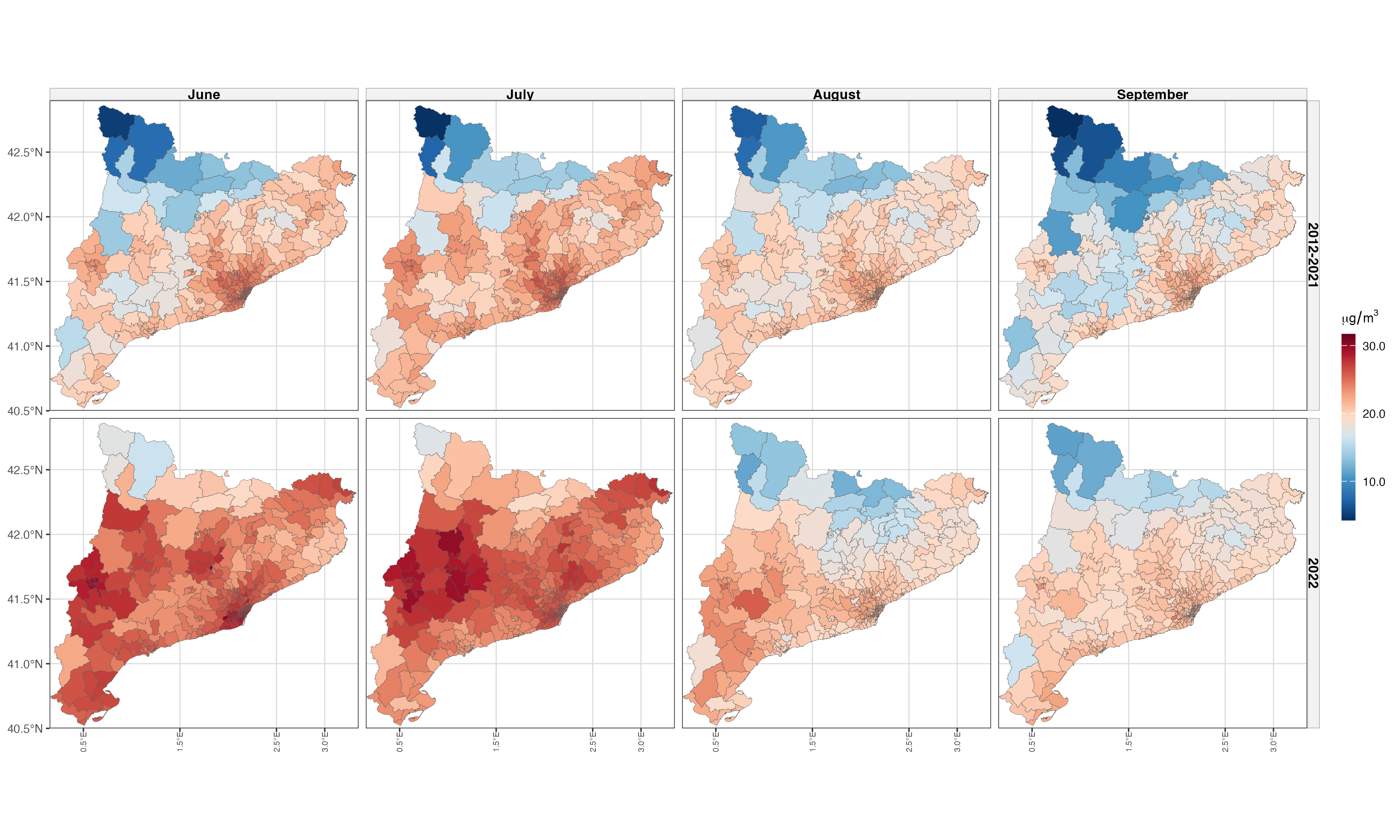}
\caption{Spatial distribution of median daily PM$_{10}$ by ABS during summer months (June--September): 2012--2021 (top row) and 2022 (bottom row). Spatiotemporal predictions \citep{saez2022spatial}.}
\label{fig:pm10-maps}
\end{figure}

\begin{figure}[htbp]
\centering
\includegraphics[width=\textwidth]{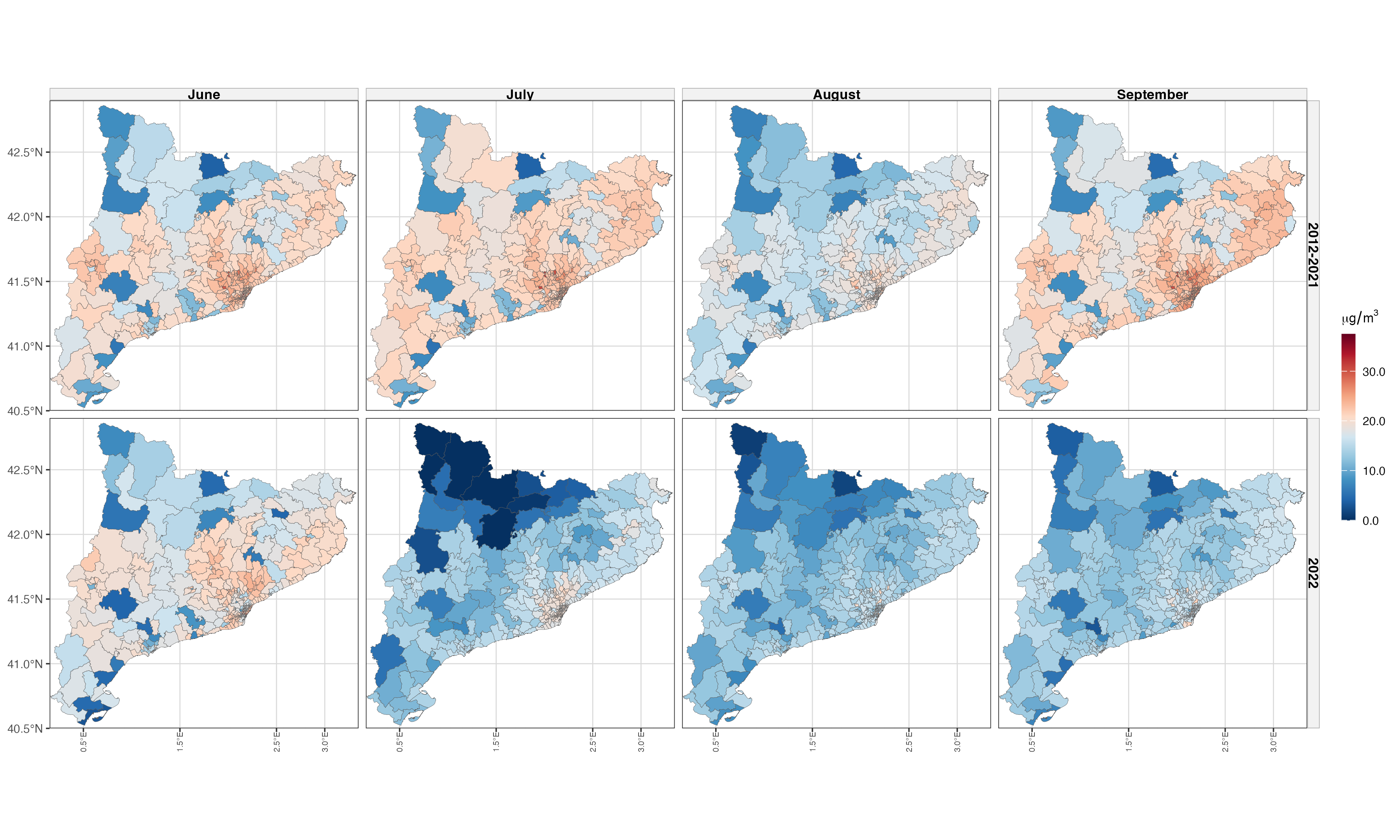}
\caption{Spatial distribution of median daily NO$_2$ by ABS during summer months (June--September): 2012--2021 (top row) and 2022 (bottom row). Spatiotemporal predictions \citep{saez2022spatial}.}
\label{fig:no2-maps}
\end{figure}

\begin{figure}[htbp]
\centering
\includegraphics[width=\textwidth]{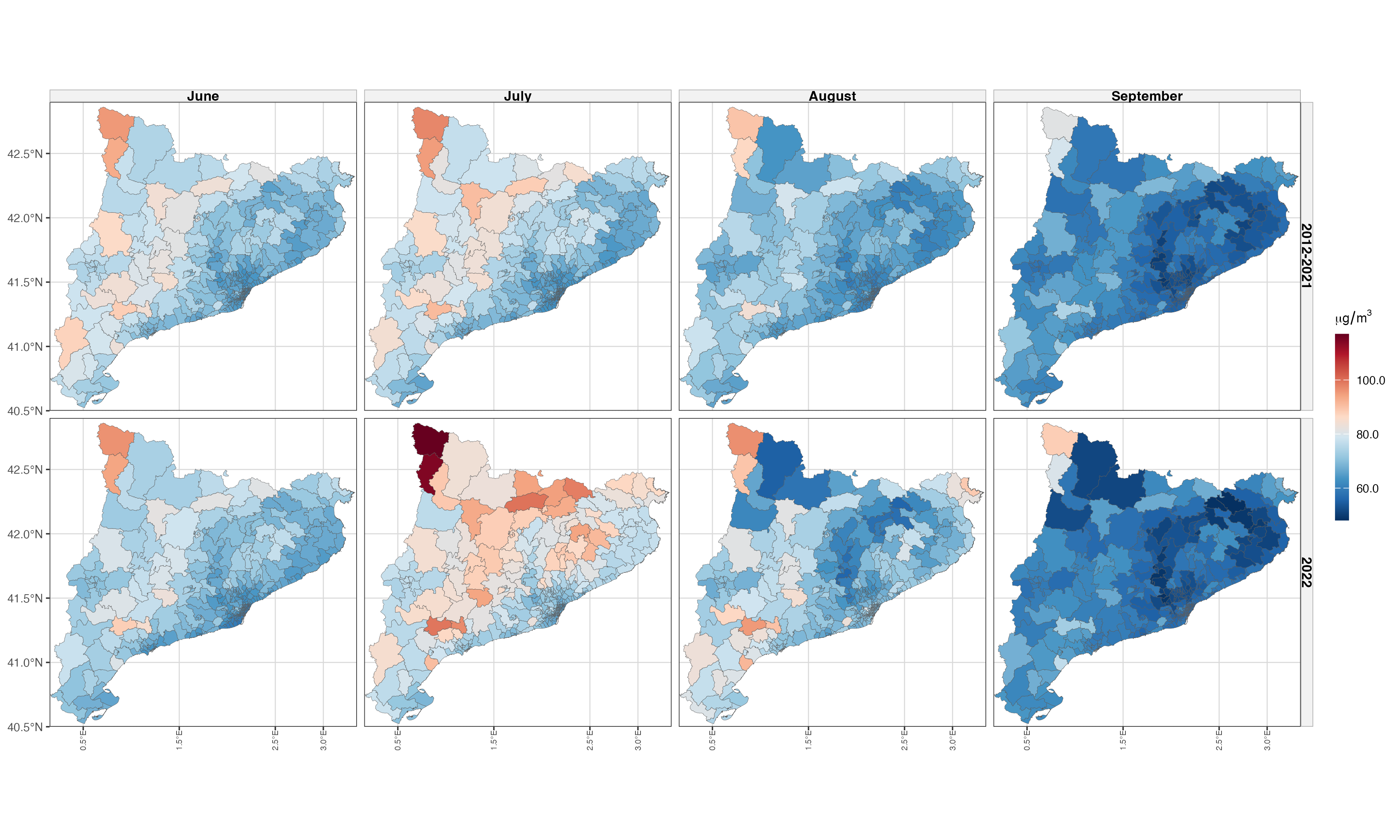}
\caption{Spatial distribution of median daily O$_3$ by ABS during summer months (June--September): 2012--2021 (top row) and 2022 (bottom row). Spatiotemporal predictions \citep{saez2022spatial}.}
\label{fig:o3-maps}
\end{figure}

Spatially, nitrogen dioxide concentrations decreased across most ABSs relative to the previous decade, particularly in densely urbanised areas. Conversely, ozone concentrations increased across many ABSs, especially in rural and peri-urban areas (Figures~\ref{fig:pm10-maps}--\ref{fig:o3-maps}).

\subsubsection{Mortality characteristics}

Between 2012 and 2022, 730{,}634 deaths were registered in Catalonia (Table~\ref{tab:2a}). The median age at death was 84 years (interquartile range, IQR: 73--90), 50.25\% were male, and most were married (42.57\%) or widowed (42.88\%). The majority had at least primary education (79.65\%) and were pensioners (69.47\%).

The leading causes of death were tumours (26.29\%), circulatory diseases (26.0\%), and respiratory diseases (9.95\%). COVID-19 accounted for 28{,}162 deaths (3.86\%) between 2020 and 2022 (Table~\ref{tab:2a}). Additional causes are detailed in Table~\ref{tab:2b}.

In the complete environmental analysis sample used for modelling (498{,}385 summer ABS-days), 210{,}860 deaths were recorded during summer months (180{,}305 aged $\geq 65$; 96{,}133 aged $\geq 85$; Table~\ref{tab:2summer}).

\subsection{Results of the estimation of the spatiotemporal models}

To interpret the coefficient estimates, we calculated relative risks ($\mathrm{RR}=e^{\hat{\gamma}}$) and 95\% credible intervals (95\% CrI), together with the one-tailed posterior probability that the coefficient differs from zero (Probs). Posterior probabilities greater than 0.90 were treated as indicating substantial evidence of association, while inference emphasised the full posterior distributions rather than a dichotomous threshold alone.

\subsubsection{Main-effects models: marginal heat associations}

Extreme heat and heatwaves were positively associated with mortality after adjustment for pollutants, humidity and socioeconomic quartiles (Tables~\ref{tab:3a}--\ref{tab:3b} and~\ref{tab:4a}--\ref{tab:4b}).

\paragraph{Extreme maximum temperatures}
Extreme heat at lag~$2$ was associated with higher all-cause mortality ($\mathrm{RR}=1.105$; 95\% CrI: $1.086$--$1.125$; Prob~$\approx 1$), with similar estimates at ages $\geq 65$ ($\mathrm{RR}=1.120$; 95\% CrI: $1.099$--$1.142$) and $\geq 85$ ($\mathrm{RR}=1.123$; 95\% CrI: $1.095$--$1.153$) (Table~\ref{tab:3a}).

Among pollutants, ozone $\geq 100\,\mu$g/m$^3$ showed a consistent positive association (all ages: $\mathrm{RR}=1.074$; 95\% CrI: $1.019$--$1.132$; Prob~$0.996$), increasing further in older groups. PM$_{10}$ categories $15$--$44.9$ and $\geq 45\,\mu$g/m$^3$, and NO$_2\geq 25\,\mu$g/m$^3$, were also positively associated with mortality (Table~\ref{tab:3a}). Higher Gini and percentage aged $\geq 65$ quartiles were associated with elevated mortality (Table~\ref{tab:3b}).

\paragraph{Heatwaves}
Heatwaves at lag~$3$ showed stronger marginal associations than extreme-heat days: all ages $\mathrm{RR}=1.142$ (95\% CrI: $1.116$--$1.167$), $\geq 65$ $\mathrm{RR}=1.152$ (95\% CrI: $1.124$--$1.180$), and $\geq 85$ $\mathrm{RR}=1.170$ (95\% CrI: $1.133$--$1.208$), all with Prob~$\approx 1$ (Table~\ref{tab:4a}). Ozone associations were milder under the heatwave lag window than under the extreme-heat models, whereas NO$_2\geq 25\,\mu$g/m$^3$ remained positively associated. Socioeconomic gradients were broadly consistent with the extreme-heat models (Table~\ref{tab:4b}).

\subsubsection{Interaction models and stratum-specific combined heat effects}

Interaction models (Tables~\ref{tab:3c}--\ref{tab:3d} and~\ref{tab:4c}--\ref{tab:4d}) include heat$\times$modifier terms. The heat main-effect coefficient is the association in the \emph{reference covariate stratum} (lowest pollutant categories, humidity not in Q4, SES Q1) and was compatible with the null across age groups---for example, extreme heat all ages $\mathrm{RR}=0.904$ (95\% CrI: $0.743$--$1.101$; Prob~$0.84$).

Stratum-specific combined heat effects $\exp(\beta_H+\delta_j)$ (Tables~\ref{tab:3e}--\ref{tab:3f} and~\ref{tab:4e}--\ref{tab:4f}) give the heat association within each modifier level while holding other modifiers at reference. For extreme heat, the clearest positive combined associations were with intermediate PM$_{10}$ ($15$--$44.9\,\mu$g/m$^3$) at ages $\geq 65$ ($\mathrm{RR}=1.115$; 95\% CrI: $1.010$--$1.230$; Prob~$0.985$) and $\geq 85$ ($\mathrm{RR}=1.107$; Prob~$0.917$), and with O$_3\geq 100\,\mu$g/m$^3$ at ages $\geq 65$ ($\mathrm{RR}=1.153$; Prob~$0.888$). Heatwave combined effects by pollutant stratum were generally closer to the null than the corresponding main-effects heatwave RRs, consistent with wide uncertainty when partitioning heat across sparse modifier cells.

\paragraph{Extreme heat interactions (modification ratios)}
Extreme heat$\times$ozone interactions were positive across age groups: $\times$O$_3$ $60$--$99.9\,\mu$g/m$^3$ yielded RR~$1.06$--$1.11$ (Prob~$\geq 0.985$), and $\times$O$_3\geq 100\,\mu$g/m$^3$ yielded RR~$1.17$--$1.19$ (Prob~$\geq 0.980$) (Table~\ref{tab:3c}). These interaction RRs describe \emph{ratios} of heat effects across ozone levels, not the absolute heat effect within the high-ozone stratum (Table~\ref{tab:3e}). Extreme heat$\times$PM$_{10}$ $15$--$44.9\,\mu$g/m$^3$ was elevated particularly at ages $\geq 85$. Heat$\times$high humidity was mildly inverse in the all-ages model. Income Q4$\times$extreme heat was positively associated (Table~\ref{tab:3d}).

\paragraph{Heatwave interactions}
Heatwave$\times$NO$_2\geq 25\,\mu$g/m$^3$ showed positive interactions across ages (RR~$1.09$--$1.17$; Prob~$\geq 0.97$), with intermediate NO$_2$ also elevated (Table~\ref{tab:4c}). Heatwave$\times$ozone interactions were weaker and consistent with the null. Socioeconomic interactions were generally modest (Table~\ref{tab:4d}).

In summary, main-effects models show that heat indicators raise mortality on average; among interaction \emph{ratios}, extreme heat$\times$O$_3$ and heatwave$\times$NO$_2$ had the most consistent posterior support across age strata, whereas absolute stratum-combined heat effects were clearest for extreme heat with intermediate PM$_{10}$ (older ages) and elevated O$_3$ ($\geq 65$).

\subsection{Additional diagnostics}

Lag-selection WAIC diagnostics and pollutant$\times$heat cell counts are reported in Supplementary Tables~S1a--S2b; Bayesian FDR discoveries for interaction terms are summarised in Supplementary Table~S3.

\section{Discussion}
\label{sec:discussion}

Our findings confirm that the summer of 2022 in Catalonia presented exceptional characteristics in terms of the frequency, intensity and duration of extreme heat episodes, consistent with reports at the European level by Barriopedro et al.\ \citep{barriopedro2023heat}, Beck et al.\ \citep{beck2024mortality} and Batibeniz et al.\ \citep{batibeniz2025rapid}, who underline that this type of phenomena will continue to intensify under climate change scenarios.

From a public health perspective, an important question concerns the number of premature deaths that can be attributed to heat exposure. Estimating attributable mortality, however, requires a clearly defined causal framework, including an individual-level exposure--response function and a counterfactual temperature scenario. Given the ecological and contextual nature of our design, our models estimate small-area-level associations rather than individual causal effects. Therefore, our findings should not be interpreted as providing direct estimates of heat-attributable mortality. Instead, they identify patterns of vulnerability and contextual modifiers of mortality risk during extreme heat episodes.

We identified a significant role for air pollutants, especially ozone. Our results show an increased risk of mortality at ozone concentrations $\geq 100\,\mu$g/m$^3$, particularly among those aged 85 years and older, during episodes of extreme maximum temperatures and, to a lesser extent, during heatwaves. Importantly, extreme temperature and heatwave indicators remained positively associated with mortality in main-effects models. Interaction coefficients describe modification ratios across strata, whereas stratum-specific combined contrasts $\exp(\beta_H+\delta)$ give the absolute heat association within each modifier level.

Among pollutant modifiers, extreme heat$\times$O$_3$ showed clearer posterior support than heatwave$\times$O$_3$, whereas heatwave$\times$NO$_2\geq 25\,\mu$g/m$^3$ was the more consistent heatwave interaction. We therefore emphasise these epidemiological contrasts rather than a photochemical interpretation of day-type ozone differences.

The interaction we observed between heatwaves and higher coarse particulate concentrations may also reflect the influence of episodic Saharan dust intrusions that occur over the Iberian Peninsula and Mediterranean basin during the summer months. Saharan dust events (`calima') can elevate concentrations of coarse particles over wide areas and have been documented to affect air quality in Spain and neighbouring regions. Previous research in Spanish cities has investigated the role of mineral dust events in the association between particulate matter and short-term mortality among older adults. Moreover, Spanish researchers engaged in climate-health studies have highlighted the complexity of heat-related health risks and the importance of considering compound environmental exposures such as temperature, air pollution and social vulnerability \citep{jimenez2010role,tobias2011short}. In this context, the joint presence of heatwave conditions and higher coarse particulate levels in our analysis may reflect compound exposure scenarios that are climatologically and regionally plausible. Nevertheless, given the ecological design of our study, these findings should be interpreted as contextual associations rather than evidence of a specific causal pathway. The limited and inconsistent interaction evidence for coarse particulate matter may also reflect correlated exposure structures and reduced statistical power in stratified interaction models. Given the complexity of multipollutant environments and the ecological nature of the analysis, distinguishing true biological interaction from statistical artefact remains challenging.

Furthermore, our results highlight the importance of social and demographic determinants in heat vulnerability. Across all models, ABSs with greater inequality (high Gini coefficients) and a higher proportion of the population aged 65 years or older were found to have a significantly higher risk of mortality. These results reinforce the literature on socioeconomic inequalities in environmental health \citep{murray2020global,brauer2024global,fuller2022pollution}, as well as studies highlighting how population aging amplifies the effects of extreme heat \citep{luthi2023rapid,achebak2023drivers}. Furthermore, they are in line with research that underscores the need to consider compound heat and pollution exposures in unequal urban populations \citep{chen2025trends,feron2023compound}.

Taken together, our findings suggest that excess mortality during extreme heat periods reflects both a marginal heat association and concurrent ozone exposure and socioeconomic vulnerability. This perspective complements burden-of-disease studies estimating heat-attributable mortality by highlighting where and in whom vulnerability is concentrated, and by clarifying that compound exposures amplify---rather than fully replace---heat-related risk. Such information is essential for designing targeted adaptation strategies under climate change, particularly in ageing and socially unequal populations.

Consistent with Ballester et al.\ \citep{ballester2023heat}, Beck et al.\ \citep{beck2024mortality} and Zhao et al.\ \citep{zhao2024global}, we find that extreme heat (both extreme maximum temperatures and heatwaves) predicts all-cause mortality after multipollutant and socioeconomic adjustment, in the total population and in older age groups ($\geq 65$ or $\geq 85$ years). Relative to those studies, our estimates are obtained at fine spatial resolution with explicit multipollutant/SES control and spatiotemporal random effects; heat associations remain after this adjustment.

Only Analitis et al.\ \citep{analitis2014effects} consider the role air pollutants play as possible confounders. Specifically, they showed that exposure to ozone (as well as PM$_{10}$) could partially confound the relationship between heatwaves and mortality, with the effect of heatwaves reduced by 15\% to 30\% after adjustment for O$_3$ and PM$_{10}$.

It is important to distinguish ozone as a co-exposure from ozone as an effect modifier. In our fully adjusted models, O$_3\geq 100\,\mu$g/m$^3$ was positively associated with mortality in its own right, and extreme-heat$\times$ozone interaction ratios indicated that the heat association differed across ozone levels. These are distinct estimands from a marginal heat RR after multipollutant adjustment.

The interpretation of interaction terms requires caution. Among pollutant interaction \emph{ratios}, extreme heat$\times$O$_3$ (Prob~$\geq 0.98$ across ages) and heatwave$\times$NO$_2\geq 25\,\mu$g/m$^3$ (Prob~$\geq 0.97$) showed the most consistent posterior support; other heat$\times$pollutant ratios were weaker or age-specific. Absolute stratum-combined heat effects $\exp(\beta_H+\delta)$ were clearest for extreme heat with intermediate PM$_{10}$ in older adults. Interactions should therefore be read as patterns of modification, not as proof of a single synergistic pathway.

In several models, pollutant levels appeared to modify the association between extreme heat and mortality. For extreme maximum temperatures, high ozone (and, in the oldest group, intermediate PM$_{10}$) increased the heat-associated risk. For heatwaves, higher nitrogen dioxide concentrations ($\geq 25.0\,\mu$g/m$^3$) increased risk across age groups. These patterns are broadly consistent with previous studies suggesting potential synergistic effects between high temperatures and ozone or traffic-related pollution on cardiovascular and respiratory mortality \citep{qi2023modification,xu2023extreme,du2024exposure,alari2023role,rahman2022effects,analitis2014effects}. Differences relative to London or France \citep{dear2005effects,gao2025synergistic} may reflect contextual heterogeneity in climate, baseline pollution, demography and adaptation.

From a mechanistic perspective, the combination of extreme heat and ozone could be explained by increased oxidative stress and systemic inflammation, with adverse effects on the cardiovascular and respiratory systems \citep{xu2023extreme,deng2024short}.

As in Barcel\'o and Saez \citep{barcelo2025excess}, we found that some socioeconomic variables could be effect modifiers. In this sense, inequality measured by the Gini index was associated with a greater effect of extreme heat on all-cause mortality, although not for those under 65 years of age.

Higher levels of social vulnerability and population aging imply a reduced capacity for physiological adaptation and unequal access to protective measures, such as air conditioning or green infrastructure, which intensifies risks \citep{fuller2022pollution}. This pattern aligns with findings from global and regional studies that have pointed out the disproportionate impact of social determinants and the need for an intersectional approach to prevention \citep{barriopedro2023heat,brauer2024global}.

Overall, our findings point to the vulnerability of older age groups (65 years and older) to both social and environmental stressors. The selected lag structure was extreme heat at lag~$2$ and heatwaves at lag~$3$. A mortality-displacement (harvesting) interpretation remains possible but was not formally tested here \citep{barcelo2025excess}.

Note that intermediate nitrogen dioxide categories were not interpreted as protective effects. Apparent inverse associations for NO$_2$ in some strata may reflect residual collinearity or atmospheric coupling with ozone rather than a true protective effect \citep{miteco2025ozone}, and should not be read causally.

Finally, the potential influence of the COVID-19 pandemic on mortality patterns during 2020--2022 deserves consideration. COVID-19 accounted for a modest share of registered deaths in Catalonia over 2020--2022 (Table~\ref{tab:2a}), and summer pandemic mortality was substantially lower than in other seasons. Indirect effects cannot be ruled out: individuals who experienced COVID-19 infection (particularly older adults and those with cardiovascular or respiratory comorbidities) may have become more vulnerable to subsequent environmental stressors. Formal re-estimation of the nested models after subtracting COVID-19 deaths from the outcome was not repeated in this revision; pandemic-related changes in population frailty may still have influenced susceptibility patterns.

These findings have several practical implications. Heat-health prevention plans may benefit from integrating real-time air quality indicators (particularly ozone forecasts) into early warning systems, rather than relying exclusively on temperature thresholds. In addition, small-area indicators of socioeconomic inequality and population ageing could be incorporated into targeted intervention strategies, prioritising outreach and protective measures in the most vulnerable health areas. Given the compound nature of environmental exposures observed in our analysis, adaptation policies should combine heat mitigation (e.g., cooling strategies, urban greening) with air pollution control and social vulnerability reduction.

\section{Conclusion}
\label{sec:conclusion}

Our study highlights the complex interplay between extreme heat, air pollution, and social determinants in shaping mortality risks during summer months in Catalonia (2012--2022), with particular intensity in 2022. While previous research has consistently documented the impact of heatwaves on mortality \citep{ballester2023heat,beck2024mortality,zhao2024global}, our findings show that heat indicators remain associated with higher mortality after multipollutant and socioeconomic adjustment, and that ozone---especially $\geq 100\,\mu$g/m$^3$ and extreme heat$\times$ozone interactions---further amplifies risk, in line with evidence from Analitis et al.\ \citep{analitis2014effects} and other combined-exposure studies \citep{qi2023modification,xu2023extreme,du2024exposure,alari2023role,rahman2022effects}.

Equally important, we observed that inequality and population ageing magnify contextual vulnerability. These results add to evidence that socioeconomic conditions and demographic structures play a central role in environmental health risks \citep{murray2020global,luthi2023rapid,brauer2024global,fuller2022pollution,achebak2023drivers}.

From a methodological perspective, we report three complementary heat estimands: marginal associations from main-effects models; modification ratios from interaction coefficients; and stratum-specific combined effects $\exp(\beta_H+\delta_j)$ from joint posterior draws.

Taken together, these results support heat-health prevention plans that retain temperature-based warnings while also integrating ozone (and, for heatwaves, NO$_2$) forecasts and small-area indicators of inequality and ageing. In line with Batibeniz et al.\ \citep{batibeniz2025rapid} and Chen et al.\ \citep{chen2025trends}, compound heat--pollution exposures and social adaptation measures should be addressed jointly.

Our research is subject to several limitations. Its ecological and observational nature limits individual-level and causal inference; residual ecological bias may remain despite structured and unstructured random effects. The 95th percentile maximum-temperature threshold is a meteorological cut-off rather than a physiological one \citep{ministeriosanidad2023plan,tobias2023heat}. Some exposure misclassification from residential ABS assignment is inevitable and expected to be largely non-differential. Lag selection used main-effects-only nested models (same random-effect structure and likelihood as the finals) on a common sample; interaction and stratum-specific contrasts were estimated afterwards at the selected lags. Finally, some ZINB fits produced numerical warnings under INLA; estimates were returned but sensitivity checks remain advisable.

Despite these caveats, strengths include complete ABS coverage, hierarchical spatiotemporal dependence, and explicit reporting of marginal, interaction, and stratum-combined heat estimands.

In conclusion, mortality during extreme heat in Catalonia is associated with heat indicators themselves, amplified by ozone (and heatwave$\times$NO$_2$) and shaped by contextual socioeconomic vulnerability. Heat-health adaptation should therefore combine temperature warnings with air quality management and equity-oriented targeting of ageing and disadvantaged areas.

\section*{Declarations}

\paragraph{Lead contact}
Requests for further information and resources should be directed to and will be fulfilled by the lead contact, Marc Saez (\texttt{marc.saez@udg.edu}).

\paragraph{Data and code availability}
Regarding mortality data, in accordance with Article 2 of Regulation 223/2009 of the Council and the European Parliament on European Statistics; Articles 13, 17.3, and 17.4 of the Spanish Law on the Public Statistical Function; Article 25.1 of Spanish Organic Law 3/2018, of December 5, on the Protection of Personal Data and Guarantee of Digital Rights; and Regulation (EU) 2016/679 General Data Protection Regulation, the mortality databases of the National Institute of Statistics (INE) are subject to Statistical Confidentiality. Therefore, there are restrictions on their transfer to third parties, and data are not publicly available. However, following the approval of a research proposal and the signing of a data access agreement, anonymized data will be made available upon reasonable request to the corresponding author.

The remaining data are open data:
\begin{itemize}
\item Meteorological data \citep{xema2025data}.
\item Air pollution data \citep{xema2025airquality}.
\item Socioeconomic data \citep{ine2022household}.
\item Population data \citep{ine2025continuous}.
\end{itemize}
The code will be available at \url{https://www.researchprojects.es}.

Any additional information required to reanalyse the data reported in this paper is available from the lead contact upon request.

\paragraph{Acknowledgements}
This study was carried out within the `Collaboration agreement between Dipsalut and the University of Girona to promote and develop joint work in relation to health inequalities generated by the social determinants of health and environmental factors'; and within the `Health Inequalities and COVID-19' and `Atlas of Social and Environmental Determinants of Health' subprograms of CIBER of Epidemiology and Public Health (CIBERESP). We are grateful to the editor and to two anonymous reviewers for their constructive comments on an earlier version of this work, which have undoubtedly helped us to improve it. The usual disclaimer applies.

\paragraph{Authors' contributions}
Conceptualization: MAB and MS; Methodology: MS and MAB; Resources: MS and MAB; Data curation: MAB and MS; Analysis: DS, MSo, MS and MAB; Writing---original draft preparation: MS, MAB, DS and MSo; Writing---review and editing: MS, MAB, DS, AR-C and MSo; Project administration: MAB and MS; Funding acquisition: MAB and MS.

\paragraph{Declaration of interests}
The authors declare no competing interests.

The manuscript is an original contribution that has not been published before, whole or in part, in any format, including electronically. All authors will disclose any actual or potential conflict of interest including any financial, personal, or other relationships with other people or organizations that could inappropriately influence or be perceived to influence their work, within three years of beginning the submitted work.

\paragraph{Declaration of generative AI and AI-assisted technologies in the writing process}
During the preparation of this work the authors used generative AI in order to assist with manuscript drafting, editing, and formatting of results tables. After using this tool, the authors reviewed and edited the content as needed and take full responsibility for the content of the publication.

\paragraph{Supplemental information}
Supplemental information: Tables S1a--S1b (lag WAIC), S2a--S2b (pollutant$\times$heat cell counts), S3 (Bayesian FDR); Figures S1--S4 (minimum temperature, relative humidity, PM$_{10}$, NO$_2$ time series).

% Primary tables: 1a temperature; 1b humidity/pollutants; 1d heat-event prevalence;
% 2summer modelling-sample mortality; 2a--2b all-year demographics (unchanged);
% 3a--3d extreme heat (main pollutants/SES; interactions pollutants/SES);
% 4a--4d heatwave (same split).

\paragraph{Ethics approval and consent to participate}
The study used an ecological design, with data aggregated at the primary health care area (ABS) level. Therefore, ethics approval and consent to participate are not applicable.

\paragraph{Consent for publication}
Not applicable.

\paragraph{Funding}
This work was partially financed by AGAUR, the Department of Climate Action, Food and Rural Agenda, and by the Department of Research and Universities, both part of the Government of Catalonia (\textit{Generalitat de Catalunya}) (grant number 2023 CLIMA 00037). The funding bodies did not participate in the design or conduct of the study; the collection, management, analysis, or interpretation of the data; or the preparation, review, and approval of the manuscript.

\bibliographystyle{elsarticle-num}
\bibliography{bibfile}

\newpage
\input{tables/table1a}

\input{tables/table1b}

\input{tables/table1d}

\input{tables/table2summer}

\input{tables/table2a}
\input{tables/table2b}
\input{tables/table3a}

\input{tables/table3b}

\input{tables/table3c}

\input{tables/table3d}

\input{tables/table3e}

\input{tables/table3f}

\input{tables/table4a}

\input{tables/table4b}

\input{tables/table4c}

\input{tables/table4d}

\input{tables/table4e}

\input{tables/table4f}

\begin{figure}[htbp]
\centering
\begin{subfigure}{0.48\textwidth}
\includegraphics[width=\linewidth]{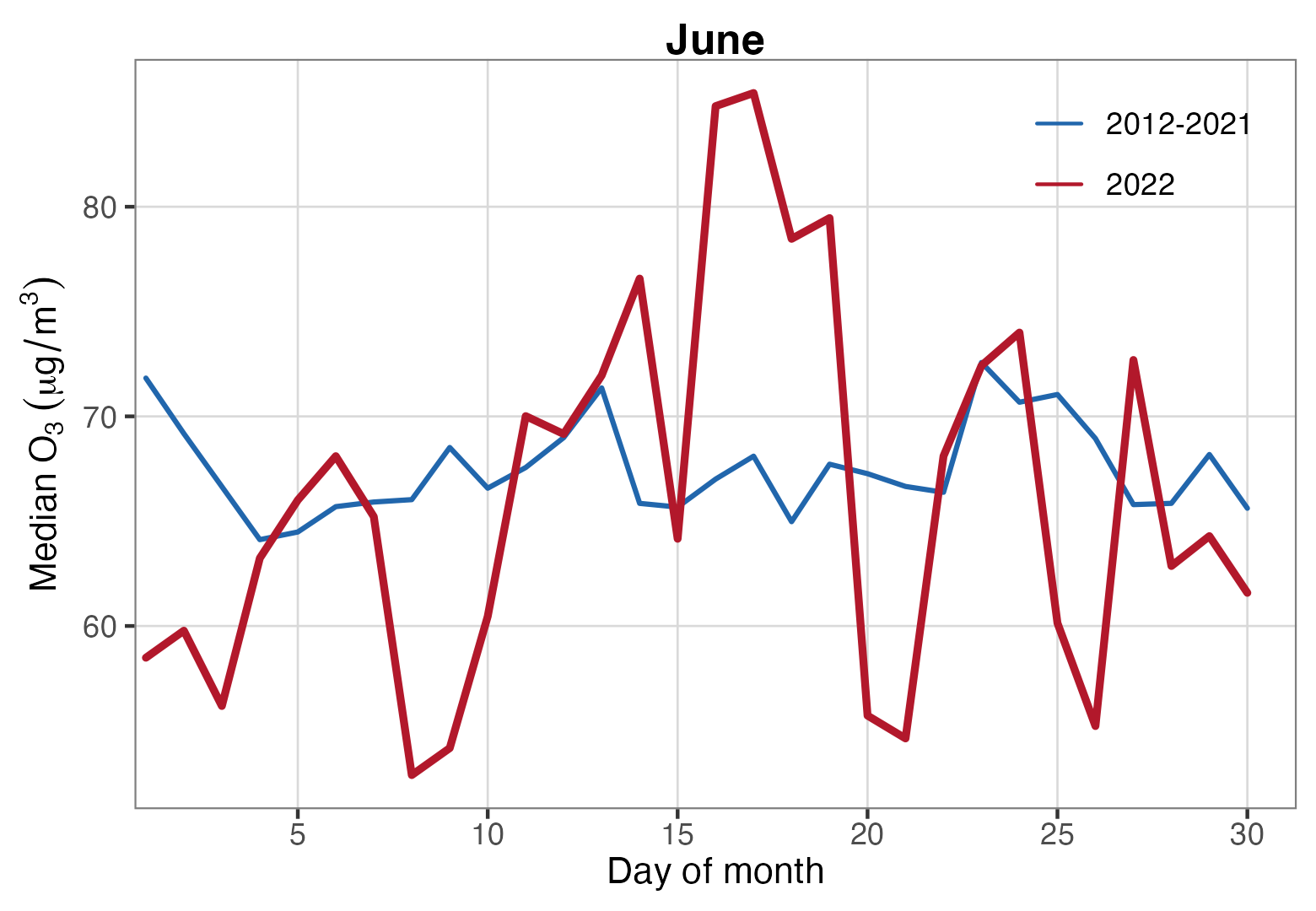}
\end{subfigure}\hfill
\begin{subfigure}{0.48\textwidth}
\includegraphics[width=\linewidth]{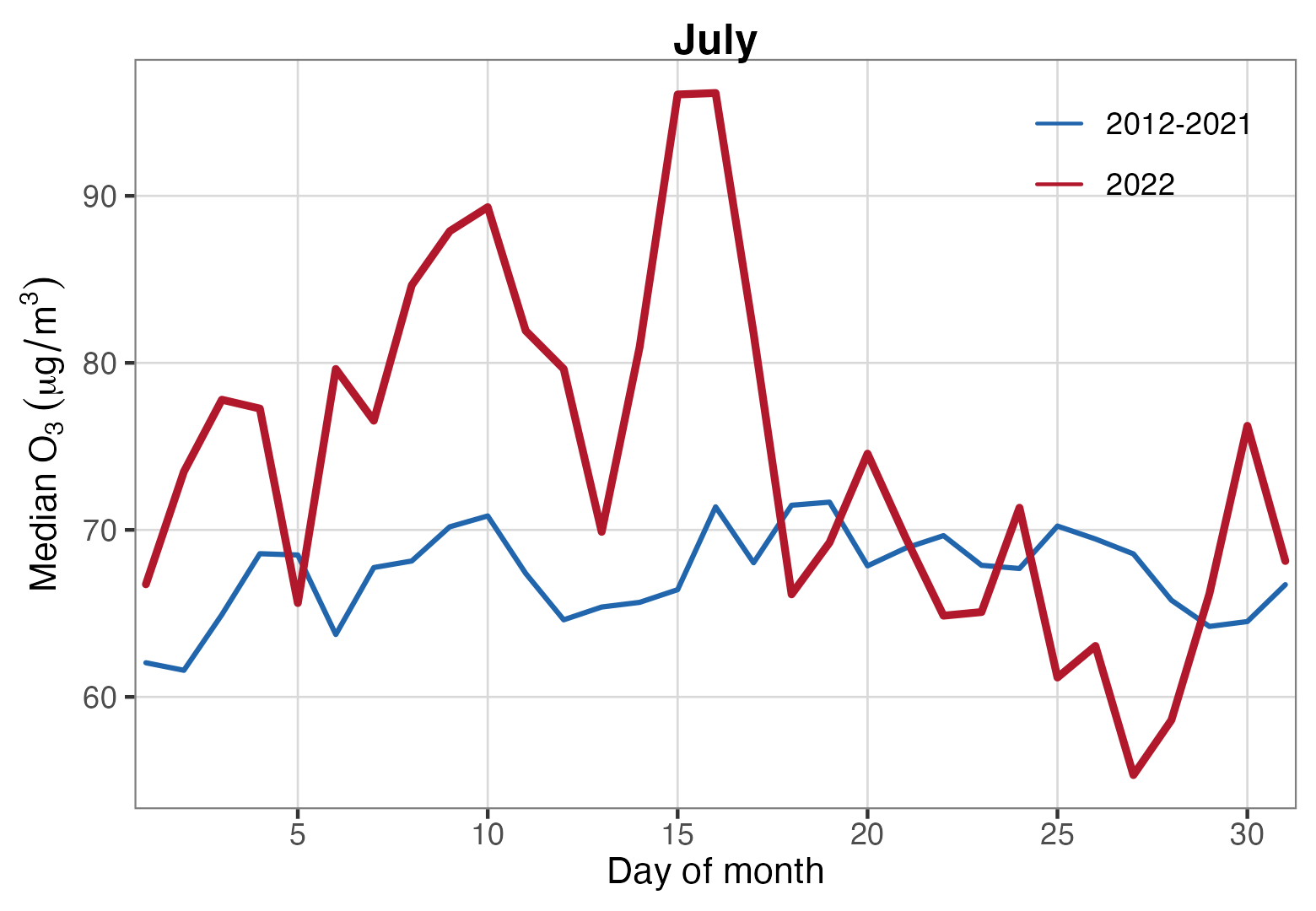}
\end{subfigure}\\[0.6em]
\begin{subfigure}{0.48\textwidth}
\includegraphics[width=\linewidth]{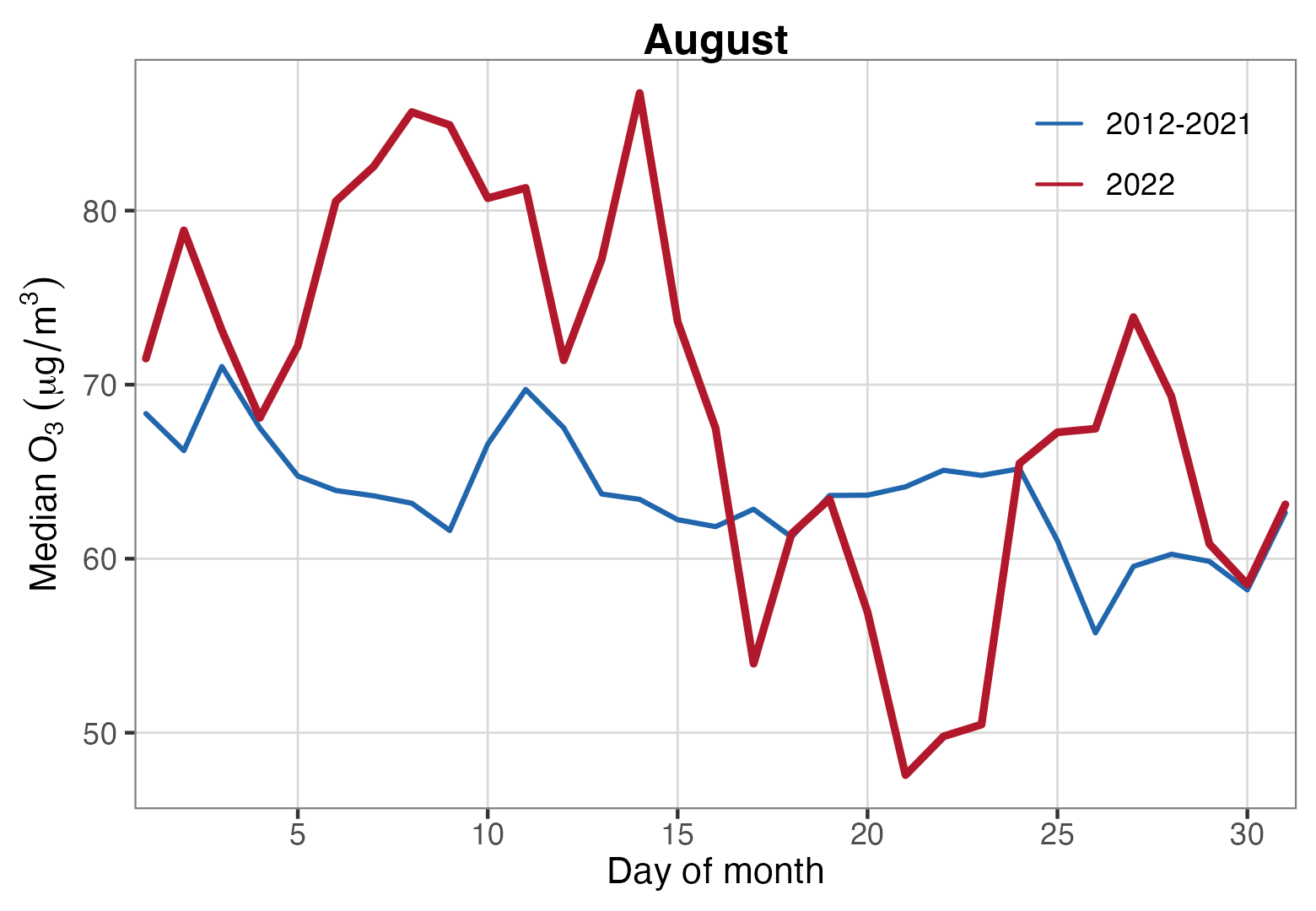}
\end{subfigure}\hfill
\begin{subfigure}{0.48\textwidth}
\includegraphics[width=\linewidth]{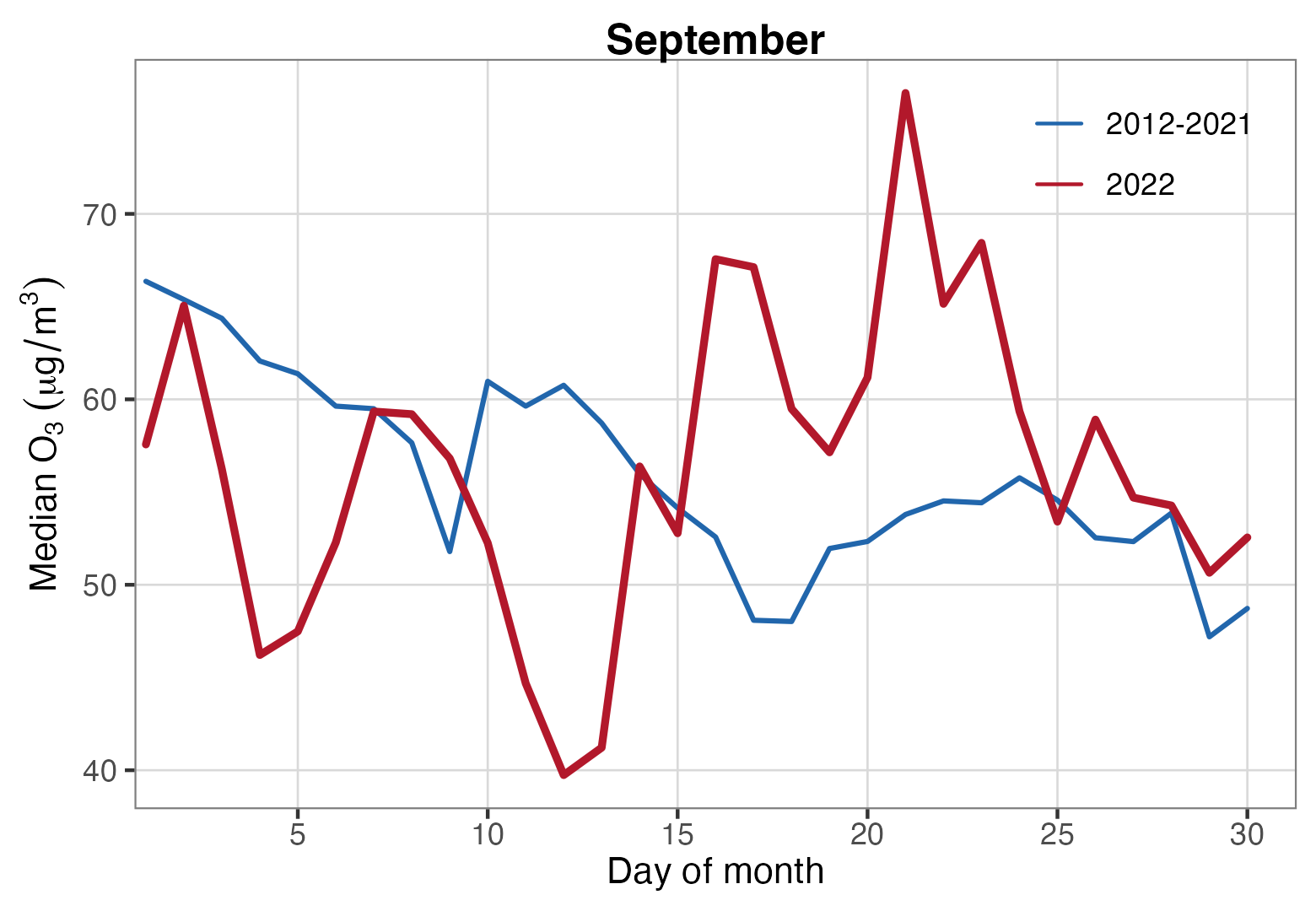}
\end{subfigure}
\caption{Temporal evolution of median ozone levels in 2012--2021 and in 2022.}
\label{fig:o3-ts}
\end{figure}

\end{document}

\begin{frontmatter}

\title{Supplementary information:\\
A spatiotemporal Bayesian hierarchical model of heat-related mortality in Catalonia, Spain (2012--2022)}

\author{David Solano}
\author{Marta Solans}
\author{Xavier Perafita}
\author{Anna Ruiz-Comellas}
\author{Marc Saez}
\author{Maria A. Barcel\'o}

\begin{abstract}
This supplement accompanies the nested space--time zero-inflated negative binomial analysis of summer heat-related mortality in Catalonia's basic health areas (ABS), 2012--2022. It reports lag-selection diagnostics, pollutant$\times$heat cell counts, Bayesian FDR summaries for interaction terms, and additional environmental time-series figures.
\end{abstract}

\end{frontmatter}

\section*{Contents}
\begin{itemize}
\item Tables~S1a--S1b: lag selection by WAIC
\item Tables~S2a--S2b: pollutant$\times$heat cell counts
\item Table~S3: Bayesian FDR discoveries (interactions, $\alpha=0.05$)
\item Figures~S1--S4: day-of-month medians for minimum temperature, relative humidity, PM$_{10}$ and NO$_2$
\end{itemize}

\section{Lag selection}
Lag screening used main-effects-only nested spatiotemporal ZINB models (Mat\'ern SPDE$\times$year, month nested within year, IID ABS frailty; outcome: all-age summer deaths), fitted on a common summer sample so that WAIC is comparable across lags. The minimum-WAIC lag was retained for the final models (extreme heat lag~2; heatwave lag~3).

\renewcommand{\thetable}{S1a}
\input{si_tables/tableS1a}

\renewcommand{\thetable}{S1b}
\input{si_tables/tableS1b}

\section{Pollutant$\times$heat cell counts}
Sparse co-occurrence of heat with the highest ozone category motivated collapsing O$_3$ at $\geq 100\,\mu$g/m$^3$ (WHO one-day AQG) in the analysis models.

\renewcommand{\thetable}{S2a}
\input{si_tables/tableS2a}

\renewcommand{\thetable}{S2b}
\input{si_tables/tableS2b}

\section{Multiplicity (Bayesian FDR)}
Posterior probabilities $\mathrm{Prob}(|\log\mathrm{RR}|>0)$ for interaction coefficients were converted to local false-discovery rates and thresholded at target FDR $\alpha=0.05$.

\renewcommand{\thetable}{S3}
\input{si_tables/tableS3}

\section{Additional environmental time series}
Figures~S1--S4 show ABS-median day-of-month trajectories in 2012--2021 versus 2022 for minimum temperature, relative humidity, PM$_{10}$ and NO$_2$ (June--September).

\begin{figure}[htbp]
\centering
\begin{subfigure}{0.48\textwidth}
\includegraphics[width=\linewidth]{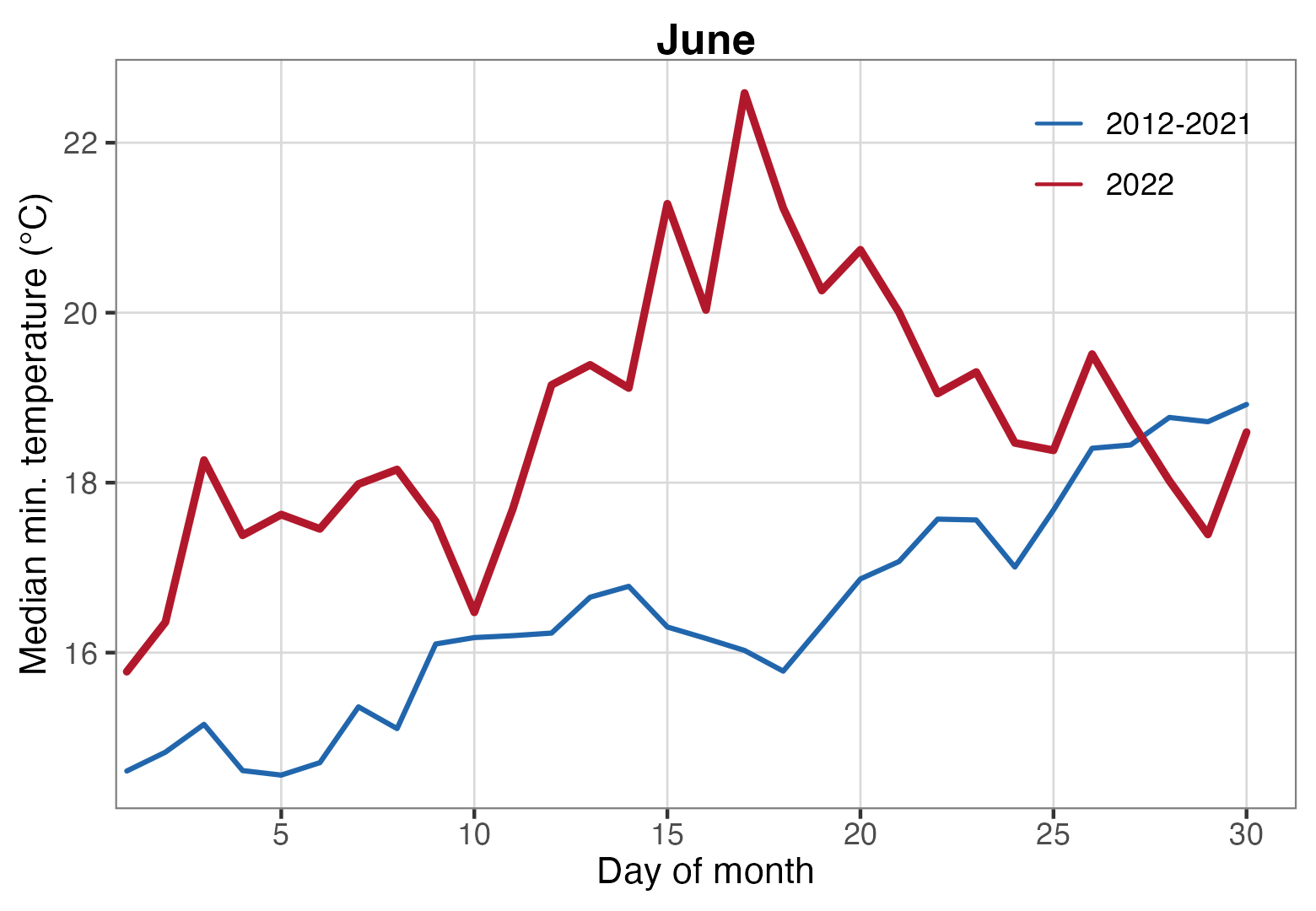}
\end{subfigure}\hfill
\begin{subfigure}{0.48\textwidth}
\includegraphics[width=\linewidth]{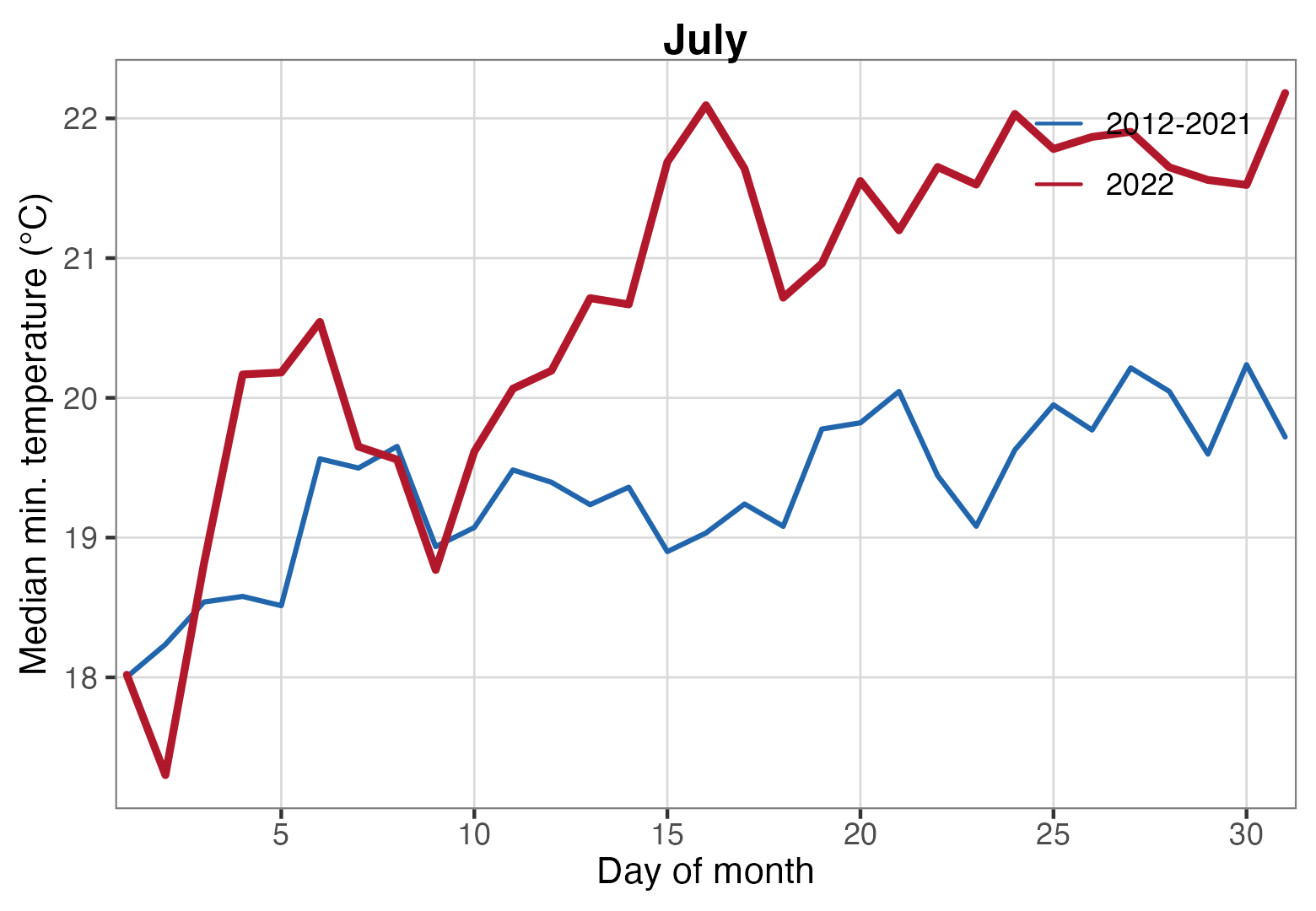}
\end{subfigure}\\[0.5em]
\begin{subfigure}{0.48\textwidth}
\includegraphics[width=\linewidth]{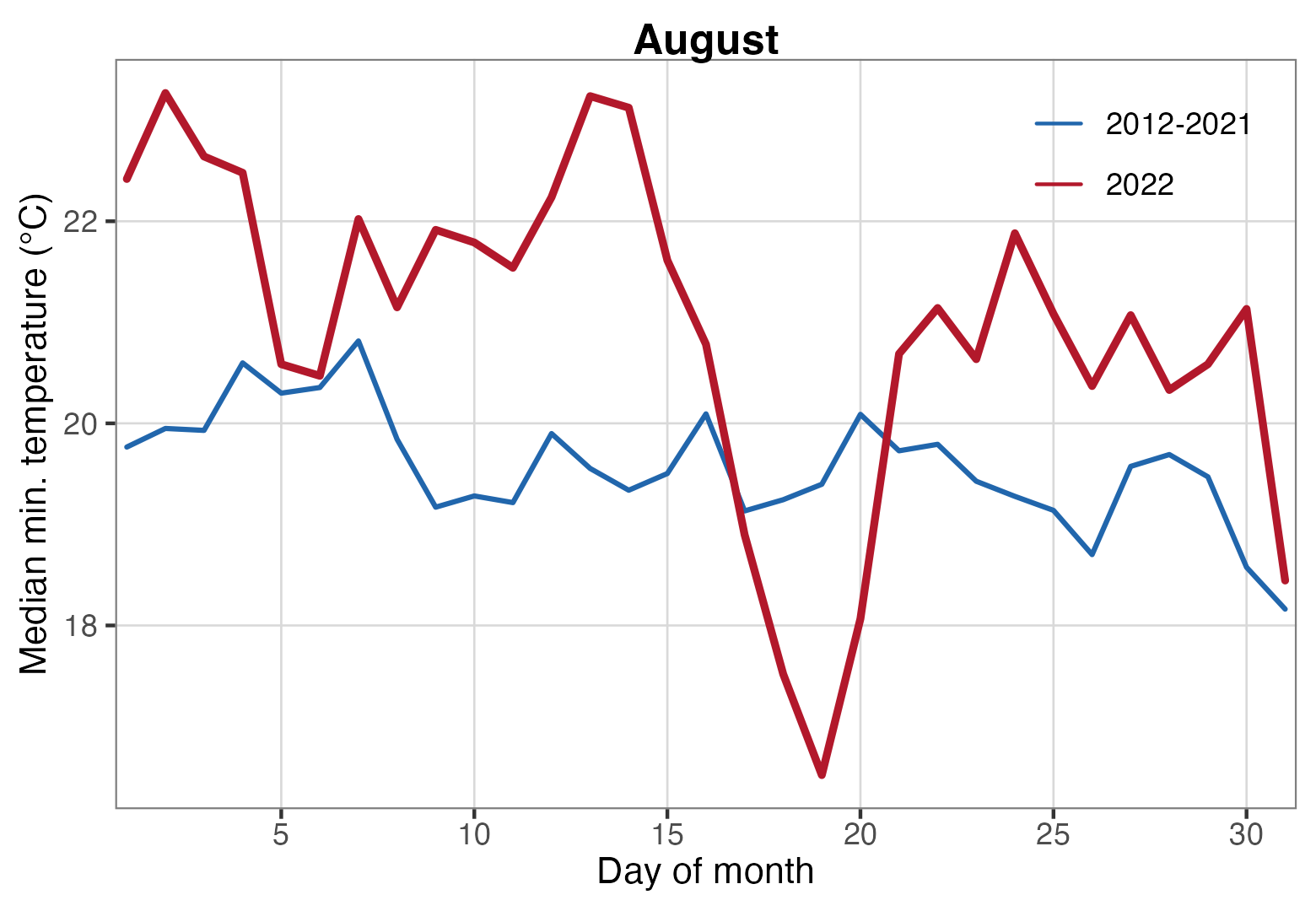}
\end{subfigure}\hfill
\begin{subfigure}{0.48\textwidth}
\includegraphics[width=\linewidth]{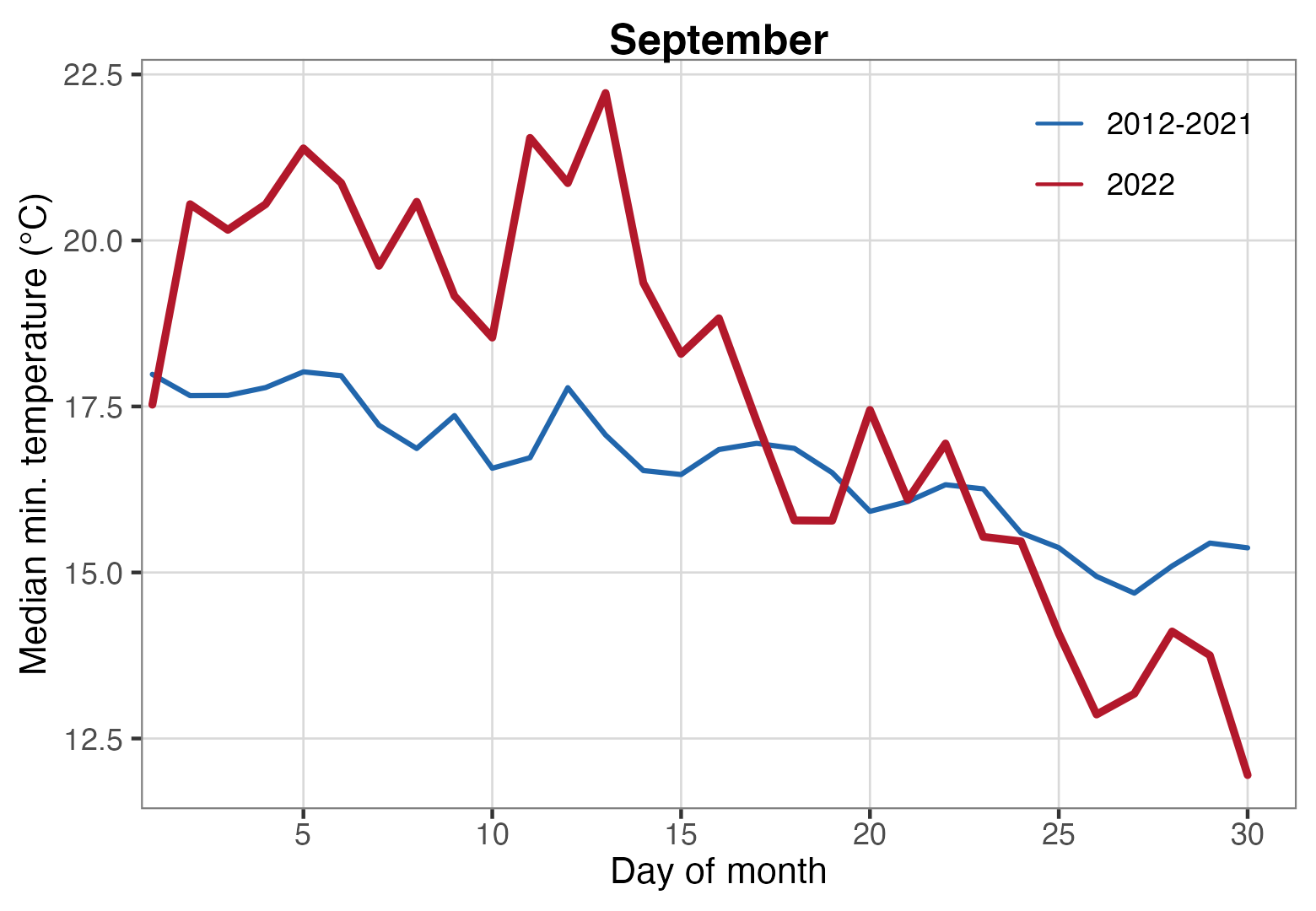}
\end{subfigure}
\caption{Temporal evolution of median daily minimum temperature in 2012--2021 and in 2022.}
\label{fig:S1}
\end{figure}

\begin{figure}[htbp]
\centering
\begin{subfigure}{0.48\textwidth}
\includegraphics[width=\linewidth]{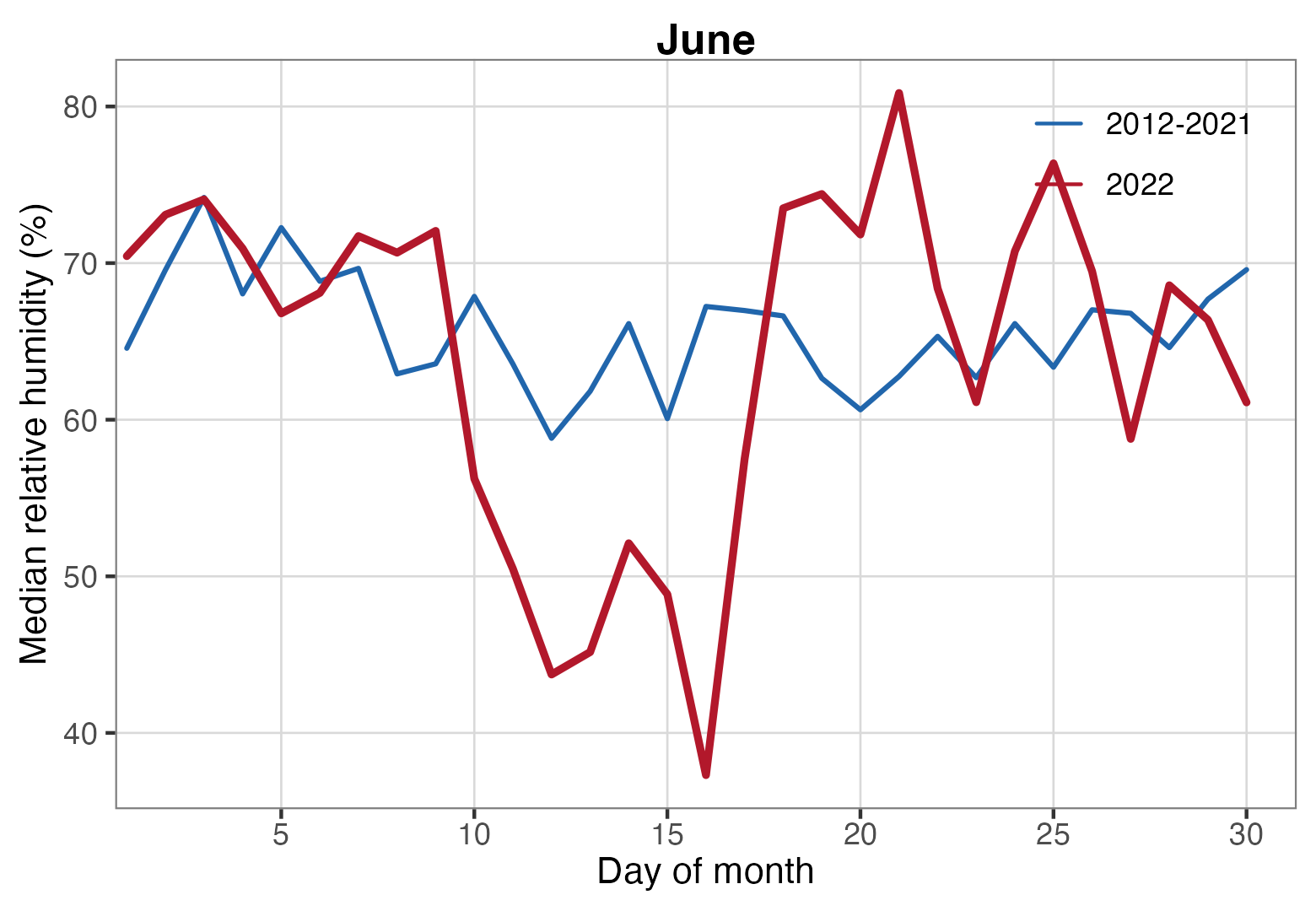}
\end{subfigure}\hfill
\begin{subfigure}{0.48\textwidth}
\includegraphics[width=\linewidth]{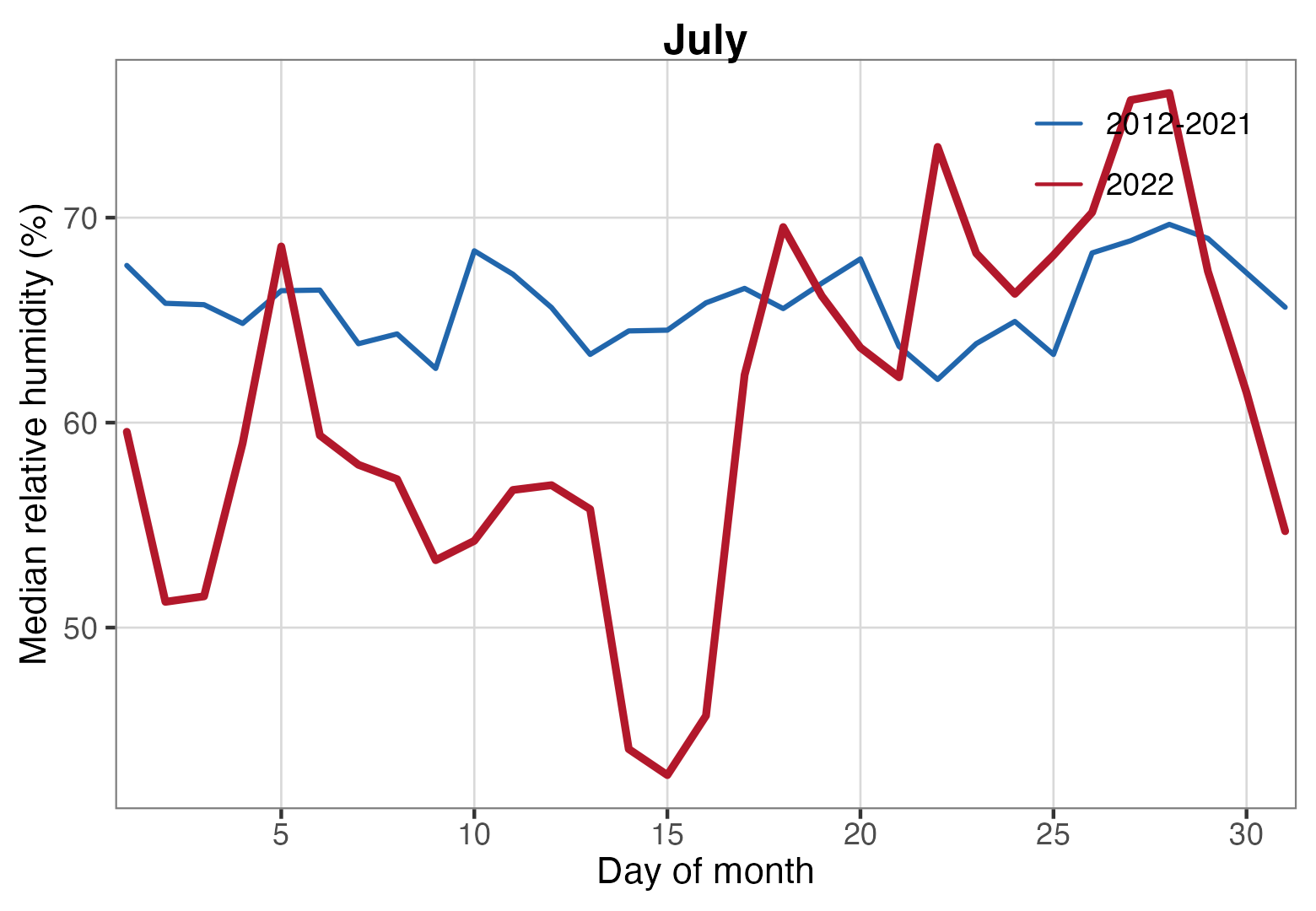}
\end{subfigure}\\[0.5em]
\begin{subfigure}{0.48\textwidth}
\includegraphics[width=\linewidth]{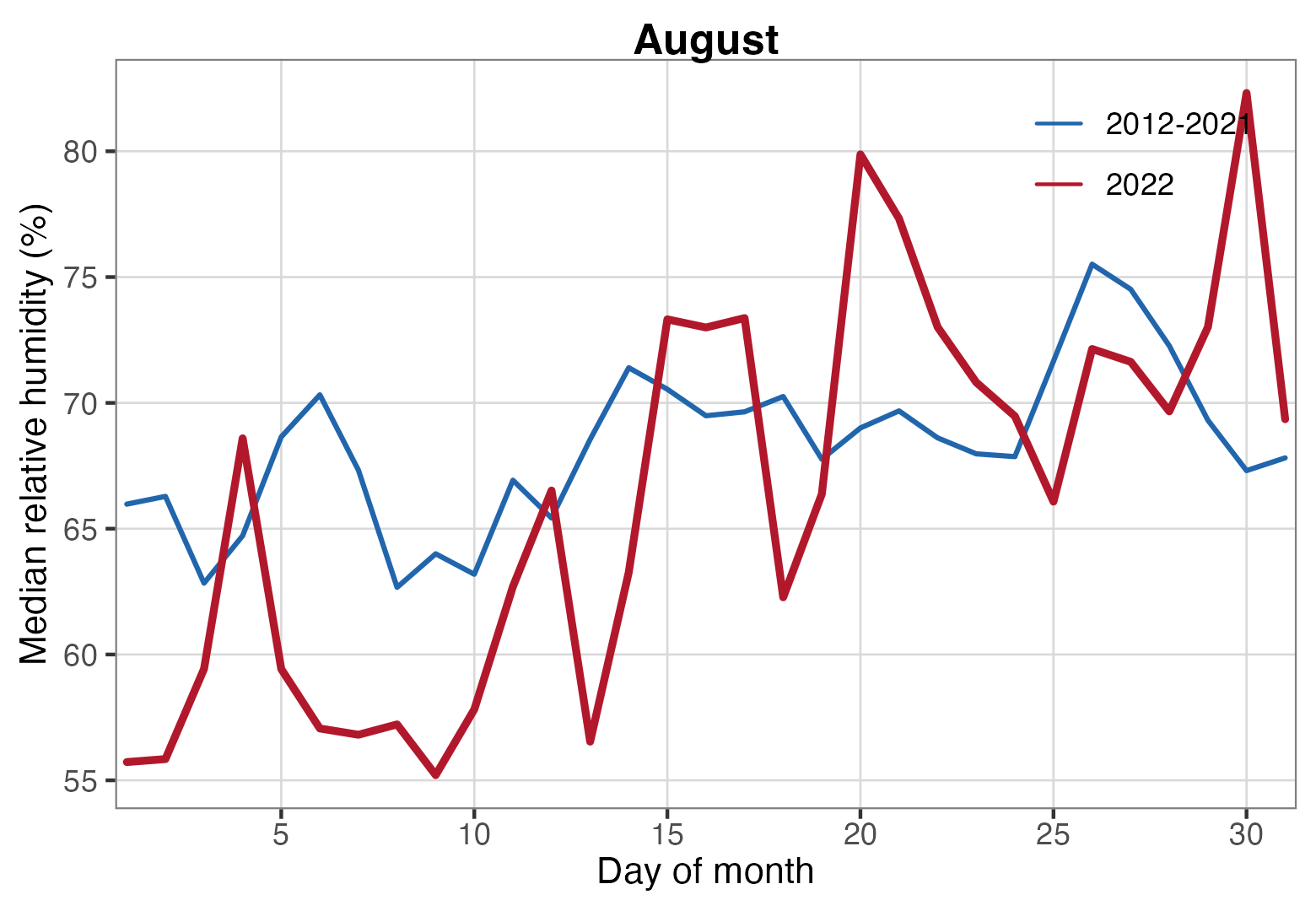}
\end{subfigure}\hfill
\begin{subfigure}{0.48\textwidth}
\includegraphics[width=\linewidth]{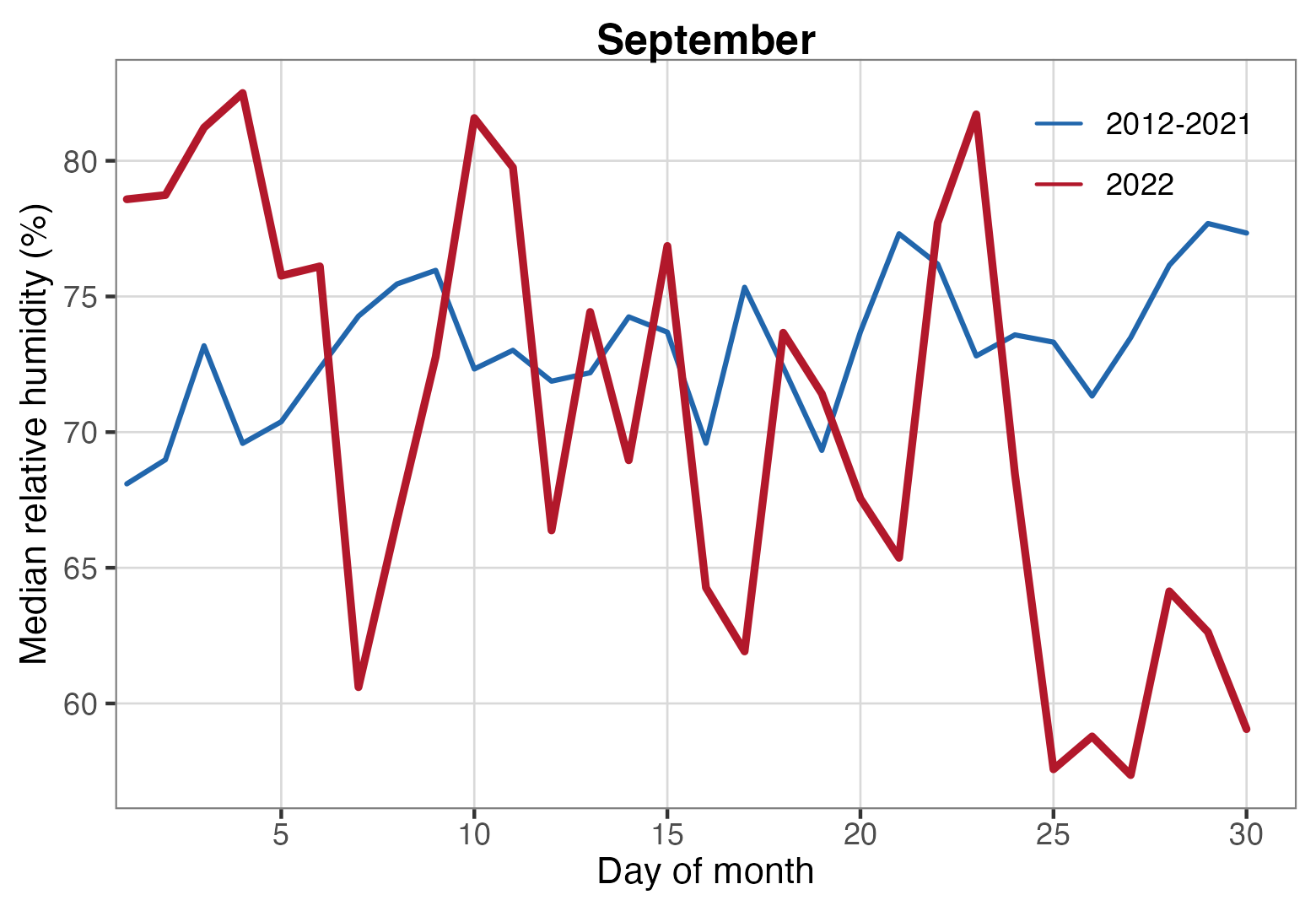}
\end{subfigure}
\caption{Temporal evolution of median relative humidity in 2012--2021 and in 2022.}
\label{fig:S2}
\end{figure}

\begin{figure}[htbp]
\centering
\begin{subfigure}{0.48\textwidth}
\includegraphics[width=\linewidth]{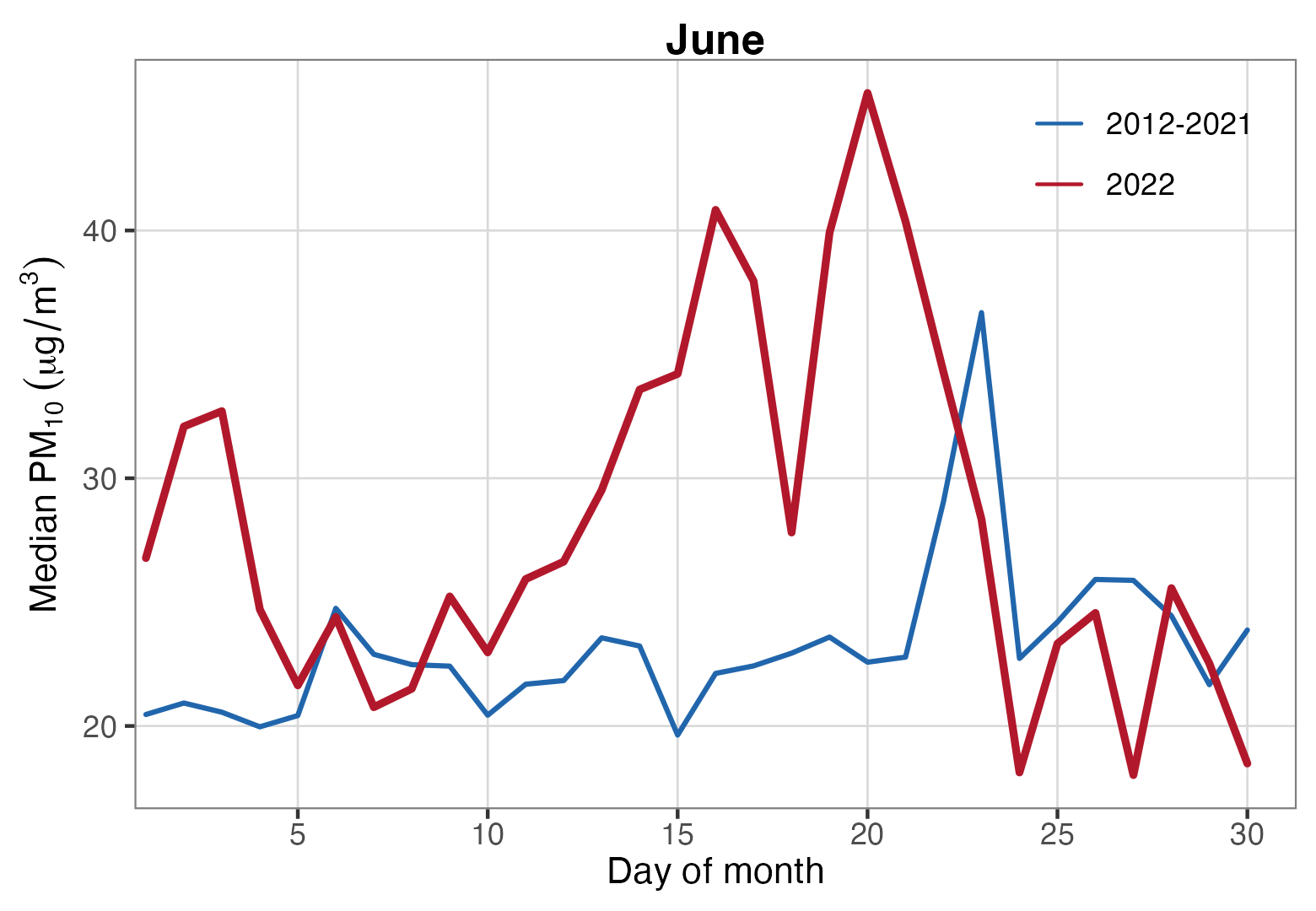}
\end{subfigure}\hfill
\begin{subfigure}{0.48\textwidth}
\includegraphics[width=\linewidth]{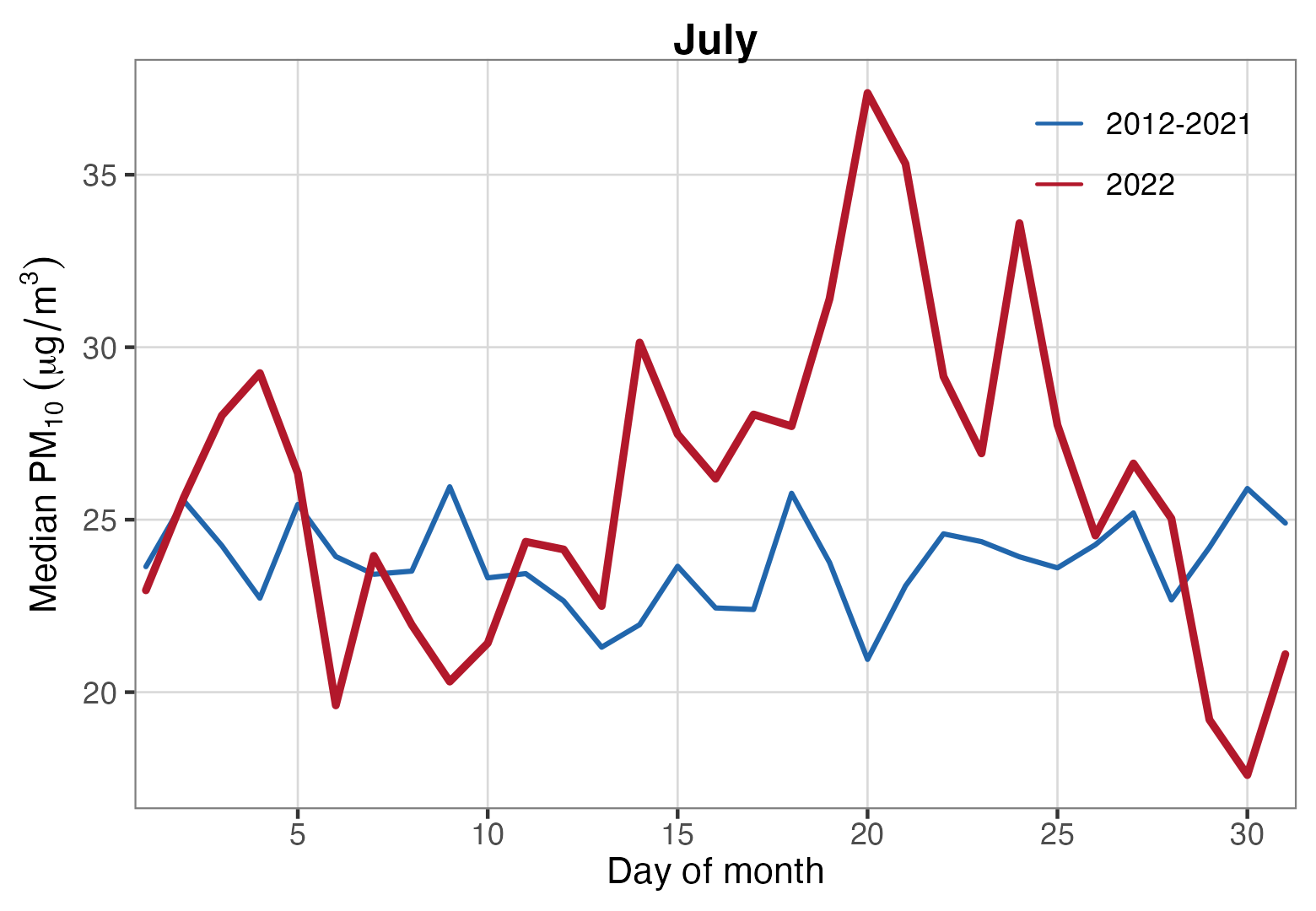}
\end{subfigure}\\[0.5em]
\begin{subfigure}{0.48\textwidth}
\includegraphics[width=\linewidth]{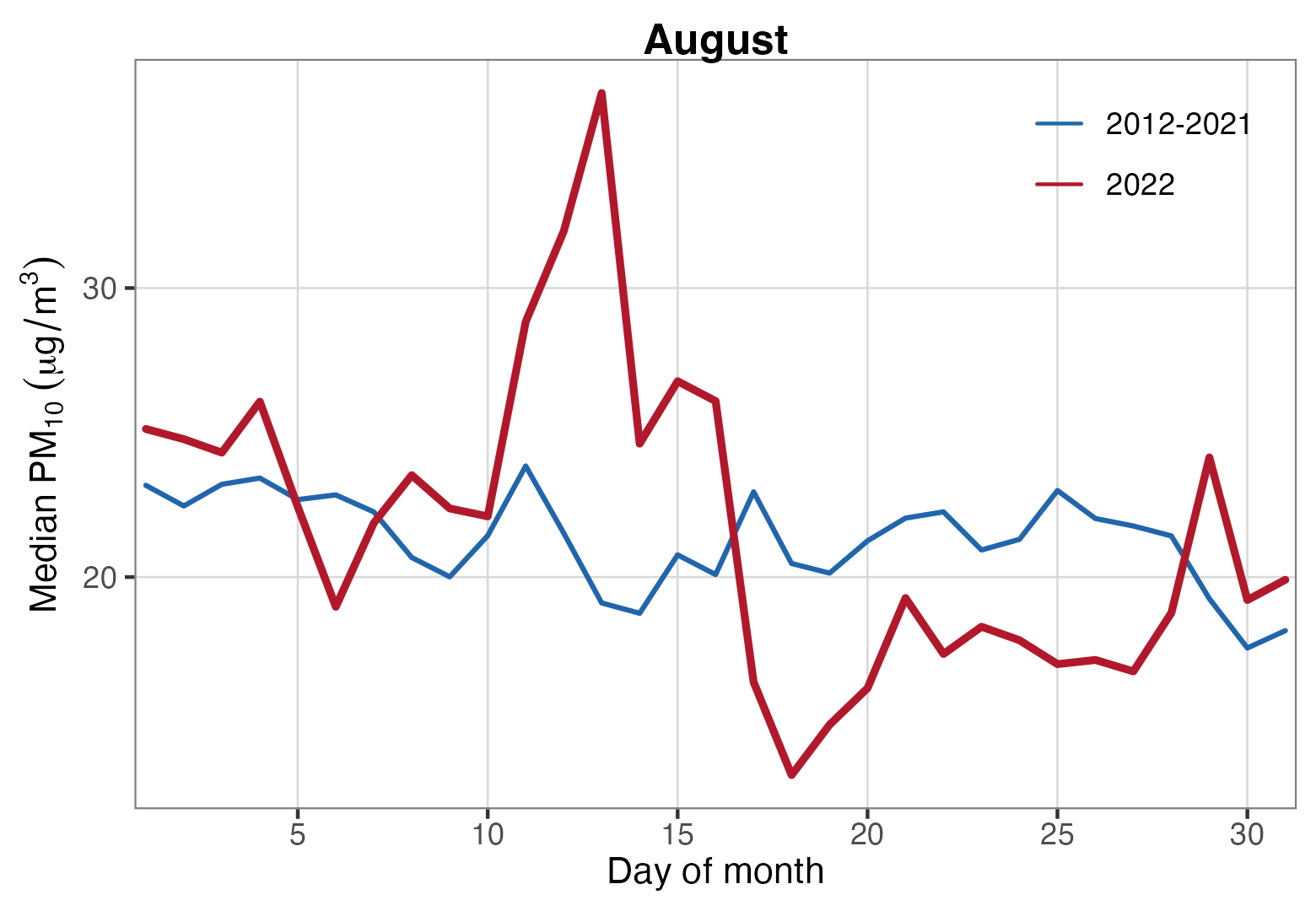}
\end{subfigure}\hfill
\begin{subfigure}{0.48\textwidth}
\includegraphics[width=\linewidth]{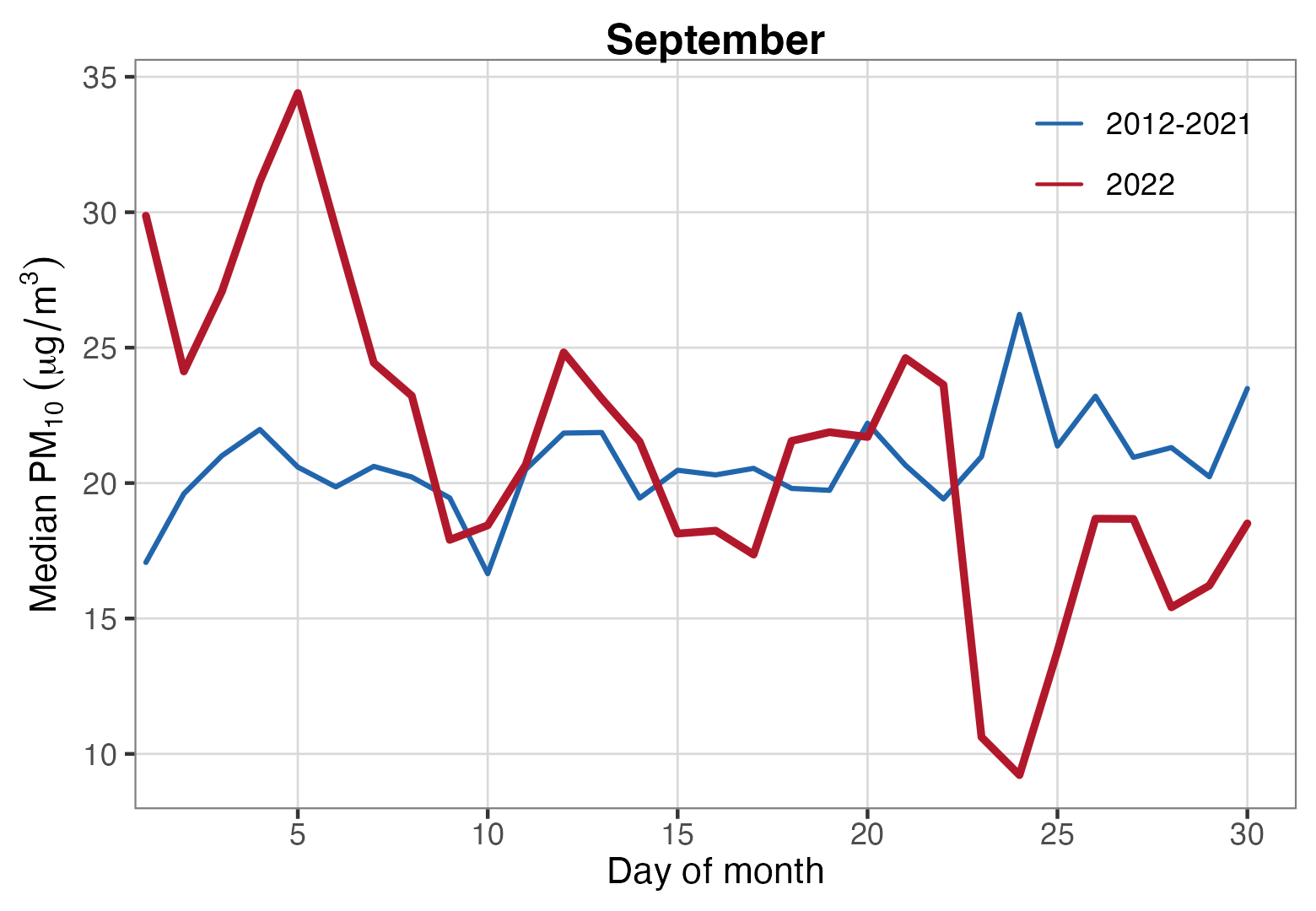}
\end{subfigure}
\caption{Temporal evolution of median PM$_{10}$ in 2012--2021 and in 2022.}
\label{fig:S3}
\end{figure}

\begin{figure}[htbp]
\centering
\begin{subfigure}{0.48\textwidth}
\includegraphics[width=\linewidth]{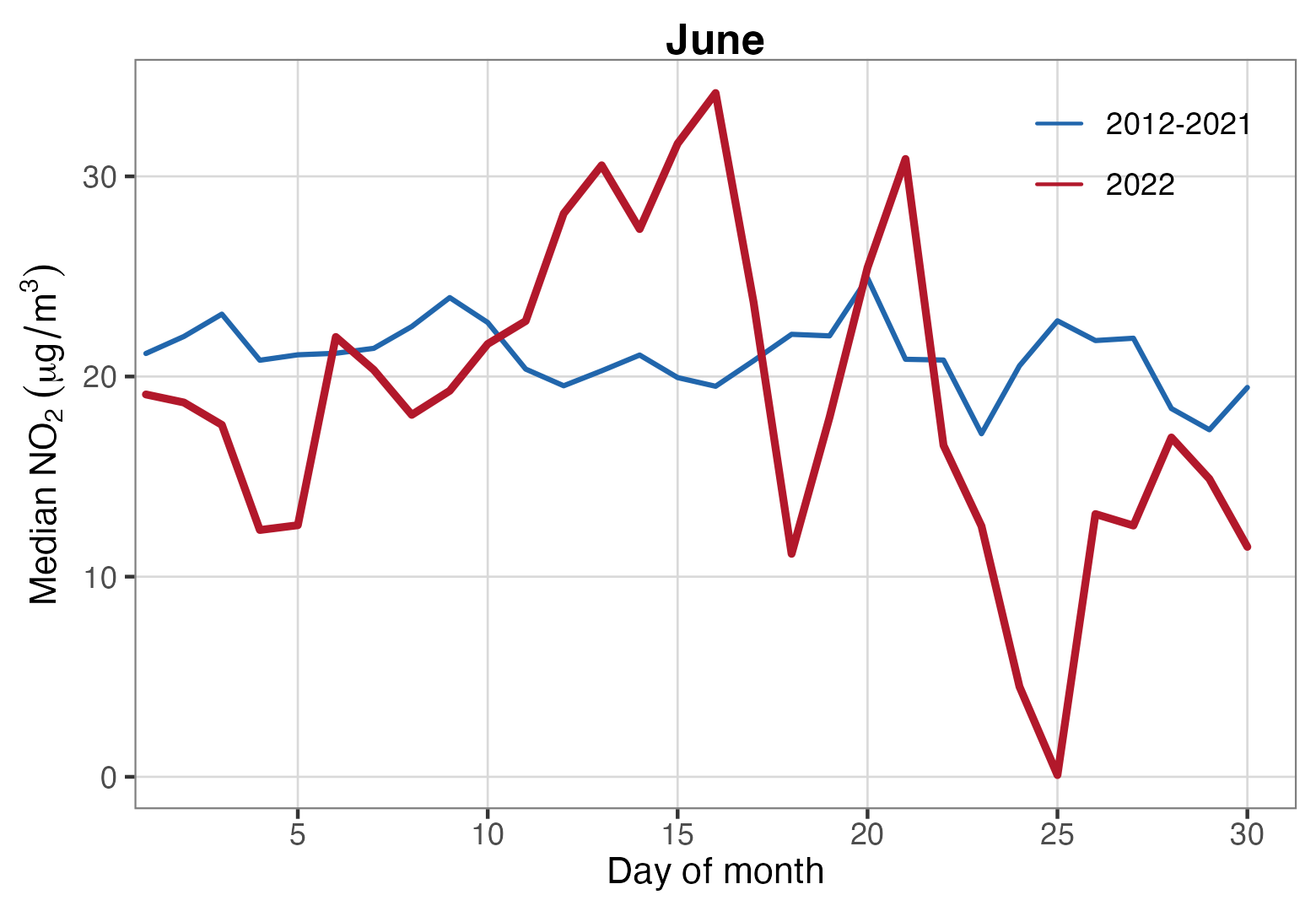}
\end{subfigure}\hfill
\begin{subfigure}{0.48\textwidth}
\includegraphics[width=\linewidth]{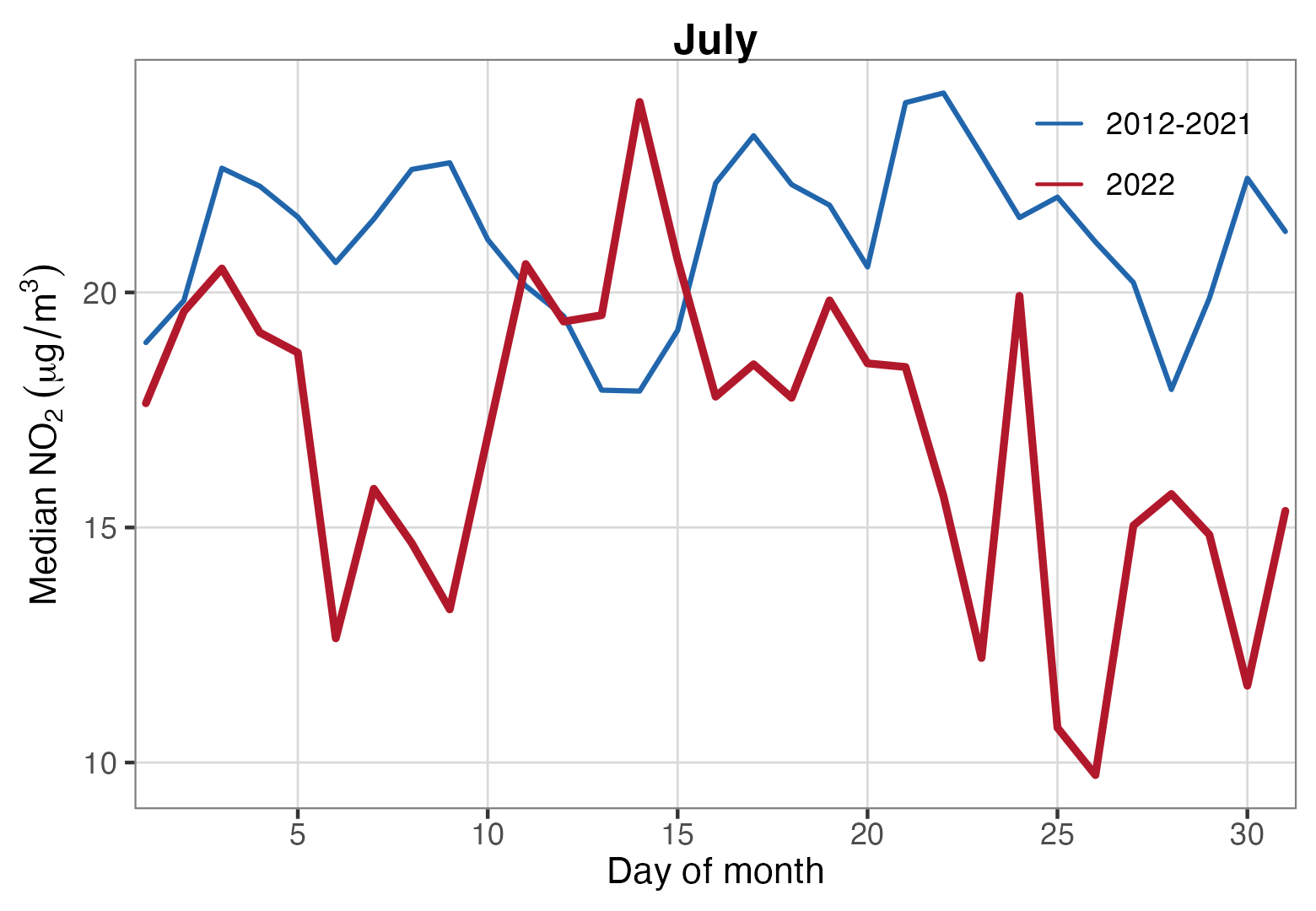}
\end{subfigure}\\[0.5em]
\begin{subfigure}{0.48\textwidth}
\includegraphics[width=\linewidth]{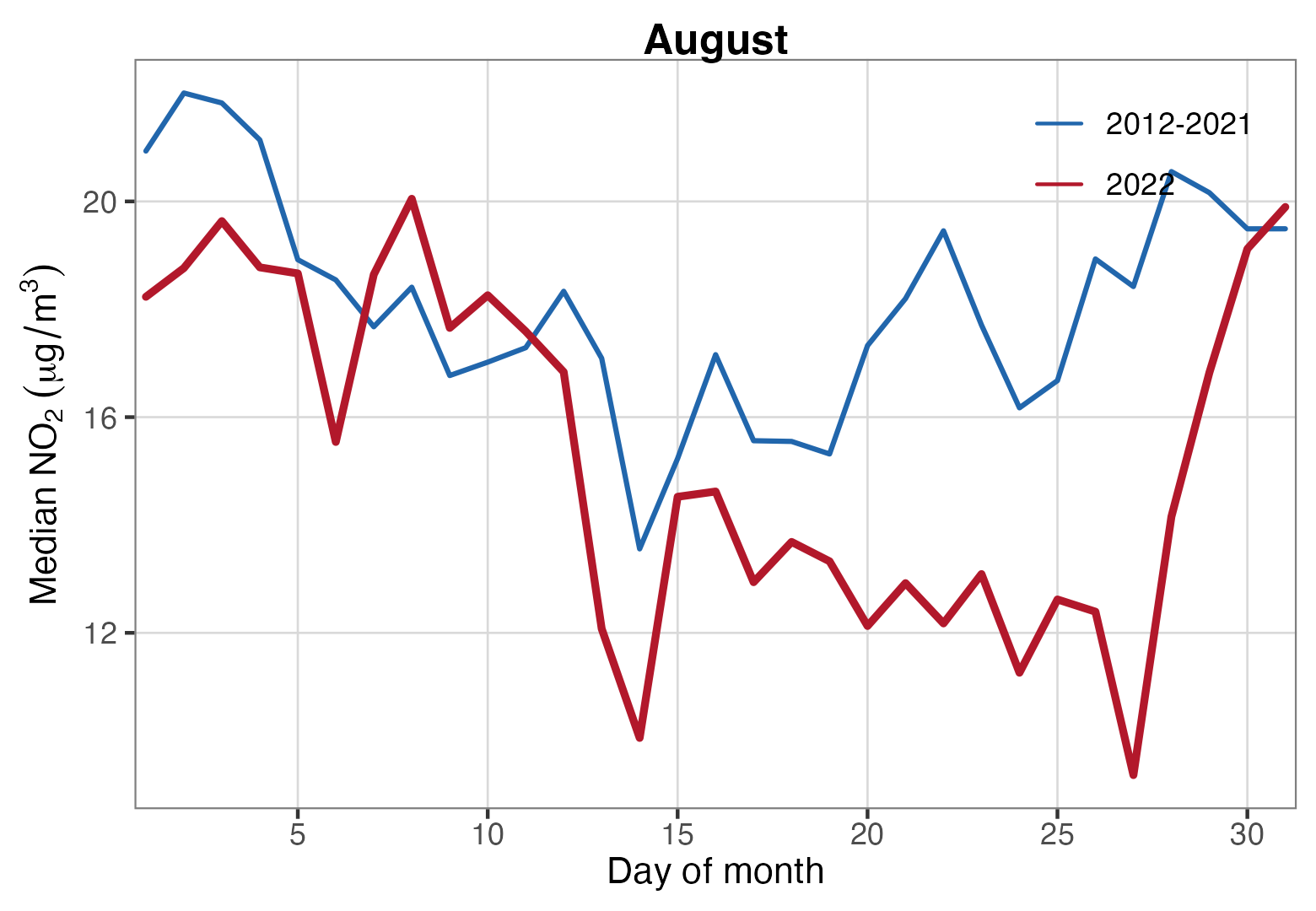}
\end{subfigure}\hfill
\begin{subfigure}{0.48\textwidth}
\includegraphics[width=\linewidth]{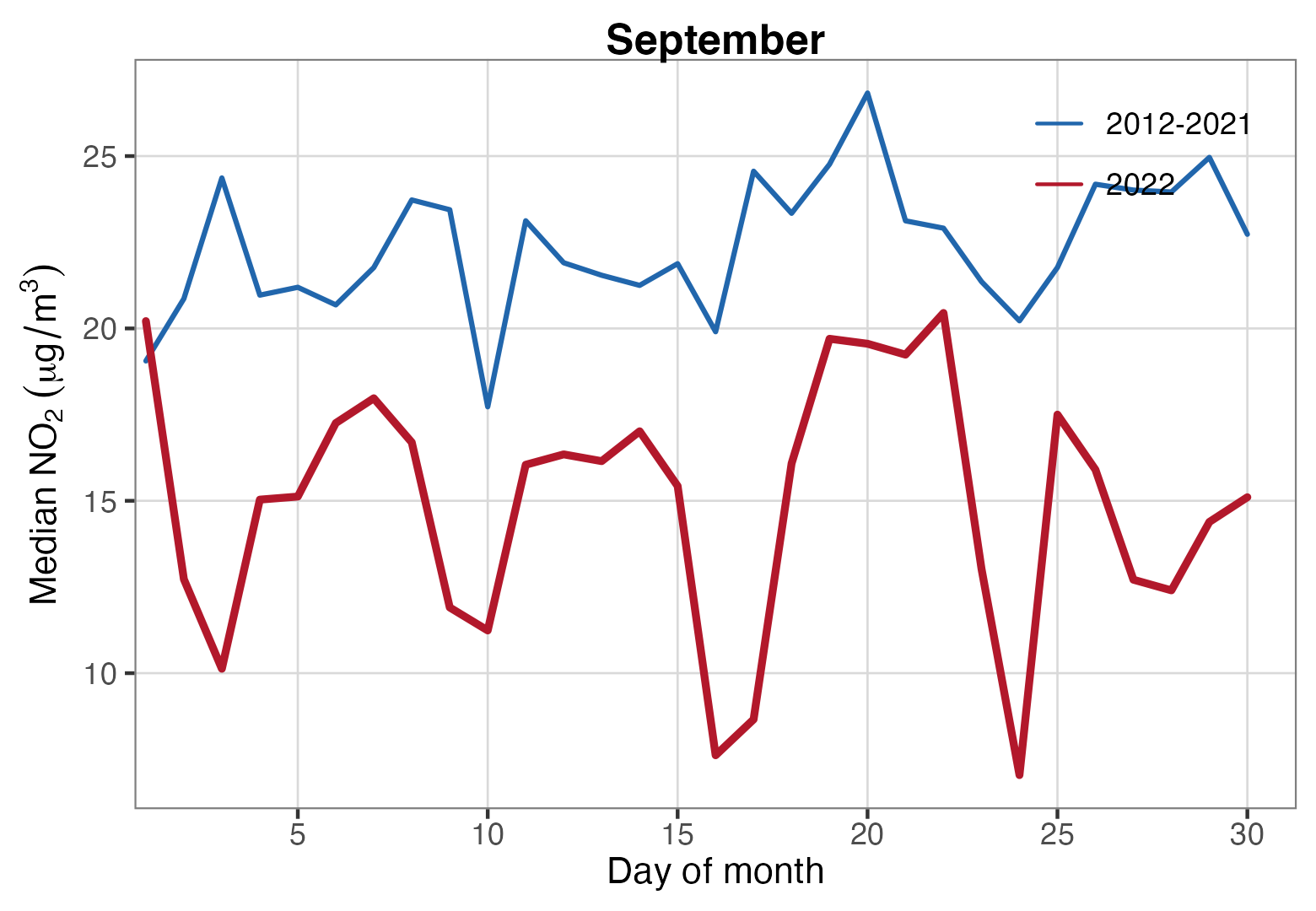}
\end{subfigure}
\caption{Temporal evolution of median NO$_2$ in 2012--2021 and in 2022.}
\label{fig:S4}
\end{figure}

%% file: tables/table1a.tex
\begin{sidewaystable}[htbp]
\centering
\caption{Descriptive statistics of temperature. Catalonia, summer months 2012--2022 (complete environmental sample).}
\label{tab:1a}
\scriptsize
\setlength{\tabcolsep}{2.5pt}

\resizebox{0.95\textheight}{!}{\begin{tabular}{@{}lcccccccc@{}}
\toprule
 & \multicolumn{2}{c}{June} & \multicolumn{2}{c}{July} & \multicolumn{2}{c}{August} & \multicolumn{2}{c}{September} \\
\cmidrule(lr){2-3}\cmidrule(lr){4-5}\cmidrule(lr){6-7}\cmidrule(lr){8-9}
 & 2012--2021 & 2022 & 2012--2021 & 2022 & 2012--2021 & 2022 & 2012--2021 & 2022 \\
\midrule
\multicolumn{9}{l}{\textit{Maximum temperature ($^\circ$C)}} \\
All days & \makecell[c]{26.94 (3.66) \\ 26.94 [24.75, 29.19]} & \makecell[c]{29.73 (3.03) \\ 29.33 [27.92, 31.47]} & \makecell[c]{29.77 (2.86) \\ 29.90 [28.11, 31.54]} & \makecell[c]{31.61 (2.55) \\ 31.61 [30.16, 33.15]} & \makecell[c]{29.64 (3.13) \\ 29.75 [27.95, 31.53]} & \makecell[c]{31.86 (2.75) \\ 31.91 [30.43, 33.65]} & \makecell[c]{26.10 (3.22) \\ 26.39 [24.38, 28.28]} & \makecell[c]{26.74 (3.51) \\ 26.90 [24.52, 29.48]} \\
Extreme max.\ temp.\ days & \makecell[c]{34.47 (2.39) \\ 34.11 [32.86, 35.85]} & \makecell[c]{34.65 (2.20) \\ 34.57 [33.11, 36.05]} & \makecell[c]{33.99 (1.98) \\ 33.81 [32.82, 35.13]} & \makecell[c]{33.96 (2.05) \\ 33.75 [32.75, 35.29]} & \makecell[c]{34.24 (2.05) \\ 34.20 [32.87, 35.59]} & \makecell[c]{33.98 (1.76) \\ 33.89 [32.84, 35.03]} & \makecell[c]{33.66 (1.71) \\ 33.53 [32.54, 34.77]} & \makecell[c]{31.86 (0.92) \\ 31.67 [31.40, 32.23]} \\
Heatwave days & \makecell[c]{35.32 (2.56) \\ 34.94 [33.34, 36.83]} & \makecell[c]{35.18 (2.22) \\ 35.16 [33.74, 36.57]} & \makecell[c]{34.47 (2.19) \\ 34.33 [33.16, 35.81]} & \makecell[c]{34.10 (2.15) \\ 33.92 [32.78, 35.51]} & \makecell[c]{34.69 (2.02) \\ 34.68 [33.38, 36.05]} & \makecell[c]{34.10 (1.69) \\ 33.99 [33.01, 35.07]} & \makecell[c]{33.03 (1.56) \\ 32.68 [31.98, 34.02]} & --- \\
\addlinespace
\multicolumn{9}{l}{\textit{Minimum temperature ($^\circ$C)}} \\
All days & \makecell[c]{16.26 (3.27) \\ 16.42 [14.24, 18.51]} & \makecell[c]{18.42 (2.68) \\ 18.62 [16.85, 20.13]} & \makecell[c]{19.03 (2.87) \\ 19.34 [17.35, 21.07]} & \makecell[c]{20.45 (2.53) \\ 20.67 [18.94, 22.26]} & \makecell[c]{19.27 (3.01) \\ 19.59 [17.43, 21.40]} & \makecell[c]{20.53 (2.92) \\ 20.86 [18.74, 22.60]} & \makecell[c]{16.27 (3.31) \\ 16.62 [14.29, 18.65]} & \makecell[c]{17.20 (3.84) \\ 17.49 [14.66, 20.20]} \\
Extreme max.\ temp.\ days & \makecell[c]{20.07 (2.18) \\ 20.07 [18.66, 21.49]} & \makecell[c]{20.74 (2.29) \\ 20.71 [19.27, 22.11]} & \makecell[c]{20.66 (2.22) \\ 20.75 [19.18, 22.27]} & \makecell[c]{21.25 (2.32) \\ 21.34 [19.61, 22.98]} & \makecell[c]{21.76 (2.36) \\ 21.81 [20.26, 23.39]} & \makecell[c]{22.35 (2.17) \\ 22.55 [21.03, 23.83]} & \makecell[c]{20.91 (1.99) \\ 20.94 [19.66, 22.54]} & \makecell[c]{23.45 (1.83) \\ 23.81 [22.81, 24.91]} \\
Heatwave days & \makecell[c]{20.28 (2.20) \\ 20.24 [18.97, 21.66]} & \makecell[c]{20.52 (1.98) \\ 20.53 [19.28, 21.82]} & \makecell[c]{20.80 (2.14) \\ 20.82 [19.30, 22.36]} & \makecell[c]{21.23 (2.37) \\ 21.30 [19.54, 22.92]} & \makecell[c]{22.12 (2.30) \\ 22.15 [20.63, 23.69]} & \makecell[c]{22.65 (1.98) \\ 22.71 [21.39, 24.02]} & \makecell[c]{21.47 (1.83) \\ 22.04 [20.24, 22.81]} & --- \\
\addlinespace
\multicolumn{9}{l}{\textit{Mean temperature ($^\circ$C)}} \\
All days & \makecell[c]{21.43 (3.13) \\ 21.60 [19.64, 23.46]} & \makecell[c]{23.87 (2.41) \\ 23.86 [22.48, 25.27]} & \makecell[c]{24.18 (2.42) \\ 24.57 [22.99, 25.79]} & \makecell[c]{25.91 (2.04) \\ 26.22 [25.05, 27.18]} & \makecell[c]{24.08 (2.68) \\ 24.44 [22.74, 25.79]} & \makecell[c]{25.75 (2.68) \\ 26.18 [24.41, 27.61]} & \makecell[c]{20.72 (2.93) \\ 21.07 [19.11, 22.78]} & \makecell[c]{21.54 (3.53) \\ 21.95 [19.25, 24.49]} \\
Extreme max.\ temp.\ days & \makecell[c]{27.17 (1.64) \\ 27.08 [26.25, 28.01]} & \makecell[c]{27.51 (1.56) \\ 27.52 [26.82, 28.45]} & \makecell[c]{27.01 (1.43) \\ 27.21 [26.42, 27.89]} & \makecell[c]{27.49 (1.53) \\ 27.70 [26.91, 28.40]} & \makecell[c]{27.54 (1.63) \\ 27.69 [26.72, 28.58]} & \makecell[c]{27.85 (1.37) \\ 28.05 [27.14, 28.83]} & \makecell[c]{26.68 (1.00) \\ 26.84 [26.19, 27.35]} & \makecell[c]{26.86 (0.69) \\ 27.06 [26.74, 27.28]} \\
Heatwave days & \makecell[c]{27.81 (1.69) \\ 27.75 [26.81, 29.07]} & \makecell[c]{27.56 (1.59) \\ 27.67 [26.91, 28.52]} & \makecell[c]{27.42 (1.49) \\ 27.61 [26.83, 28.36]} & \makecell[c]{27.60 (1.63) \\ 27.85 [27.00, 28.57]} & \makecell[c]{27.98 (1.52) \\ 28.13 [27.27, 28.89]} & \makecell[c]{28.12 (1.11) \\ 28.22 [27.46, 28.94]} & \makecell[c]{26.77 (0.95) \\ 26.99 [26.23, 27.32]} & --- \\
\addlinespace
Extreme-heat ABS-days (\%) & 3.2 & 12.7 & 7.5 & 25.6 & 9.6 & 37.2 & 0.5 & 0.7 \\
Heatwave ABS-days (\%) & 1.6 & 8.9 & 2.9 & 14.9 & 6.4 & 31.5 & 0.1 & 0.0 \\
\bottomrule
\end{tabular}}
\vspace{0.4em}\begin{minipage}{\linewidth}\footnotesize
\noindent In each cell the first line is the mean (standard deviation) and the second the median [Q1, Q3].
\noindent Extreme max.\ temp.\ days / heatwave days: restricted to ABS-days with the indicator equal to 1.
\noindent Analysis sample: summer ABS-days with complete environmental fields ($n=498{,}385$; 379 ABS).
\end{minipage}

\end{sidewaystable}

%% file: tables/table1b.tex
\begin{sidewaystable}[htbp]
\centering
\caption{Descriptive statistics of relative humidity and air pollutants. Catalonia, summer months 2012--2022 (complete environmental sample).}
\label{tab:1b}
\scriptsize
\setlength{\tabcolsep}{2.5pt}

\resizebox{0.95\textheight}{!}{\begin{tabular}{@{}lcccccccc@{}}
\toprule
 & \multicolumn{2}{c}{June} & \multicolumn{2}{c}{July} & \multicolumn{2}{c}{August} & \multicolumn{2}{c}{September} \\
\cmidrule(lr){2-3}\cmidrule(lr){4-5}\cmidrule(lr){6-7}\cmidrule(lr){8-9}
 & 2012--2021 & 2022 & 2012--2021 & 2022 & 2012--2021 & 2022 & 2012--2021 & 2022 \\
\midrule
\multicolumn{9}{l}{\textit{Relative humidity, mean (\%)}} \\
All days & \makecell[c]{64.90 (11.13) \\ 65.50 [57.24, 72.93]} & \makecell[c]{63.46 (12.55) \\ 65.57 [55.36, 72.61]} & \makecell[c]{64.91 (9.38) \\ 65.86 [58.72, 71.77]} & \makecell[c]{60.37 (10.98) \\ 59.89 [52.98, 68.52]} & \makecell[c]{67.39 (9.65) \\ 68.53 [61.24, 74.44]} & \makecell[c]{66.35 (9.10) \\ 67.58 [59.00, 72.87]} & \makecell[c]{72.23 (9.81) \\ 73.36 [66.40, 79.34]} & \makecell[c]{70.84 (9.23) \\ 71.38 [63.88, 78.39]} \\
Extreme max.\ temp.\ days & \makecell[c]{52.96 (11.48) \\ 51.80 [44.99, 60.07]} & \makecell[c]{48.82 (12.37) \\ 46.45 [38.18, 57.45]} & \makecell[c]{58.54 (10.93) \\ 59.35 [51.08, 66.41]} & \makecell[c]{56.38 (12.10) \\ 55.25 [46.03, 66.54]} & \makecell[c]{60.30 (10.67) \\ 60.16 [52.48, 68.68]} & \makecell[c]{60.75 (7.79) \\ 59.40 [54.98, 66.81]} & \makecell[c]{59.26 (7.15) \\ 59.44 [55.87, 63.13]} & \makecell[c]{72.38 (3.98) \\ 70.44 [69.61, 76.84]} \\
Heatwave days & \makecell[c]{51.01 (13.12) \\ 48.30 [41.08, 59.20]} & \makecell[c]{48.35 (11.99) \\ 46.41 [38.04, 55.96]} & \makecell[c]{53.94 (11.17) \\ 53.84 [44.33, 63.19]} & \makecell[c]{52.14 (11.37) \\ 50.87 [42.95, 59.45]} & \makecell[c]{57.91 (10.45) \\ 57.33 [50.33, 66.13]} & \makecell[c]{60.18 (7.27) \\ 58.95 [54.86, 65.22]} & \makecell[c]{58.36 (9.04) \\ 60.86 [57.28, 63.30]} & --- \\
\addlinespace
\multicolumn{9}{l}{\textit{PM$_{10}$ ($\mu$g/m$^3$)}} \\
All days & \makecell[c]{23.60 (8.80) \\ 22.87 [18.07, 27.87]} & \makecell[c]{28.23 (8.54) \\ 26.24 [22.11, 33.71]} & \makecell[c]{23.98 (6.57) \\ 23.63 [19.81, 27.94]} & \makecell[c]{25.96 (5.70) \\ 25.46 [21.80, 29.39]} & \makecell[c]{22.30 (7.71) \\ 21.35 [17.46, 26.01]} & \makecell[c]{21.38 (5.78) \\ 20.15 [17.31, 24.77]} & \makecell[c]{20.76 (6.44) \\ 20.54 [16.75, 24.64]} & \makecell[c]{20.79 (6.17) \\ 20.70 [17.49, 24.64]} \\
Extreme max.\ temp.\ days & \makecell[c]{37.02 (11.00) \\ 35.08 [29.30, 44.80]} & \makecell[c]{36.41 (7.83) \\ 35.62 [30.53, 40.73]} & \makecell[c]{30.07 (7.14) \\ 30.35 [26.17, 34.09]} & \makecell[c]{28.94 (5.25) \\ 28.62 [25.43, 33.17]} & \makecell[c]{31.77 (10.91) \\ 29.40 [24.06, 37.31]} & \makecell[c]{25.23 (5.74) \\ 24.28 [22.00, 27.10]} & \makecell[c]{30.91 (7.54) \\ 32.77 [25.49, 36.66]} & \makecell[c]{26.55 (3.27) \\ 25.45 [25.10, 30.63]} \\
Heatwave days & \makecell[c]{42.01 (10.60) \\ 42.62 [34.46, 48.24]} & \makecell[c]{38.64 (7.55) \\ 38.48 [33.87, 42.22]} & \makecell[c]{32.98 (6.22) \\ 33.09 [29.92, 35.85]} & \makecell[c]{28.17 (5.47) \\ 27.51 [24.79, 31.56]} & \makecell[c]{34.69 (11.03) \\ 32.42 [26.93, 40.47]} & \makecell[c]{25.87 (5.58) \\ 24.65 [22.62, 27.44]} & \makecell[c]{26.91 (8.21) \\ 26.59 [21.16, 34.59]} & --- \\
\addlinespace
\multicolumn{9}{l}{\textit{NO$_2$ ($\mu$g/m$^3$)}} \\
All days & \makecell[c]{22.30 (10.05) \\ 21.05 [15.17, 27.83]} & \makecell[c]{20.17 (15.26) \\ 19.00 [12.68, 24.89]} & \makecell[c]{21.56 (8.36) \\ 21.02 [15.92, 26.57]} & \makecell[c]{16.42 (5.77) \\ 16.51 [12.67, 20.03]} & \makecell[c]{20.03 (20.16) \\ 17.95 [13.47, 23.47]} & \makecell[c]{14.95 (4.68) \\ 14.69 [11.84, 18.22]} & \makecell[c]{22.68 (8.34) \\ 22.25 [17.18, 27.80]} & \makecell[c]{14.67 (4.77) \\ 14.99 [11.68, 17.81]} \\
Extreme max.\ temp.\ days & \makecell[c]{32.22 (12.59) \\ 30.67 [23.59, 39.71]} & \makecell[c]{26.57 (9.27) \\ 27.56 [21.07, 32.25]} & \makecell[c]{26.07 (9.53) \\ 24.98 [19.88, 31.88]} & \makecell[c]{17.34 (5.82) \\ 17.72 [14.52, 20.95]} & \makecell[c]{23.37 (11.34) \\ 22.54 [15.08, 30.95]} & \makecell[c]{17.05 (4.75) \\ 17.68 [13.98, 20.07]} & \makecell[c]{30.90 (8.51) \\ 30.95 [25.51, 37.22]} & \makecell[c]{18.25 (2.54) \\ 18.12 [17.21, 19.27]} \\
Heatwave days & \makecell[c]{35.20 (13.62) \\ 34.97 [25.67, 43.38]} & \makecell[c]{25.82 (9.51) \\ 26.72 [20.42, 31.97]} & \makecell[c]{29.09 (10.29) \\ 28.34 [22.42, 34.93]} & \makecell[c]{17.52 (6.33) \\ 18.21 [14.88, 20.99]} & \makecell[c]{25.70 (11.58) \\ 26.69 [16.87, 33.37]} & \makecell[c]{17.43 (4.76) \\ 18.10 [14.71, 20.27]} & \makecell[c]{27.72 (9.15) \\ 26.54 [19.98, 37.23]} & --- \\
\addlinespace
\multicolumn{9}{l}{\textit{O$_3$ ($\mu$g/m$^3$)}} \\
All days & \makecell[c]{68.96 (13.08) \\ 67.38 [59.91, 77.15]} & \makecell[c]{67.90 (11.34) \\ 66.44 [58.97, 75.61]} & \makecell[c]{68.92 (12.94) \\ 67.59 [60.05, 76.80]} & \makecell[c]{75.72 (13.39) \\ 74.33 [65.16, 84.28]} & \makecell[c]{64.79 (11.78) \\ 63.82 [56.71, 71.41]} & \makecell[c]{70.02 (12.56) \\ 70.04 [61.19, 79.16]} & \makecell[c]{56.79 (11.69) \\ 56.42 [48.91, 63.97]} & \makecell[c]{57.10 (9.62) \\ 56.74 [51.12, 62.83]} \\
Extreme max.\ temp.\ days & \makecell[c]{81.89 (13.24) \\ 80.98 [72.25, 91.94]} & \makecell[c]{80.77 (12.23) \\ 81.44 [75.45, 88.25]} & \makecell[c]{74.25 (14.22) \\ 72.46 [64.12, 84.19]} & \makecell[c]{81.58 (16.34) \\ 80.31 [67.63, 93.20]} & \makecell[c]{72.26 (15.12) \\ 69.33 [61.40, 80.13]} & \makecell[c]{77.92 (8.98) \\ 78.31 [71.20, 83.67]} & \makecell[c]{66.75 (13.25) \\ 63.58 [58.35, 73.68]} & \makecell[c]{44.93 (6.69) \\ 41.71 [39.74, 48.60]} \\
Heatwave days & \makecell[c]{83.24 (14.68) \\ 83.15 [72.85, 94.98]} & \makecell[c]{83.19 (12.21) \\ 84.11 [77.93, 90.65]} & \makecell[c]{78.10 (15.27) \\ 77.50 [66.53, 90.12]} & \makecell[c]{87.54 (15.63) \\ 88.97 [73.93, 97.72]} & \makecell[c]{75.02 (15.72) \\ 71.68 [63.57, 83.43]} & \makecell[c]{78.53 (8.19) \\ 78.87 [72.00, 83.91]} & \makecell[c]{67.77 (16.36) \\ 63.58 [57.76, 73.29]} & --- \\
\addlinespace
\bottomrule
\end{tabular}}
\vspace{0.4em}\begin{minipage}{\linewidth}\footnotesize
\noindent In each cell the first line is the mean (standard deviation) and the second the median [Q1, Q3].
\noindent Extreme max.\ temp.\ days / heatwave days: restricted to ABS-days with the indicator equal to 1.
\noindent Analysis sample: summer ABS-days with complete environmental fields ($n=498{,}385$; 379 ABS).
\end{minipage}

\end{sidewaystable}

%% file: tables/table1d.tex
\begin{table}[htbp]
\centering
\begin{threeparttable}
\caption{Share of ABS-days with extreme heat or heatwave conditions by month and period. Catalonia, summer 2012--2022.}
\label{tab:1d}
\begin{tabular}{@{}llcc@{}}
\toprule
Month & Period & Extreme heat (\%) & Heatwave (\%) \\
\midrule
June & 2012--2021 & 3.2 & 1.6 \\
June & 2022 & 12.7 & 8.9 \\
July & 2012--2021 & 7.5 & 2.9 \\
July & 2022 & 25.6 & 14.9 \\
August & 2012--2021 & 9.6 & 6.4 \\
August & 2022 & 37.2 & 31.5 \\
September & 2012--2021 & 0.5 & 0.1 \\
September & 2022 & 0.7 & 0.0 \\
\bottomrule
\end{tabular}
\begin{tablenotes}[flushleft]
\footnotesize
\item[] Complete environmental sample ($n=498{,}385$ ABS-days).
\item[] Extreme heat: daily maximum temperature above the ABS-month 95th percentile (2012--2021, excluding 2015).
\item[] Heatwave: national definition applied with ABS-specific triggers.
\end{tablenotes}
\end{threeparttable}
\end{table}

%% file: tables/table2summer.tex
\begin{table}[htbp]
\centering
\begin{threeparttable}
\caption{Summer mortality totals in the complete environmental analysis sample. Catalonia, June--September 2012--2022.}
\label{tab:2summer}
\begin{tabular}{@{}lr@{}}
\toprule
Quantity & Value \\
\midrule
ABS-days (complete environmental fields) & 498,385 \\
Deaths, all ages & 210,860 \\
Deaths, $\geq 65$ years & 180,305 \\
Deaths, $\geq 85$ years & 96,133 \\
Number of ABS & 379 \\
\bottomrule
\end{tabular}
\begin{tablenotes}[flushleft]
\footnotesize
\item[] Analysis sample: summer ABS-days with complete environmental fields ($n=498{,}385$; 379 ABS).
\end{tablenotes}
\end{threeparttable}
\end{table}

%% file: tables/table2a.tex
\begin{sidewaystable}[htbp]
\centering
\caption{Descriptive statistics on mortality. All-cause mortality and causes accounting for more than 3\% of deaths. Catalonia 2012--2022.}
\label{tab:2a}
\tiny
\setlength{\tabcolsep}{2.2pt}

\resizebox{0.95\textheight}{!}{\begin{tabular}{@{}lccccccccccc@{}}
\toprule
 & Total & Tumours & Circulatory & Respiratory & Nervous system & Mental disorders & Digestive & External causes & COVID-19 & Endocrine/metabolic & Genitourinary \\
\midrule
\textit{Sex}, $n=730{,}364$ & & & & & & & & & & & \\
Female & 363356 (49.75\%) & 76918 (40.06\%) & 101983 (53.69\%) & 31358 (43.16\%) & 31248 (62.13\%) & 29883 (66.33\%) & 17020 (48.80\%) & 12849 (41.62\%) & 13421 (47.66\%) & 13186 (55.84\%) & 12794 (56.08\%) \\
Male & 367008 (50.25\%) & 115097 (59.94\%) & 87979 (46.31\%) & 41299 (56.84\%) & 19049 (37.87\%) & 15166 (33.67\%) & 17858 (51.20\%) & 18024 (58.38\%) & 14741 (52.34\%) & 10427 (44.16\%) & 10018 (43.92\%) \\
\addlinespace
\textit{Age}, $n=730{,}364$ & \makecell[c]{79.9 (14.32)\\84.0 [73, 90]} & \makecell[c]{73.8 (13.46)\\76.0 [65, 84]} & \makecell[c]{82.8 (12.09)\\86.0 [78, 91]} & \makecell[c]{83.2 (11.38)\\86.0 [79, 91]} & \makecell[c]{82.7 (12.03)\\85.0 [79, 90]} & \makecell[c]{88.0 (7.89)\\89.0 [84, 93]} & \makecell[c]{78.7 (13.64)\\82.0 [71, 89]} & \makecell[c]{70.1 (22.46)\\79.0 [52, 88]} & \makecell[c]{82.6 (11.03)\\85.0 [77, 90]} & \makecell[c]{82.4 (12.41)\\85.0 [78, 90]} & \makecell[c]{86.0 (8.94)\\87.0 [82, 92]} \\
\addlinespace
\textit{Marital status}, $n=510{,}532$ & & & & & & & & & & & \\
Divorced & 23495 (4.60\%) & 8795 (6.28\%) & 5575 (4.03\%) & 2058 (3.73\%) & 1033 (2.78\%) & 674 (2.00\%) & 1622 (6.48\%) & 1593 (7.15\%) & 629 (3.86\%) & 364 (2.42\%) & --- \\
Married & 212239 (41.57\%) & 79129 (56.54\%) & 50523 (36.55\%) & 21730 (39.37\%) & 14266 (38.38\%) & 9070 (26.87\%) & 9998 (39.92\%) & 7947 (35.65\%) & 5842 (35.81\%) & 5011 (33.33\%) & --- \\
Single & 55879 (10.95\%) & 14018 (10.02\%) & 13433 (9.72\%) & 5725 (10.37\%) & 3477 (9.36\%) & 3291 (9.75\%) & 3014 (12.04\%) & 4834 (21.68\%) & 1685 (10.33\%) & 1394 (9.27\%) & --- \\
Widowed & 218919 (42.88\%) & 37998 (27.15\%) & 68717 (49.71\%) & 25685 (46.53\%) & 18391 (49.48\%) & 20720 (61.38\%) & 10408 (41.56\%) & 7918 (35.52\%) & 8157 (50.00\%) & 8267 (54.98\%) & --- \\
\addlinespace
\textit{Education}, $n=715{,}842$ & & & & & & & & & & & \\
Insufficient & 154833 (21.63\%) & 28502 (15.08\%) & 44357 (23.71\%) & 18606 (25.99\%) & 11950 (24.17\%) & 13383 (30.12\%) & 7525 (22.00\%) & 5107 (17.50\%) & 5196 (18.75\%) & 5925 (25.54\%) & 5748 (25.48\%) \\
Primary & 415338 (58.02\%) & 107162 (56.70\%) & 109261 (58.40\%) & 41744 (58.32\%) & 28990 (58.63\%) & 25593 (57.60\%) & 20062 (58.64\%) & 16070 (55.05\%) & 16769 (60.50\%) & 13908 (59.95\%) & 13525 (59.95\%) \\
Secondary & 73369 (10.25\%) & 25943 (13.73\%) & 16964 (9.07\%) & 5836 (8.15\%) & 4207 (8.51\%) & 2962 (6.67\%) & 3520 (10.29\%) & 4138 (14.18\%) & 2894 (10.44\%) & 1781 (7.68\%) & 1659 (7.35\%) \\
University & 53723 (7.50\%) & 20366 (10.78\%) & 12373 (6.61\%) & 4060 (5.67\%) & 3278 (6.63\%) & 1938 (4.36\%) & 2198 (6.42\%) & 2697 (9.24\%) & 2106 (7.60\%) & 1151 (4.96\%) & 1191 (5.28\%) \\
Vocational & 18579 (2.60\%) & 7026 (3.72\%) & 4122 (2.20\%) & 1333 (1.86\%) & 1019 (2.06\%) & 553 (1.24\%) & 906 (2.65\%) & 1177 (4.03\%) & 752 (2.71\%) & 433 (1.87\%) & 437 (1.94\%) \\
\addlinespace
\textit{Occupation}, $n=555{,}791$ & & & & & & & & & & & \\
Occupied & 22429 (4.04\%) & 10683 (7.54\%) & 4718 (3.30\%) & 755 (1.37\%) & 373 (0.97\%) & 120 (0.34\%) & 831 (3.17\%) & 2919 (13.72\%) & 863 (3.11\%) & 250 (1.32\%) & 110 (0.60\%) \\
Other & 112599 (20.26\%) & 21028 (14.84\%) & 31000 (21.71\%) & 10589 (19.15\%) & 9400 (24.46\%) & 9221 (26.40\%) & 5467 (20.84\%) & 4538 (21.32\%) & 5169 (18.64\%) & 4451 (23.52\%) & 4230 (23.10\%) \\
Pensioner & 386111 (69.47\%) & 96500 (68.12\%) & 100396 (70.30\%) & 41549 (75.14\%) & 26749 (69.61\%) & 24616 (70.46\%) & 17646 (67.28\%) & 11329 (53.23\%) & 20530 (74.03\%) & 13362 (70.61\%) & 13405 (73.19\%) \\
Permanent inactivity & 26250 (4.72\%) & 10066 (7.11\%) & 5001 (3.50\%) & 1996 (3.61\%) & 1737 (4.52\%) & 926 (2.65\%) & 1585 (6.04\%) & 1333 (6.26\%) & 940 (3.39\%) & 729 (3.85\%) & 530 (2.89\%) \\
Student & 788 (0.14\%) & 232 (0.16\%) & 63 (0.04\%) & 33 (0.06\%) & 70 (0.18\%) & 7 (0.02\%) & 10 (0.04\%) & 259 (1.22\%) & 6 (0.02\%) & 18 (0.10\%) & 7 (0.04\%) \\
Unemployed & 7614 (1.37\%) & 3160 (2.23\%) & 1628 (1.14\%) & 377 (0.68\%) & 99 (0.26\%) & 44 (0.13\%) & 690 (2.63\%) & 905 (4.25\%) & 223 (0.80\%) & 115 (0.61\%) & 33 (0.18\%) \\
\addlinespace
Total & 730364 (100\%) & 192015 (26.29\%) & 189962 (26.01\%) & 72657 (9.95\%) & 50297 (6.89\%) & 45049 (6.17\%) & 34878 (4.78\%) & 30873 (4.23\%) & 28162 (3.86\%) & 23613 (3.23\%) & 22812 (3.12\%) \\
\bottomrule
\end{tabular}}
\vspace{0.4em}\begin{minipage}{\linewidth}\footnotesize
\footnotesize
\noindent Values are $n$ (\%) unless otherwise indicated. Age is reported as mean (SD) and median [Q1, Q3].
\noindent Percentages for cause-specific columns are within-cause. The bottom row shows the share of all deaths.
\end{minipage}

\end{sidewaystable}

%% file: tables/table2b.tex
\begin{sidewaystable}[htbp]
\centering
\caption{Descriptive statistics on mortality. Causes accounting for less than 3\% of deaths. Catalonia 2012--2022.}
\label{tab:2b}
\tiny
\setlength{\tabcolsep}{3pt}

\resizebox{0.95\textheight}{!}{\begin{tabular}{@{}lcccccccc@{}}
\toprule
 & Symptoms/abnormal findings & Infectious diseases & Musculoskeletal & Haematological/immune & Skin diseases & Congenital malformations & Perinatal conditions & Pregnancy/childbirth \\
\midrule
\textit{Sex}, $n=730{,}364$ & & & & & & & & \\
Female & 6445 (53.85\%) & 5661 (50.95\%) & 6195 (69.33\%) & 2041 (55.04\%) & 1305 (66.04\%) & 618 (47.43\%) & 411 (40.06\%) & 20 (100\%) \\
Male & 5523 (46.15\%) & 5449 (49.05\%) & 2740 (30.67\%) & 1667 (44.96\%) & 671 (33.96\%) & 685 (52.57\%) & 615 (59.94\%) & --- \\
\addlinespace
\textit{Age} & \makecell[c]{70.0 (18.84)\\86.0 [73, 92]} & \makecell[c]{77.9 (15.31)\\82.0 [71, 88]} & \makecell[c]{86.4 (10.31)\\88.0 [83, 93]} & \makecell[c]{81.6 (16.02)\\86.0 [78, 91]} & \makecell[c]{36.1 (5.95)\\35.5 [31.5, 41]} & \makecell[c]{38.0 (32.55)\\46.0 [0, 64]} & \makecell[c]{0.94 (6.89)\\0.0 [0, 0]} & \makecell[c]{36.1 (5.95)\\35.5 [31.5, 41]} \\
\addlinespace
Total & 11968 (1.64\%) & 11110 (1.52\%) & 8935 (1.22\%) & 3708 (0.51\%) & 1976 (0.27\%) & 1303 (0.18\%) & 1026 (0.14\%) & 20 (0.00\%) \\
\bottomrule
\end{tabular}}
\vspace{0.4em}
\begin{minipage}{0.95\textheight}
\footnotesize
\noindent Values are $n$ (\%) unless otherwise indicated. Age is reported as mean (SD) and median [Q1, Q3].
\end{minipage}

\end{sidewaystable}

%% file: tables/table3a.tex
\begin{table}[htbp]
\centering
\caption{Main-effects models including extreme maximum temperatures: heat indicator, air pollutants and humidity. \textbf{Cells:} RR (95\% CrI) and Probs $= \mathrm{Prob}(|\log(\mathrm{RR})|>0)$. Reference category in brackets. Spatiotemporal random effects and ABS population offset as described in Methods. \textbf{Shading:} Prob $\geq 0.90$ (light) or $95\%$ CrI excludes unity (darker). Marginal heat RR after adjustment for pollutants and SES (SES shown in Table~\ref{tab:3b}). O$_3$ categories collapsed at $\geq 100\,\mu$g/m$^3$.}
\label{tab:3a}
\scriptsize
\setlength{\tabcolsep}{3pt}

\begin{tabular}{@{}lcccccc@{}}
\toprule
 & All & Probs & $\geq 65$ years & Probs & $\geq 85$ years & Probs \\
\midrule
Extreme heat$_{t-2}$ & \cellcolor{sig95}1.105 (1.086--1.125) & 1.000 & \cellcolor{sig95}1.120 (1.099--1.142) & 1.000 & \cellcolor{sig95}1.123 (1.095--1.153) & 1.000 \\
mean(PM$_{10}$, $t-1$ to $t-2$) [$<15.0\,\mu$g/m$^3$] & & & & & & \\
\quad $15.0$--$44.9\,\mu$g/m$^3$ & \cellcolor{sig95}1.044 (1.026--1.062) & 1.000 & \cellcolor{sig95}1.040 (1.020--1.060) & 1.000 & \cellcolor{sig95}1.046 (1.019--1.073) & 1.000 \\
\quad $\geq 45.0\,\mu$g/m$^3$ & \cellcolor{sig95}1.102 (1.048--1.158) & 1.000 & \cellcolor{sig95}1.098 (1.041--1.159) & 1.000 & \cellcolor{sig95}1.097 (1.020--1.181) & 0.994 \\
mean(O$_3$, $t-1$ to $t-2$) [$<60.0\,\mu$g/m$^3$] & & & & & & \\
\quad $60.0$--$99.9\,\mu$g/m$^3$ & \cellcolor{sig90}1.011 (1.000--1.021) & 0.973 & \cellcolor{sig95}1.015 (1.003--1.027) & 0.994 & \cellcolor{sig95}1.019 (1.003--1.035) & 0.989 \\
\quad $\geq 100.0\,\mu$g/m$^3$ & \cellcolor{sig95}1.074 (1.019--1.132) & 0.996 & \cellcolor{sig95}1.088 (1.029--1.151) & 0.998 & \cellcolor{sig95}1.103 (1.023--1.188) & 0.995 \\
mean(NO$_2$, $t-1$ to $t-2$) [$<10.0\,\mu$g/m$^3$] & & & & & & \\
\quad $10.0$--$24.9\,\mu$g/m$^3$ & 1.013 (0.992--1.035) & 0.882 & 1.013 (0.991--1.037) & 0.874 & 0.999 (0.969--1.030) & 0.515 \\
\quad $\geq 25.0\,\mu$g/m$^3$ & \cellcolor{sig95}1.051 (1.026--1.075) & 1.000 & \cellcolor{sig95}1.052 (1.026--1.079) & 1.000 & \cellcolor{sig95}1.035 (1.001--1.071) & 0.978 \\
Q4 relative humidity$_{t-2}$ [No] & 1.001 (0.990--1.013) & 0.583 & 1.002 (0.990--1.015) & 0.637 & 0.999 (0.982--1.016) & 0.543 \\
\bottomrule
\end{tabular}
\end{table}

%% file: tables/table3b.tex
\begin{table}[htbp]
\centering
\caption{Main-effects models including extreme maximum temperatures: contextual socioeconomic and demographic quartiles. \textbf{Cells:} RR (95\% CrI) and Probs $= \mathrm{Prob}(|\log(\mathrm{RR})|>0)$. Reference category in brackets. Spatiotemporal random effects and ABS population offset as described in Methods. \textbf{Shading:} Prob $\geq 0.90$ (light) or $95\%$ CrI excludes unity (darker). Same models as Table~\ref{tab:3a}; heat and pollutant coefficients reported there.}
\label{tab:3b}
\scriptsize
\setlength{\tabcolsep}{3pt}

\begin{tabular}{@{}lcccccc@{}}
\toprule
 & All & Probs & $\geq 65$ years & Probs & $\geq 85$ years & Probs \\
\midrule
Average net income per person [Q1] & & & & & & \\
\quad Q2 & 1.006 (0.968--1.045) & 0.619 & 1.009 (0.968--1.051) & 0.663 & 1.019 (0.964--1.077) & 0.745 \\
\quad Q3 & 1.001 (0.948--1.057) & 0.517 & 0.995 (0.939--1.054) & 0.571 & 0.966 (0.893--1.046) & 0.802 \\
\quad Q4 & 1.017 (0.946--1.093) & 0.675 & 1.013 (0.938--1.094) & 0.626 & 0.974 (0.877--1.081) & 0.689 \\
Gini index [Q1] & & & & & & \\
\quad Q2 & \cellcolor{sig95}1.051 (1.019--1.083) & 0.999 & \cellcolor{sig95}1.062 (1.028--1.097) & 1.000 & \cellcolor{sig95}1.059 (1.012--1.107) & 0.994 \\
\quad Q3 & \cellcolor{sig95}1.076 (1.035--1.118) & 1.000 & \cellcolor{sig95}1.090 (1.046--1.136) & 1.000 & \cellcolor{sig95}1.082 (1.023--1.145) & 0.997 \\
\quad Q4 & \cellcolor{sig95}1.072 (1.020--1.125) & 0.997 & \cellcolor{sig95}1.077 (1.023--1.134) & 0.998 & \cellcolor{sig90}1.048 (0.977--1.125) & 0.904 \\
Population aged 65 or over [Q1] & & & & & & \\
\quad Q2 & 1.022 (0.986--1.058) & 0.880 & 1.019 (0.981--1.059) & 0.839 & 0.993 (0.942--1.047) & 0.597 \\
\quad Q3 & \cellcolor{sig95}1.053 (1.007--1.101) & 0.988 & \cellcolor{sig95}1.060 (1.010--1.112) & 0.991 & 1.010 (0.946--1.079) & 0.617 \\
\quad Q4 & \cellcolor{sig95}1.082 (1.020--1.147) & 0.996 & \cellcolor{sig95}1.094 (1.028--1.165) & 0.998 & 1.041 (0.956--1.133) & 0.823 \\
\bottomrule
\end{tabular}
\end{table}

%% file: tables/table3c.tex
\begin{table}[htbp]
\centering
\caption{Interaction models including extreme maximum temperatures: heat$\times$pollutant / humidity terms. \textbf{Cells:} RR (95\% CrI) and Probs $= \mathrm{Prob}(|\log(\mathrm{RR})|>0)$. Reference category in brackets. Spatiotemporal random effects and ABS population offset as described in Methods. \textbf{Shading:} Prob $\geq 0.90$ (light) or $95\%$ CrI excludes unity (darker). $^*$\textbf{Heat coefficient} is the association in the reference covariate stratum (lowest pollutant categories, humidity not in Q4, SES Q1); marginal heat associations are reported in Tables~\ref{tab:3a}--\ref{tab:3b}.}
\label{tab:3c}
\scriptsize
\setlength{\tabcolsep}{3pt}

\begin{tabular}{@{}lcccccc@{}}
\toprule
 & All & Probs & $\geq 65$ years & Probs & $\geq 85$ years & Probs \\
\midrule
Extreme heat$_{t-2}$ (reference stratum)$^*$ & 0.904 (0.743--1.101) & 0.843 & 0.987 (0.800--1.217) & 0.550 & 0.831 (0.615--1.124) & 0.886 \\
Extreme heat$_{t-2}$ $\times$ mean(PM$_{10}$, $t-1$ to $t-2$) & & & & & & \\
\quad $15.0$--$44.9\,\mu$g/m$^3$ & \cellcolor{sig90}1.146 (0.961--1.367) & 0.935 & 1.130 (0.937--1.363) & 0.899 & \cellcolor{sig95}1.334 (1.015--1.753) & 0.980 \\
\quad $\geq 45.0\,\mu$g/m$^3$ & 1.005 (0.824--1.225) & 0.518 & 0.982 (0.795--1.213) & 0.569 & 1.154 (0.852--1.563) & 0.822 \\
Extreme heat$_{t-2}$ $\times$ mean(O$_3$, $t-1$ to $t-2$) & & & & & & \\
\quad $60.0$--$99.9\,\mu$g/m$^3$ & \cellcolor{sig95}1.071 (1.016--1.129) & 0.995 & \cellcolor{sig95}1.064 (1.006--1.126) & 0.985 & \cellcolor{sig95}1.110 (1.027--1.200) & 0.996 \\
\quad $\geq 100.0\,\mu$g/m$^3$ & \cellcolor{sig95}1.189 (1.061--1.332) & 0.999 & \cellcolor{sig95}1.170 (1.037--1.322) & 0.995 & \cellcolor{sig95}1.186 (1.008--1.395) & 0.980 \\
Extreme heat$_{t-2}$ $\times$ mean(NO$_2$, $t-1$ to $t-2$) & & & & & & \\
\quad $10.0$--$24.9\,\mu$g/m$^3$ & 0.991 (0.927--1.059) & 0.606 & 0.960 (0.895--1.030) & 0.871 & \cellcolor{sig90}0.941 (0.857--1.033) & 0.900 \\
\quad $\geq 25.0\,\mu$g/m$^3$ & 0.992 (0.925--1.063) & 0.594 & 0.969 (0.900--1.044) & 0.797 & 0.960 (0.870--1.059) & 0.795 \\
Extreme heat$_{t-2}$ $\times$ Q4 relative humidity$_{t-2}$ & \cellcolor{sig90}0.956 (0.897--1.020) & 0.914 & 0.958 (0.894--1.026) & 0.890 & 0.992 (0.904--1.088) & 0.572 \\
\bottomrule
\end{tabular}
\end{table}

%% file: tables/table3d.tex
\begin{table}[htbp]
\centering
\caption{Interaction models including extreme maximum temperatures: heat$\times$socioeconomic / demographic terms. \textbf{Cells:} RR (95\% CrI) and Probs $= \mathrm{Prob}(|\log(\mathrm{RR})|>0)$. Reference category in brackets. Spatiotemporal random effects and ABS population offset as described in Methods. \textbf{Shading:} Prob $\geq 0.90$ (light) or $95\%$ CrI excludes unity (darker). Same interaction models as Table~\ref{tab:3c}; reference-stratum heat and pollutant interactions reported there.}
\label{tab:3d}
\scriptsize
\setlength{\tabcolsep}{3pt}

\begin{tabular}{@{}lcccccc@{}}
\toprule
 & All & Probs & $\geq 65$ years & Probs & $\geq 85$ years & Probs \\
\midrule
Extreme heat$_{t-2}$ $\times$ average net income & & & & & & \\
\quad Q2 & 1.029 (0.980--1.080) & 0.874 & 1.020 (0.968--1.076) & 0.773 & 0.998 (0.928--1.074) & 0.520 \\
\quad Q3 & \cellcolor{sig90}1.043 (0.992--1.097) & 0.950 & \cellcolor{sig90}1.037 (0.982--1.094) & 0.904 & 1.016 (0.943--1.094) & 0.663 \\
\quad Q4 & \cellcolor{sig95}1.059 (1.010--1.109) & 0.992 & \cellcolor{sig95}1.053 (1.001--1.107) & 0.978 & 1.040 (0.972--1.114) & 0.873 \\
Extreme heat$_{t-2}$ $\times$ Gini index & & & & & & \\
\quad Q2 & 0.973 (0.925--1.023) & 0.859 & \cellcolor{sig90}0.963 (0.912--1.017) & 0.912 & \cellcolor{sig90}0.950 (0.881--1.024) & 0.910 \\
\quad Q3 & 0.999 (0.949--1.052) & 0.515 & 0.992 (0.939--1.049) & 0.608 & 0.971 (0.900--1.047) & 0.779 \\
\quad Q4 & 1.011 (0.960--1.065) & 0.661 & 1.012 (0.957--1.070) & 0.657 & 0.998 (0.925--1.078) & 0.519 \\
Extreme heat$_{t-2}$ $\times$ population aged 65 or over & & & & & & \\
\quad Q2 & 0.981 (0.935--1.028) & 0.793 & 0.968 (0.920--1.018) & 0.897 & 0.991 (0.924--1.063) & 0.605 \\
\quad Q3 & 0.989 (0.943--1.037) & 0.682 & \cellcolor{sig90}0.967 (0.919--1.017) & 0.904 & 0.982 (0.916--1.053) & 0.692 \\
\quad Q4 & 0.990 (0.943--1.039) & 0.663 & \cellcolor{sig90}0.952 (0.904--1.003) & 0.969 & 0.957 (0.892--1.027) & 0.890 \\
\bottomrule
\end{tabular}
\end{table}

%% file: tables/table3e.tex
\begin{table}[htbp]
\centering
\caption{Stratum-specific combined heat effects for extreme maximum temperatures by pollutant / humidity level. \textbf{Cells:} stratum-specific heat RR $=\exp(\beta_H)$ (reference) or $\exp(\beta_H+\delta_{H\times\mathrm{modifier}})$ (non-reference), holding other modifiers at their reference level (2{,}000 INLA draws). Probs $= \mathrm{Prob}(|\log(\mathrm{RR})|>0)$. \textbf{Shading:} Prob $\geq 0.90$ (light) or $95\%$ CrI excludes unity (darker).}
\label{tab:3e}
\scriptsize
\setlength{\tabcolsep}{3pt}

\begin{tabular}{@{}lcccccc@{}}
\toprule
 & All & Probs & $\geq 65$ years & Probs & $\geq 85$ years & Probs \\
\midrule
Extreme heat$_{t-2}$ | reference stratum & 0.900 (0.739--1.101) & 0.847 & 0.985 (0.801--1.215) & 0.540 & 0.829 (0.626--1.113) & 0.889 \\
Extreme heat$_{t-2}$ by PM$_{10}$ stratum & & & & & & \\
\quad $15.0$--$44.9\,\mu$g/m$^3$ & 1.035 (0.938--1.139) & 0.761 & \cellcolor{sig95}1.115 (1.010--1.234) & 0.985 & \cellcolor{sig90}1.107 (0.964--1.274) & 0.917 \\
\quad $\geq 45.0\,\mu$g/m$^3$ & \cellcolor{sig90}0.908 (0.798--1.036) & 0.928 & 0.970 (0.841--1.111) & 0.663 & 0.957 (0.792--1.160) & 0.671 \\
Extreme heat$_{t-2}$ by O$_3$ stratum & & & & & & \\
\quad $60.0$--$99.9\,\mu$g/m$^3$ & 0.964 (0.789--1.172) & 0.650 & 1.049 (0.858--1.288) & 0.668 & 0.918 (0.695--1.217) & 0.707 \\
\quad $\geq 100.0\,\mu$g/m$^3$ & 1.071 (0.861--1.332) & 0.737 & 1.153 (0.924--1.462) & 0.888 & 0.980 (0.706--1.346) & 0.550 \\
Extreme heat$_{t-2}$ by NO$_2$ stratum & & & & & & \\
\quad $10.0$--$24.9\,\mu$g/m$^3$ & 0.892 (0.736--1.081) & 0.880 & 0.947 (0.774--1.156) & 0.700 & \cellcolor{sig90}0.780 (0.586--1.034) & 0.953 \\
\quad $\geq 25.0\,\mu$g/m$^3$ & 0.893 (0.741--1.079) & 0.875 & 0.956 (0.783--1.171) & 0.675 & \cellcolor{sig90}0.796 (0.596--1.062) & 0.938 \\
Extreme heat$_{t-2}$ | Q4 relative humidity & \cellcolor{sig90}0.861 (0.699--1.061) & 0.921 & 0.944 (0.766--1.166) & 0.693 & 0.822 (0.614--1.104) & 0.896 \\
\bottomrule
\end{tabular}
\end{table}

%% file: tables/table3f.tex
\begin{table}[htbp]
\centering
\caption{Stratum-specific combined heat effects for extreme maximum temperatures by socioeconomic / demographic quartile. \textbf{Cells:} stratum-specific heat RR $=\exp(\beta_H)$ (reference) or $\exp(\beta_H+\delta_{H\times\mathrm{modifier}})$ (non-reference), holding other modifiers at their reference level (2{,}000 INLA draws). Probs $= \mathrm{Prob}(|\log(\mathrm{RR})|>0)$. \textbf{Shading:} Prob $\geq 0.90$ (light) or $95\%$ CrI excludes unity (darker).}
\label{tab:3f}
\scriptsize
\setlength{\tabcolsep}{3pt}

\begin{tabular}{@{}lcccccc@{}}
\toprule
 & All & Probs & $\geq 65$ years & Probs & $\geq 85$ years & Probs \\
\midrule
Extreme heat$_{t-2}$ | reference stratum & 0.900 (0.739--1.101) & 0.847 & 0.985 (0.801--1.215) & 0.540 & 0.829 (0.626--1.113) & 0.889 \\
Extreme heat$_{t-2}$ by income quartile & & & & & & \\
\quad Q2 & 0.925 (0.760--1.133) & 0.769 & 1.005 (0.818--1.242) & 0.521 & 0.828 (0.624--1.105) & 0.897 \\
\quad Q3 & 0.939 (0.768--1.150) & 0.731 & 1.021 (0.831--1.257) & 0.582 & 0.843 (0.636--1.123) & 0.876 \\
\quad Q4 & 0.952 (0.777--1.167) & 0.684 & 1.038 (0.836--1.278) & 0.628 & 0.862 (0.649--1.156) & 0.831 \\
Extreme heat$_{t-2}$ by Gini quartile & & & & & & \\
\quad Q2 & \cellcolor{sig90}0.876 (0.721--1.073) & 0.901 & 0.949 (0.779--1.171) & 0.686 & \cellcolor{sig90}0.789 (0.592--1.045) & 0.942 \\
\quad Q3 & 0.900 (0.739--1.107) & 0.846 & 0.978 (0.800--1.203) & 0.578 & \cellcolor{sig90}0.805 (0.605--1.066) & 0.922 \\
\quad Q4 & 0.911 (0.747--1.113) & 0.824 & 0.997 (0.816--1.229) & 0.499 & 0.828 (0.627--1.098) & 0.898 \\
Extreme heat$_{t-2}$ by \% aged $\geq 65$ quartile & & & & & & \\
\quad Q2 & 0.883 (0.721--1.086) & 0.881 & 0.954 (0.771--1.176) & 0.656 & \cellcolor{sig90}0.822 (0.618--1.096) & 0.903 \\
\quad Q3 & 0.890 (0.730--1.090) & 0.876 & 0.952 (0.770--1.168) & 0.663 & \cellcolor{sig90}0.815 (0.611--1.092) & 0.914 \\
\quad Q4 & 0.891 (0.729--1.094) & 0.871 & 0.938 (0.760--1.152) & 0.717 & \cellcolor{sig90}0.794 (0.601--1.066) & 0.937 \\
\bottomrule
\end{tabular}
\end{table}

%% file: tables/table4a.tex
\begin{table}[htbp]
\centering
\caption{Main-effects models including heatwaves: heat indicator, air pollutants and humidity. \textbf{Cells:} RR (95\% CrI) and Probs $= \mathrm{Prob}(|\log(\mathrm{RR})|>0)$. Reference category in brackets. Spatiotemporal random effects and ABS population offset as described in Methods. \textbf{Shading:} Prob $\geq 0.90$ (light) or $95\%$ CrI excludes unity (darker). Marginal heat RR after adjustment for pollutants and SES (SES shown in Table~\ref{tab:4b}). O$_3$ categories collapsed at $\geq 100\,\mu$g/m$^3$.}
\label{tab:4a}
\scriptsize
\setlength{\tabcolsep}{3pt}

\begin{tabular}{@{}lcccccc@{}}
\toprule
 & All & Probs & $\geq 65$ years & Probs & $\geq 85$ years & Probs \\
\midrule
Heatwave$_{t-3}$ & \cellcolor{sig95}1.142 (1.116--1.167) & 1.000 & \cellcolor{sig95}1.152 (1.124--1.180) & 1.000 & \cellcolor{sig95}1.170 (1.133--1.208) & 1.000 \\
mean(PM$_{10}$, $t-1$ to $t-3$) [$<15.0\,\mu$g/m$^3$] & & & & & & \\
\quad $15.0$--$44.9\,\mu$g/m$^3$ & \cellcolor{sig95}1.037 (1.018--1.057) & 1.000 & \cellcolor{sig95}1.037 (1.016--1.059) & 1.000 & \cellcolor{sig95}1.045 (1.016--1.074) & 0.999 \\
\quad $\geq 45.0\,\mu$g/m$^3$ & \cellcolor{sig90}1.046 (0.988--1.107) & 0.939 & 1.029 (0.968--1.094) & 0.818 & 1.009 (0.929--1.097) & 0.586 \\
mean(O$_3$, $t-1$ to $t-3$) [$<60.0\,\mu$g/m$^3$] & & & & & & \\
\quad $60.0$--$99.9\,\mu$g/m$^3$ & 1.004 (0.993--1.015) & 0.772 & 1.007 (0.995--1.019) & 0.864 & 1.010 (0.994--1.026) & 0.879 \\
\quad $\geq 100.0\,\mu$g/m$^3$ & \cellcolor{sig90}1.045 (0.983--1.111) & 0.922 & \cellcolor{sig90}1.046 (0.980--1.116) & 0.910 & 1.054 (0.966--1.149) & 0.882 \\
mean(NO$_2$, $t-1$ to $t-3$) [$<10.0\,\mu$g/m$^3$] & & & & & & \\
\quad $10.0$--$24.9\,\mu$g/m$^3$ & 0.999 (0.975--1.022) & 0.550 & 0.998 (0.973--1.023) & 0.569 & 0.998 (0.964--1.032) & 0.553 \\
\quad $\geq 25.0\,\mu$g/m$^3$ & \cellcolor{sig95}1.050 (1.023--1.077) & 1.000 & \cellcolor{sig95}1.051 (1.023--1.081) & 1.000 & \cellcolor{sig95}1.051 (1.013--1.091) & 0.996 \\
Q4 relative humidity$_{t-3}$ [No] & 0.998 (0.987--1.009) & 0.632 & 0.997 (0.985--1.009) & 0.701 & 0.995 (0.979--1.012) & 0.712 \\
\bottomrule
\end{tabular}
\end{table}

%% file: tables/table4b.tex
\begin{table}[htbp]
\centering
\caption{Main-effects models including heatwaves: contextual socioeconomic and demographic quartiles. \textbf{Cells:} RR (95\% CrI) and Probs $= \mathrm{Prob}(|\log(\mathrm{RR})|>0)$. Reference category in brackets. Spatiotemporal random effects and ABS population offset as described in Methods. \textbf{Shading:} Prob $\geq 0.90$ (light) or $95\%$ CrI excludes unity (darker). Same models as Table~\ref{tab:4a}; heat and pollutant coefficients reported there.}
\label{tab:4b}
\scriptsize
\setlength{\tabcolsep}{3pt}

\begin{tabular}{@{}lcccccc@{}}
\toprule
 & All & Probs & $\geq 65$ years & Probs & $\geq 85$ years & Probs \\
\midrule
Average net income per person [Q1] & & & & & & \\
\quad Q2 & 1.006 (0.968--1.045) & 0.616 & 1.008 (0.968--1.050) & 0.653 & 1.018 (0.963--1.076) & 0.738 \\
\quad Q3 & 1.002 (0.949--1.057) & 0.523 & 0.994 (0.938--1.054) & 0.577 & 0.967 (0.893--1.046) & 0.801 \\
\quad Q4 & 1.017 (0.946--1.094) & 0.680 & 1.012 (0.937--1.094) & 0.623 & 0.973 (0.877--1.080) & 0.694 \\
Gini index [Q1] & & & & & & \\
\quad Q2 & \cellcolor{sig95}1.050 (1.019--1.083) & 0.999 & \cellcolor{sig95}1.061 (1.027--1.096) & 1.000 & \cellcolor{sig95}1.058 (1.012--1.106) & 0.993 \\
\quad Q3 & \cellcolor{sig95}1.075 (1.034--1.118) & 1.000 & \cellcolor{sig95}1.089 (1.045--1.134) & 1.000 & \cellcolor{sig95}1.081 (1.022--1.144) & 0.997 \\
\quad Q4 & \cellcolor{sig95}1.071 (1.020--1.125) & 0.997 & \cellcolor{sig95}1.076 (1.022--1.133) & 0.997 & 1.047 (0.975--1.123) & 0.897 \\
Population aged 65 or over [Q1] & & & & & & \\
\quad Q2 & 1.022 (0.986--1.059) & 0.881 & 1.019 (0.981--1.059) & 0.838 & 0.994 (0.943--1.047) & 0.594 \\
\quad Q3 & \cellcolor{sig95}1.054 (1.007--1.102) & 0.989 & \cellcolor{sig95}1.060 (1.010--1.112) & 0.991 & 1.012 (0.947--1.081) & 0.633 \\
\quad Q4 & \cellcolor{sig95}1.083 (1.021--1.148) & 0.996 & \cellcolor{sig95}1.094 (1.028--1.165) & 0.998 & 1.043 (0.958--1.135) & 0.836 \\
\bottomrule
\end{tabular}
\end{table}

%% file: tables/table4c.tex
\begin{table}[htbp]
\centering
\caption{Interaction models including heatwaves: heat$\times$pollutant / humidity terms. \textbf{Cells:} RR (95\% CrI) and Probs $= \mathrm{Prob}(|\log(\mathrm{RR})|>0)$. Reference category in brackets. Spatiotemporal random effects and ABS population offset as described in Methods. \textbf{Shading:} Prob $\geq 0.90$ (light) or $95\%$ CrI excludes unity (darker). $^*$\textbf{Heat coefficient} is the association in the reference covariate stratum (lowest pollutant categories, humidity not in Q4, SES Q1); marginal heat associations are reported in Tables~\ref{tab:4a}--\ref{tab:4b}.}
\label{tab:4c}
\scriptsize
\setlength{\tabcolsep}{3pt}

\begin{tabular}{@{}lcccccc@{}}
\toprule
 & All & Probs & $\geq 65$ years & Probs & $\geq 85$ years & Probs \\
\midrule
Heatwave$_{t-3}$ (reference stratum)$^*$ & 0.907 (0.628--1.312) & 0.698 & 1.031 (0.707--1.504) & 0.563 & 0.936 (0.562--1.560) & 0.601 \\
Heatwave$_{t-3}$ $\times$ mean(PM$_{10}$, $t-1$ to $t-3$) & & & & & & \\
\quad $15.0$--$44.9\,\mu$g/m$^3$ & 1.165 (0.820--1.653) & 0.803 & 1.043 (0.730--1.489) & 0.591 & 1.115 (0.689--1.804) & 0.670 \\
\quad $\geq 45.0\,\mu$g/m$^3$ & 1.010 (0.701--1.457) & 0.521 & 0.932 (0.641--1.355) & 0.645 & 0.941 (0.568--1.560) & 0.594 \\
Heatwave$_{t-3}$ $\times$ mean(O$_3$, $t-1$ to $t-3$) & & & & & & \\
\quad $60.0$--$99.9\,\mu$g/m$^3$ & 0.996 (0.921--1.077) & 0.543 & 0.992 (0.912--1.080) & 0.572 & 0.979 (0.874--1.098) & 0.640 \\
\quad $\geq 100.0\,\mu$g/m$^3$ & 1.077 (0.929--1.248) & 0.837 & 1.071 (0.915--1.255) & 0.803 & 0.994 (0.804--1.227) & 0.524 \\
Heatwave$_{t-3}$ $\times$ mean(NO$_2$, $t-1$ to $t-3$) & & & & & & \\
\quad $10.0$--$24.9\,\mu$g/m$^3$ & \cellcolor{sig90}1.064 (0.974--1.162) & 0.914 & \cellcolor{sig90}1.090 (0.991--1.200) & 0.962 & \cellcolor{sig90}1.129 (0.993--1.283) & 0.968 \\
\quad $\geq 25.0\,\mu$g/m$^3$ & \cellcolor{sig90}1.092 (0.996--1.197) & 0.970 & \cellcolor{sig95}1.114 (1.009--1.230) & 0.984 & \cellcolor{sig95}1.171 (1.026--1.338) & 0.990 \\
Heatwave$_{t-3}$ $\times$ Q4 relative humidity$_{t-3}$ & 1.014 (0.919--1.119) & 0.608 & 1.008 (0.906--1.122) & 0.558 & \cellcolor{sig90}1.098 (0.955--1.262) & 0.906 \\
\bottomrule
\end{tabular}
\end{table}

%% file: tables/table4d.tex
\begin{table}[htbp]
\centering
\caption{Interaction models including heatwaves: heat$\times$socioeconomic / demographic terms. \textbf{Cells:} RR (95\% CrI) and Probs $= \mathrm{Prob}(|\log(\mathrm{RR})|>0)$. Reference category in brackets. Spatiotemporal random effects and ABS population offset as described in Methods. \textbf{Shading:} Prob $\geq 0.90$ (light) or $95\%$ CrI excludes unity (darker). Same interaction models as Table~\ref{tab:4c}; reference-stratum heat and pollutant interactions reported there.}
\label{tab:4d}
\scriptsize
\setlength{\tabcolsep}{3pt}

\begin{tabular}{@{}lcccccc@{}}
\toprule
 & All & Probs & $\geq 65$ years & Probs & $\geq 85$ years & Probs \\
\midrule
Heatwave$_{t-3}$ $\times$ average net income & & & & & & \\
\quad Q2 & 1.037 (0.975--1.102) & 0.874 & 1.020 (0.954--1.090) & 0.720 & 1.032 (0.942--1.130) & 0.750 \\
\quad Q3 & \cellcolor{sig90}1.062 (0.996--1.131) & 0.968 & \cellcolor{sig90}1.057 (0.988--1.132) & 0.945 & 1.056 (0.963--1.159) & 0.876 \\
\quad Q4 & \cellcolor{sig90}1.061 (1.000--1.127) & 0.975 & \cellcolor{sig90}1.058 (0.992--1.128) & 0.957 & 1.019 (0.934--1.112) & 0.665 \\
Heatwave$_{t-3}$ $\times$ Gini index & & & & & & \\
\quad Q2 & 1.015 (0.953--1.082) & 0.677 & 0.996 (0.931--1.067) & 0.543 & 1.011 (0.921--1.111) & 0.592 \\
\quad Q3 & 0.992 (0.929--1.058) & 0.598 & 0.984 (0.918--1.055) & 0.672 & 0.986 (0.896--1.085) & 0.617 \\
\quad Q4 & 1.002 (0.938--1.071) & 0.524 & 0.983 (0.916--1.056) & 0.678 & 1.031 (0.936--1.136) & 0.733 \\
Heatwave$_{t-3}$ $\times$ population aged 65 or over & & & & & & \\
\quad Q2 & \cellcolor{sig90}0.947 (0.892--1.005) & 0.964 & \cellcolor{sig90}0.940 (0.882--1.003) & 0.969 & 0.977 (0.895--1.066) & 0.701 \\
\quad Q3 & 0.974 (0.917--1.034) & 0.808 & 0.959 (0.899--1.024) & 0.896 & 0.967 (0.885--1.056) & 0.775 \\
\quad Q4 & 0.986 (0.928--1.048) & 0.673 & 0.967 (0.906--1.032) & 0.842 & 0.966 (0.884--1.055) & 0.782 \\
\bottomrule
\end{tabular}
\end{table}

%% file: tables/table4e.tex
\begin{table}[htbp]
\centering
\caption{Stratum-specific combined heat effects for heatwaves by pollutant / humidity level. \textbf{Cells:} stratum-specific heat RR $=\exp(\beta_H)$ (reference) or $\exp(\beta_H+\delta_{H\times\mathrm{modifier}})$ (non-reference), holding other modifiers at their reference level (2{,}000 INLA draws). Probs $= \mathrm{Prob}(|\log(\mathrm{RR})|>0)$. \textbf{Shading:} Prob $\geq 0.90$ (light) or $95\%$ CrI excludes unity (darker).}
\label{tab:4e}
\scriptsize
\setlength{\tabcolsep}{3pt}

\begin{tabular}{@{}lcccccc@{}}
\toprule
 & All & Probs & $\geq 65$ years & Probs & $\geq 85$ years & Probs \\
\midrule
Heatwave$_{t-3}$ | reference stratum & 0.909 (0.628--1.321) & 0.691 & 1.027 (0.715--1.471) & 0.561 & 0.939 (0.555--1.558) & 0.592 \\
Heatwave$_{t-3}$ by PM$_{10}$ stratum & & & & & & \\
\quad $15.0$--$44.9\,\mu$g/m$^3$ & 1.055 (0.927--1.193) & 0.802 & 1.075 (0.936--1.225) & 0.850 & 1.046 (0.859--1.267) & 0.683 \\
\quad $\geq 45.0\,\mu$g/m$^3$ & 0.915 (0.773--1.072) & 0.845 & 0.960 (0.808--1.144) & 0.669 & 0.883 (0.681--1.131) & 0.836 \\
Heatwave$_{t-3}$ by O$_3$ stratum & & & & & & \\
\quad $60.0$--$99.9\,\mu$g/m$^3$ & 0.906 (0.633--1.300) & 0.698 & 1.020 (0.709--1.461) & 0.543 & 0.922 (0.553--1.524) & 0.617 \\
\quad $\geq 100.0\,\mu$g/m$^3$ & 0.983 (0.677--1.437) & 0.544 & 1.100 (0.744--1.603) & 0.693 & 0.933 (0.552--1.589) & 0.596 \\
Heatwave$_{t-3}$ by NO$_2$ stratum & & & & & & \\
\quad $10.0$--$24.9\,\mu$g/m$^3$ & 0.968 (0.668--1.401) & 0.570 & 1.117 (0.785--1.598) & 0.728 & 1.057 (0.637--1.750) & 0.581 \\
\quad $\geq 25.0\,\mu$g/m$^3$ & 0.994 (0.686--1.440) & 0.514 & 1.141 (0.792--1.635) & 0.764 & 1.097 (0.662--1.809) & 0.632 \\
Heatwave$_{t-3}$ | Q4 relative humidity & 0.922 (0.631--1.332) & 0.661 & 1.034 (0.714--1.479) & 0.572 & 1.028 (0.597--1.742) & 0.543 \\
\bottomrule
\end{tabular}
\end{table}

%% file: tables/table4f.tex
\begin{table}[htbp]
\centering
\caption{Stratum-specific combined heat effects for heatwaves by socioeconomic / demographic quartile. \textbf{Cells:} stratum-specific heat RR $=\exp(\beta_H)$ (reference) or $\exp(\beta_H+\delta_{H\times\mathrm{modifier}})$ (non-reference), holding other modifiers at their reference level (2{,}000 INLA draws). Probs $= \mathrm{Prob}(|\log(\mathrm{RR})|>0)$. \textbf{Shading:} Prob $\geq 0.90$ (light) or $95\%$ CrI excludes unity (darker).}
\label{tab:4f}
\scriptsize
\setlength{\tabcolsep}{3pt}

\begin{tabular}{@{}lcccccc@{}}
\toprule
 & All & Probs & $\geq 65$ years & Probs & $\geq 85$ years & Probs \\
\midrule
Heatwave$_{t-3}$ | reference stratum & 0.909 (0.628--1.321) & 0.691 & 1.027 (0.715--1.471) & 0.561 & 0.939 (0.555--1.558) & 0.592 \\
Heatwave$_{t-3}$ by income quartile & & & & & & \\
\quad Q2 & 0.943 (0.649--1.387) & 0.624 & 1.047 (0.732--1.501) & 0.598 & 0.967 (0.576--1.609) & 0.553 \\
\quad Q3 & 0.966 (0.668--1.404) & 0.574 & 1.086 (0.754--1.550) & 0.670 & 0.992 (0.597--1.641) & 0.515 \\
\quad Q4 & 0.966 (0.663--1.407) & 0.569 & 1.086 (0.755--1.548) & 0.668 & 0.957 (0.567--1.592) & 0.570 \\
Heatwave$_{t-3}$ by Gini quartile & & & & & & \\
\quad Q2 & 0.924 (0.642--1.341) & 0.671 & 1.024 (0.714--1.459) & 0.547 & 0.948 (0.566--1.562) & 0.576 \\
\quad Q3 & 0.902 (0.626--1.321) & 0.707 & 1.010 (0.704--1.458) & 0.515 & 0.925 (0.556--1.533) & 0.607 \\
\quad Q4 & 0.911 (0.632--1.328) & 0.692 & 1.010 (0.699--1.444) & 0.512 & 0.966 (0.579--1.598) & 0.555 \\
Heatwave$_{t-3}$ by \% aged $\geq 65$ quartile & & & & & & \\
\quad Q2 & 0.861 (0.589--1.253) & 0.791 & 0.966 (0.670--1.392) & 0.569 & 0.915 (0.547--1.510) & 0.626 \\
\quad Q3 & 0.886 (0.612--1.283) & 0.745 & 0.985 (0.685--1.402) & 0.536 & 0.906 (0.540--1.512) & 0.640 \\
\quad Q4 & 0.898 (0.625--1.309) & 0.711 & 0.993 (0.692--1.426) & 0.512 & 0.906 (0.531--1.506) & 0.645 \\
\bottomrule
\end{tabular}
\end{table}

%% file: si_tables/tableS1a.tex
\begin{table}[H]
\centering
\caption{Lag selection for extreme maximum temperature indicators (outcome: all-age summer deaths). Screening models used main effects only with the nested spatiotemporal structure (Mat\'ern SPDE field evolving over years; month nested within year; IID ABS frailty) and a zero-inflated negative binomial likelihood, on a common summer sample so that WAIC is comparable. Selected lag = minimum WAIC. Final nested ZINB models reused the selected lag.}
\label{tab:S1a}
\begin{tabular}{@{}cccccc@{}}
\toprule
Lag & $n$ & WAIC & $\Delta$WAIC & DIC & Selected \\
\midrule
1 & 495,732 & 807463.5 & 5.9 & 807463.6 &  \\
2 & 495,732 & 807457.6 & 0.0 & 807457.5 & Yes \\
3 & 495,732 & 807481.7 & 24.0 & 807481.7 &  \\
4 & 495,732 & 807535.4 & 77.8 & 807534.6 &  \\
5 & 495,732 & 807577.6 & 120.0 & 807577.8 &  \\
6 & 495,732 & 807611.7 & 154.1 & 807611.4 &  \\
7 & 495,732 & 807628.5 & 170.9 & 807628.6 &  \\
\bottomrule
\end{tabular}
\end{table}

%% file: si_tables/tableS1b.tex
\begin{table}[H]
\centering
\caption{Lag selection for heatwave indicators (outcome: all-age summer deaths). Screening models used main effects only with the nested spatiotemporal structure (Mat\'ern SPDE field evolving over years; month nested within year; IID ABS frailty) and a zero-inflated negative binomial likelihood, on a common summer sample so that WAIC is comparable. Selected lag = minimum WAIC. Final nested ZINB models reused the selected lag.}
\label{tab:S1b}
\begin{tabular}{@{}cccccc@{}}
\toprule
Lag & $n$ & WAIC & $\Delta$WAIC & DIC & Selected \\
\midrule
1 & 495,732 & 807465.5 & 15.1 & 807464.8 &  \\
2 & 495,732 & 807457.1 & 6.7 & 807457.2 &  \\
3 & 495,732 & 807450.4 & 0.0 & 807449.6 & Yes \\
4 & 495,732 & 807513.6 & 63.2 & 807513.7 &  \\
5 & 495,732 & 807548.1 & 97.7 & 807548.2 &  \\
6 & 495,732 & 807602.9 & 152.4 & 807602.8 &  \\
7 & 495,732 & 807623.0 & 172.6 & 807622.4 &  \\
\bottomrule
\end{tabular}
\end{table}

%% file: si_tables/tableS2a.tex
\begin{table}[H]
\centering
\caption{Pollutant category $\times$ extreme heat (lag~2) cell counts in the complete environmental summer sample (all-cause deaths). Sparse high-ozone cells motivated collapsing O$_3\geq 100\,\mu$g/m$^3$ in the analysis models.}
\label{tab:S2a}
\scriptsize
\begin{tabular}{@{}llrrr@{}}
\toprule
Pollutant stratum & Heat & ABS-days & Deaths & Deaths/day \\
\midrule
NO$_2$ (10,25] & No & 306,305 & 126,623 & 0.413 \\
NO$_2$ (10,25] & Yes & 18,442 & 9,210 & 0.499 \\
NO$_2$ (25,661] & No & 126,293 & 55,382 & 0.439 \\
NO$_2$ (25,661] & Yes & 11,927 & 6,017 & 0.504 \\
NO$_2$ [0,10] & No & 32,371 & 12,204 & 0.377 \\
NO$_2$ [0,10] & Yes & 2,289 & 1,118 & 0.488 \\
O$_3$ (100,163] & No & 3,433 & 1,002 & 0.292 \\
O$_3$ (100,163] & Yes & 1,427 & 633 & 0.444 \\
O$_3$ (60,100] & No & 288,827 & 118,955 & 0.412 \\
O$_3$ (60,100] & Yes & 27,598 & 13,973 & 0.506 \\
O$_3$ [16.6,60] & No & 172,709 & 74,252 & 0.430 \\
O$_3$ [16.6,60] & Yes & 3,633 & 1,739 & 0.479 \\
PM$_{10}$ (15,45] & No & 412,669 & 175,857 & 0.426 \\
PM$_{10}$ (15,45] & Yes & 30,312 & 15,323 & 0.506 \\
PM$_{10}$ (45,81] & No & 2,099 & 1,009 & 0.481 \\
PM$_{10}$ (45,81] & Yes & 1,905 & 894 & 0.469 \\
PM$_{10}$ [0,15] & No & 50,201 & 17,343 & 0.345 \\
PM$_{10}$ [0,15] & Yes & 441 & 128 & 0.290 \\
\bottomrule
\end{tabular}
\end{table}

%% file: si_tables/tableS2b.tex
\begin{table}[H]
\centering
\caption{Pollutant category $\times$ heatwave (lag~3) cell counts in the complete environmental summer sample (all-cause deaths).}
\label{tab:S2b}
\scriptsize
\begin{tabular}{@{}llrrr@{}}
\toprule
Pollutant stratum & Heat & ABS-days & Deaths & Deaths/day \\
\midrule
NO$_2$ (10,25] & No & 319,909 & 132,383 & 0.414 \\
NO$_2$ (10,25] & Yes & 10,801 & 5,634 & 0.522 \\
NO$_2$ (25,586] & No & 129,806 & 57,383 & 0.442 \\
NO$_2$ (25,586] & Yes & 6,926 & 3,683 & 0.532 \\
NO$_2$ [0,10] & No & 28,547 & 10,706 & 0.375 \\
NO$_2$ [0,10] & Yes & 1,259 & 610 & 0.485 \\
O$_3$ (100,151] & No & 3,104 & 859 & 0.277 \\
O$_3$ (100,151] & Yes & 826 & 354 & 0.429 \\
O$_3$ (60,100] & No & 304,580 & 126,095 & 0.414 \\
O$_3$ (60,100] & Yes & 16,750 & 8,832 & 0.527 \\
O$_3$ [22.1,60] & No & 170,578 & 73,518 & 0.431 \\
O$_3$ [22.1,60] & Yes & 1,410 & 741 & 0.526 \\
PM$_{10}$ (15,45] & No & 432,182 & 184,594 & 0.427 \\
PM$_{10}$ (15,45] & Yes & 17,658 & 9,351 & 0.530 \\
PM$_{10}$ (45,79.1] & No & 1,941 & 928 & 0.478 \\
PM$_{10}$ (45,79.1] & Yes & 1,211 & 544 & 0.449 \\
PM$_{10}$ [0,15] & No & 44,139 & 14,950 & 0.339 \\
PM$_{10}$ [0,15] & Yes & 117 & 32 & 0.274 \\
\bottomrule
\end{tabular}
\end{table}

%% file: si_tables/tableS3.tex
\begin{table}[H]
\centering
\caption{Bayesian false-discovery control for interaction-model terms (interaction coefficients only) at target FDR $\alpha=0.05$ under the nested space--time ZINB fits. Discoveries ranked by posterior Prob($|\log\mathrm{RR}|>0$). At most 40 leading discoveries shown.}
\label{tab:S3}
\scriptsize
\setlength{\tabcolsep}{3pt}
\begin{tabular}{@{}llllrr@{}}
\toprule
Exposure & Age & Term & RR & Prob & local FDR ($\psi$) \\
\midrule
Extreme heat & All & EH$_{t-2}$ $\times$ O$_3$(100,163] & 1.189 & 0.999 & 0.001 \\
Extreme heat & $\geq 85$ & EH$_{t-2}$ $\times$ O$_3$(60,100] & 1.110 & 0.996 & 0.004 \\
Extreme heat & All & EH$_{t-2}$ $\times$ O$_3$(60,100] & 1.071 & 0.995 & 0.005 \\
Extreme heat & $\geq 65$ & EH$_{t-2}$ $\times$ O$_3$(100,163] & 1.170 & 0.994 & 0.006 \\
Extreme heat & All & EH$_{t-2}$ $\times$ Income Q4 & 1.059 & 0.992 & 0.008 \\
Heatwave & $\geq 85$ & HW$_{t-3}$ $\times$ NO$_2$(25,586] & 1.171 & 0.990 & 0.010 \\
Extreme heat & $\geq 65$ & EH$_{t-2}$ $\times$ O$_3$(60,100] & 1.064 & 0.985 & 0.015 \\
Heatwave & $\geq 65$ & HW$_{t-3}$ $\times$ NO$_2$(25,586] & 1.114 & 0.984 & 0.016 \\
Extreme heat & $\geq 85$ & PM$_{10}$(15,45] $\times$ EH$_{t-2}$ & 1.334 & 0.980 & 0.020 \\
Extreme heat & $\geq 85$ & EH$_{t-2}$ $\times$ O$_3$(100,163] & 1.185 & 0.980 & 0.020 \\
Extreme heat & $\geq 65$ & EH$_{t-2}$ $\times$ Income Q4 & 1.053 & 0.978 & 0.022 \\
Heatwave & All & HW$_{t-3}$ $\times$ Income Q4 & 1.062 & 0.975 & 0.025 \\
Heatwave & All & HW$_{t-3}$ $\times$ NO$_2$(25,586] & 1.092 & 0.970 & 0.030 \\
Heatwave & $\geq 65$ & HW$_{t-3}$ $\times$ \%65 Q2 & 0.940 & 0.969 & 0.031 \\
Extreme heat & $\geq 65$ & EH$_{t-2}$ $\times$ \%65 Q4 & 0.952 & 0.969 & 0.031 \\
Heatwave & $\geq 85$ & HW$_{t-3}$ $\times$ NO$_2$(10,25] & 1.129 & 0.968 & 0.032 \\
Heatwave & All & HW$_{t-3}$ $\times$ Income Q3 & 1.062 & 0.968 & 0.032 \\
Heatwave & All & HW$_{t-3}$ $\times$ \%65 Q2 & 0.947 & 0.964 & 0.036 \\
Heatwave & $\geq 65$ & HW$_{t-3}$ $\times$ NO$_2$(10,25] & 1.090 & 0.961 & 0.039 \\
Heatwave & $\geq 65$ & HW$_{t-3}$ $\times$ Income Q4 & 1.058 & 0.956 & 0.044 \\
Extreme heat & All & EH$_{t-2}$ $\times$ Income Q3 & 1.043 & 0.950 & 0.050 \\
Heatwave & $\geq 65$ & HW$_{t-3}$ $\times$ Income Q3 & 1.057 & 0.945 & 0.055 \\
Extreme heat & All & PM$_{10}$(15,45] $\times$ EH$_{t-2}$ & 1.146 & 0.935 & 0.065 \\
Heatwave & All & HW$_{t-3}$ $\times$ NO$_2$(10,25] & 1.064 & 0.914 & 0.086 \\
Extreme heat & All & EH$_{t-2}$ $\times$ Q4 RH & 0.956 & 0.914 & 0.086 \\
Extreme heat & $\geq 65$ & EH$_{t-2}$ $\times$ Gini Q2 & 0.963 & 0.912 & 0.088 \\
Extreme heat & $\geq 85$ & EH$_{t-2}$ $\times$ Gini Q2 & 0.950 & 0.910 & 0.090 \\
Heatwave & $\geq 85$ & HW$_{t-3}$ $\times$ Q4 RH & 1.098 & 0.906 & 0.094 \\
Extreme heat & $\geq 65$ & EH$_{t-2}$ $\times$ Income Q3 & 1.037 & 0.904 & 0.096 \\
Extreme heat & $\geq 65$ & EH$_{t-2}$ $\times$ \%65 Q3 & 0.967 & 0.904 & 0.096 \\
Extreme heat & $\geq 85$ & EH$_{t-2}$ $\times$ NO$_2$(10,25] & 0.941 & 0.901 & 0.099 \\
Extreme heat & $\geq 65$ & PM$_{10}$(15,45] $\times$ EH$_{t-2}$ & 1.130 & 0.899 & 0.101 \\
Extreme heat & $\geq 65$ & EH$_{t-2}$ $\times$ \%65 Q2 & 0.968 & 0.897 & 0.103 \\
Heatwave & $\geq 65$ & HW$_{t-3}$ $\times$ \%65 Q3 & 0.959 & 0.896 & 0.104 \\
\bottomrule
\end{tabular}
\end{table}

%% file: bibfile.bib
@article{ballester2023heat,
  title={Heat-related mortality in Europe during the summer of 2022},
  author={Ballester, Joan and Quijal-Zamorano, Marcos and M{\'e}ndez Turrubiates, Ra{\'u}l Fernando and Pegenaute, Ferran and Herrmann, Fran{\c{c}}ois R and Robine, Jean Marie and Basaga{\~n}a, Xavier and Tonne, Cathryn and Ant{\'o}, Josep M and Achebak, Hicham},
  journal={Nature medicine},
  volume={29},
  number={7},
  pages={1857--1866},
  year={2023},
  publisher={Nature Publishing Group US New York},
  doi={10.1038/s41591-023-02419-z}
}

@article{beck2024mortality,
  title={Mortality burden attributed to anthropogenic warming during Europe’s 2022 record-breaking summer},
  author={Beck, Thessa M and Schumacher, Dominik L and Achebak, Hicham and Vicedo--Cabrera, Ana M and Seneviratne, Sonia I and Ballester, Joan},
  journal={NPJ climate and atmospheric science},
  volume={7},
  number={1},
  pages={245},
  year={2024},
  publisher={Nature Publishing Group UK London},
  doi={10.1038/s41612-024-00783-2}
}

@article{murray2020global,
  title={Global burden of 87 risk factors in 204 countries and territories, 1990--2019: a systematic analysis for the Global Burden of Disease Study 2019},
  author={Murray, Christopher JL and Aravkin, Aleksandr Y and Zheng, Peng and Abbafati, Cristiana and Abbas, Kaja M and Abbasi-Kangevari, Mohsen and Abd-Allah, Foad and Abdelalim, Ahmed and Abdollahi, Mohammad and Abdollahpour, Ibrahim and others},
  journal={The lancet},
  volume={396},
  number={10258},
  pages={1223--1249},
  year={2020},
  publisher={Elsevier},
  doi={10.1016/S0140-6736(20)30752-2}
}

@article{barriopedro2023heat,
  title={Heat waves: Physical understanding and scientific challenges},
  author={Barriopedro, David and Garc{\'\i}a-Herrera, R and Ord{\'o}nez, Carlos and Miralles, Diego G and Salcedo-Sanz, Sancho},
  journal={Reviews of Geophysics},
  volume={61},
  number={2},
  pages={e2022RG000780},
  year={2023},
  publisher={Wiley Online Library},
  doi={10.1029/2022RG000780}
}

@misc{epa2025climate,
  author       = {{United States Environmental Protection Agency}},
  title        = {Climate Change Indicators},
  year         = {2025},
  howpublished = {\url{https://www.epa.gov/climate-indicators/climate-change-indicators-heat-waves}},
  note         = {[Accessed: October 1, 2025]},
}

@article{batibeniz2025rapid,
  title={Rapid climate action is needed: comparing heat vs. COVID-19-related mortality},
  author={Batibeniz, Fulden and Seneviratne, Sonia I and Jha, Srinidhi and Ribeiro, Andreia and Suarez Gutierrez, Laura and Raible, Christoph C and Malhotra, Avni and Armstrong, Ben and Bell, Michelle L and Lavigne, Eric and others},
  journal={Scientific reports},
  volume={15},
  number={1},
  pages={1002},
  year={2025},
  publisher={Nature Publishing Group UK London},
  doi={10.1038/s41598-024-82788-8}
}

@article{luthi2023rapid,
  title={Rapid increase in the risk of heat-related mortality},
  author={L{\"u}thi, Samuel and Fairless, Christopher and Fischer, Erich M and Scovronick, Noah and Armstrong, Ben and Coelho, Micheline De Sousa Zanotti Stagliorio and Guo, Yue Leon and Guo, Yuming and Honda, Yasushi and Huber, Veronika and others},
  journal={Nature communications},
  volume={14},
  number={1},
  pages={4894},
  year={2023},
  publisher={Nature Publishing Group UK London},
  doi={10.1038/s41467-023-40599-x}
}

@article{brauer2024global,
  title={Global burden and strength of evidence for 88 risk factors in 204 countries and 811 subnational locations, 1990--2021: a systematic analysis for the Global Burden of Disease Study 2021},
  author={Brauer, Michael and Roth, Gregory A and Aravkin, Aleksandr Y and Zheng, Peng and Abate, Kalkidan Hassen and Abate, Yohannes Habtegiorgis and Abbafati, Cristiana and Abbasgholizadeh, Rouzbeh and Abbasi, Madineh Akram and Abbasian, Mohammadreza and others},
  journal={The Lancet},
  volume={403},
  number={10440},
  pages={2162--2203},
  year={2024},
  publisher={Elsevier},
  doi={10.1016/S0140-6736(24)00933-4}
}

@article{fuller2022pollution,
  title={Pollution and health: a progress update},
  author={Fuller, Richard and Landrigan, Philip J and Balakrishnan, Kalpana and Bathan, Glynda and Bose-O'Reilly, Stephan and Brauer, Michael and Caravanos, Jack and Chiles, Tom and Cohen, Aaron and Corra, Lilian and others},
  journal={The Lancet Planetary Health},
  volume={6},
  number={6},
  pages={e535--e547},
  year={2022},
  publisher={Elsevier},
  doi={10.1016/S2542-5196(22)00090-0}
}

@article{xu2023ozone,
  title={Ozone, heat wave, and cardiovascular disease mortality: a population-based case-crossover study},
  author={Xu, Ruijun and Sun, Hong and Zhong, Zihua and Zheng, Yi and Liu, Tingting and Li, Yingxin and Liu, Likun and Luo, Lu and Wang, Sirong and Lv, Ziquan and others},
  journal={Environmental Science \& Technology},
  volume={58},
  number={1},
  pages={171--181},
  year={2023},
  publisher={ACS Publications},
  doi={10.1021/acs.est.3c06889}
}

@article{qi2023modification,
  title={The modification effect of ozone pollution on the associations between heat wave and cardiovascular mortality},
  author={Qi, J and Wang, Y and Wang, L and Cao, R and Huang, J and Li, G and Yin, P},
  journal={Innovation},
  volume={1},
  number={3},
  pages={100043},
  year={2023},
  doi={10.59717/j.xinn-med.2023.100043}
}

@article{xu2023extreme,
  title={Extreme temperature events, fine particulate matter, and myocardial infarction mortality},
  author={Xu, Ruijun and Huang, Suli and Shi, Chunxiang and Wang, Rui and Liu, Tingting and Li, Yingxin and Zheng, Yi and Lv, Ziquan and Wei, Jing and Sun, Hong and others},
  journal={Circulation},
  volume={148},
  number={4},
  pages={312--323},
  year={2023},
  publisher={Lippincott Williams \& Wilkins Hagerstown, MD},
  doi={10.1161/CIRCULATIONAHA.122.063504}
}

@article{deng2024short,
  title={Short-term exposure to PM2. 5 constituents, extreme temperature events and stroke mortality},
  author={Deng, Boning and Zhu, Lifeng and Zhang, Yuanyuan and Tang, Ziqing and Shen, Jiajun and Zhang, Yalin and Zheng, Hao and Zhang, Yunquan},
  journal={Science of The Total Environment},
  volume={954},
  pages={176506},
  year={2024},
  publisher={Elsevier},
doi={10.1016/j.scitotenv.2024.176506}
}

@article{du2024exposure,
  title={Exposure to concurrent heatwaves and ozone pollution and associations with mortality risk: a nationwide study in China},
  author={Du, Hang and Yan, Meilin and Liu, Xin and Zhong, Yu and Ban, Jie and Lu, Kailai and Li, Tiantian},
  journal={Environmental health perspectives},
  volume={132},
  number={4},
  pages={047012},
  year={2024},
  doi={10.1289/EHP13790}
}

@article{dear2005effects,
  title={Effects of temperature and ozone on daily mortality during the August 2003 heat wave in France},
  author={Dear, Keith and Ranmuthugala, Geetha and Kjellstr{\"o}m, Tord and Skinner, Carol and Hanigan, Ivan},
  journal={Archives of environmental \& occupational health},
  volume={60},
  number={4},
  pages={205--212},
  year={2005},
  publisher={Taylor \& Francis},
  doi={10.3200/AEOH.60.4.205-212}
}

@article{alari2023role,
  title={The role of ozone as a mediator of the relationship between heat waves and mortality in 15 French urban areas},
  author={Alari, Anna and Chen, Chen and Schwarz, Lara and Hdansen, Kristen and Chaix, Basile and Benmarhnia, Tarik},
  journal={American Journal of Epidemiology},
  volume={192},
  number={6},
  pages={949--962},
  year={2023},
  publisher={Oxford University Press},
  doi={10.1093/aje/kwad032}
}

@article{rahman2022effects,
  title={The effects of coexposure to extremes of heat and particulate air pollution on mortality in California: implications for climate change},
  author={Rahman, Md Mostafijur and McConnell, Rob and Schlaerth, Hannah and Ko, Joseph and Silva, Sam and Lurmann, Frederick W and Palinkas, Lawrence and Johnston, Jill and Hurlburt, Michael and Yin, Hao and others},
  journal={American journal of respiratory and critical care medicine},
  volume={206},
  number={9},
  pages={1117--1127},
  year={2022},
  publisher={American Thoracic Society},
  doi={10.1164/rccm.202204-0657OC}
}

@article{analitis2014effects,
  title={Effects of heat waves on mortality: effect modification and confounding by air pollutants},
  author={Analitis, Antonis and Michelozzi, Paola and D’Ippoliti, Daniela and de’Donato, Francesca and Menne, Bettina and Matthies, Franziska and Atkinson, Richard W and I{\~n}iguez, Carmen and Basaga{\~n}a, Xavier and Schneider, Alexandra and others},
  journal={Epidemiology},
  volume={25},
  number={1},
  pages={15--22},
  year={2014},
  publisher={LWW},
  doi={10.1097/EDE.0b013e31828ac01b}
}

@article{gao2025synergistic,
  title={The synergistic and mediating effects of ozone on associations between high temperature, heatwaves and mortality in the Greater London area between 2010 and 2018},
  author={Gao, Juan and Wood, Dylan and Katsouyanni, Klea and Benmarhnia, Tarik and Evangelopoulos, Dimitris},
  journal={Environmental Research},
  volume={277},
  pages={121577},
  year={2025},
  publisher={Elsevier},
  doi={10.1016/j.envres.2025.121577}
}

@article{coenders2025moderation,
  title={Moderation effects and elasticities in compositional regression with a total. Application to Bayesian spatiotemporal modelling of all-cause mortality from environmental stressors},
  author={Coenders, Germ{\`a} and Palarea-Albaladejo, Javier and Saez, Marc and Barcel{\'o}, Maria A},
  journal={arXiv preprint arXiv:2505.07800},
  year={2025},
  doi={10.48550/arXiv.2505.07800}
}

@misc{miteco2025ozone,
  author       = {{Ministerio para la Transición Ecológica y el Reto Demográfico}},
  title        = {Effects on Health and Ecosystems: Ozone [in Spanish]},
  year         = {2025},
  howpublished = {Available at: \url{https://portal-miteco-stage.adobecqms.net/en/calidad-y-evaluacion-ambiental/temas/atmosfera-y-calidad-del-aire/calidad-del-aire/salud/ozono.html}},
  note         = {[Accessed: October 1, 2025]},
  institution  = {Gobierno de España}
}

@article{chen2025trends,
  title={Trends in population exposure to compound extreme-risk temperature and air pollution across 35 European countries: a modelling study},
  author={Chen, Zhao-Yue and Achebak, Hicham and Petetin, Herv{\'e} and Turrubiates, Ra{\'u}l Fernando M{\'e}ndez and Guo, Yuming and Garc{\'\i}a-Pando, Carlos P{\'e}rez and Ballester, Joan},
  journal={The Lancet Planetary Health},
  volume={9},
  number={5},
  pages={e384--e396},
  year={2025},
  publisher={Elsevier},
  doi={10.1016/S2542-5196(25)00048-8}
}

@article{achebak2023drivers,
  title={Drivers of the time-varying heat-cold-mortality association in Spain: a longitudinal observational study},
  author={Achebak, Hicham and Rey, Gregoire and Lloyd, Simon J and Quijal-Zamorano, Marcos and M{\'e}ndez-Turrubiates, Ra{\'u}l Fernando and Ballester, Joan},
  journal={Environment international},
  volume={182},
  pages={108284},
  year={2023},
  publisher={Elsevier},
  doi={10.1016/j.envint.2023.108284}
}

@article{feron2023compound,
  title={Compound climate-pollution extremes in Santiago de Chile},
  author={Feron, Sarah and Cordero, Ra{\'u}l R and Damiani, Alessandro and Oyola, Pedro and Ansari, Tabish and Pedemonte, Juan C and Wang, Chenghao and Ouyang, Zutao and Gallo, Valentina},
  journal={Scientific Reports},
  volume={13},
  number={1},
  pages={6726},
  year={2023},
  publisher={Nature Publishing Group UK London},
  doi={10.1038/s41598-023-33890-w}
}

@article{zhao2024global,
  title={Global, regional, and national burden of heatwave-related mortality from 1990 to 2019: A three-stage modelling study},
  author={Zhao, Qi and Li, Shanshan and Ye, Tingting and Wu, Yao and Gasparrini, Antonio and Tong, Shilu and Urban, Ale{\v{s}} and Vicedo-Cabrera, Ana Maria and Tobias, Aurelio and Armstrong, Ben and others},
  journal={PLoS medicine},
  volume={21},
  number={5},
  pages={e1004364},
  year={2024},
  publisher={Public Library of Science},
  doi={10.1371/journal.pmed.1004364}
}

@article{barcelo2025excess,
  author  = {Barcel{\'o}, Maria A. and Saez, Marc},
  title   = {Assessing Excess Mortality and Heat-Attributable Risk during the Summer of 2022 in Catalonia, Spain: A Bayesian Spatiotemporal Analysis},
  journal = {Journal of Geographical Systems},
  year    = {2025},
  doi     = {10.1007/s10109-025-00475-2}
}

@misc{xema2025data,
  author       = {{Departament de Territori, Habitatge i Transició Ecològica}},
  title        = {Dades meteorol{\`o}giques de la XEMA [in Catalan]},
  year         = {2025},
  howpublished = {Available at: \url{https://analisi.transparenciacatalunya.cat/ca/Medi-Ambient/Dades-meteorol-giques-de-la-XEMA/nzvn-apee/about_data}},
  note         = {[Accessed: October 1, 2025]}
}

@misc{xema2025airquality,
  author       = {{Departament de Territori, Habitatge i Transició Ecològica}},
  title        = {Qualitat de l’aire als punts de mesurament autom{\`a}tics de la Xarxa de Vigil{\`a}ncia i Previsi{\'o} de la Contaminaci{\'o} Atmosf{\`e}rica [in Catalan]},
  year         = {2025},
  howpublished = {Available at: \url{https://analisi.transparenciacatalunya.cat/Medi-Ambient/Qualitat-de-l-aire-als-punts-de-mesurament-autom-t/tasf-thgu/about_data}},
  note         = {[Accessed: October 1, 2025]}
}

@article{wannemuehler2009conditional,
  title={A conditional expectation approach for associating ambient air pollutant exposures with health outcomes},
  author={Wannemuehler, Kathleen A and Lyles, Robert H and Waller, Lance A and Hoekstra, Robert M and Klein, Mitchel and Tolbert, Paige},
  journal={Environmetrics: The official journal of the International Environmetrics Society},
  volume={20},
  number={7},
  pages={877--894},
  year={2009},
  publisher={Wiley Online Library},
  doi={10.1002/env.978}
}

@article{saez2022spatial,
  title={Spatial prediction of air pollution levels using a hierarchical Bayesian spatiotemporal model in Catalonia, Spain},
  author={S{\'a}ez Zafra, Marc and Barcel{\'o} Rado, Mar{\'\i}a Antonia},
  journal={Environmental Modelling \& Software},
  volume={151},
  pages={105369},
  year={2022},
  publisher={Elsevier},
  doi={10.1016/j.envsoft.2022.105369}
}

@misc{who2021airquality,
  author       = {{World Health Organization}},
  title        = {WHO Global Air Quality Guidelines: Particulate Matter (PM$_{2.5}$ and PM$_{10}$), Ozone, Nitrogen Dioxide, Sulphur Dioxide and Carbon Monoxide: Executive Summary},
  year         = {2021},
  address      = {Geneva},
  publisher    = {World Health Organization},
  howpublished = {Available at: \url{https://www.who.int/publications/i/item/9789240034228}},
  note         = {[Accessed: October 1, 2025]}
}

@misc{ine2022household,
  author       = {{INE}. Instituto Nacional de Estadística},
  title        = {Household Income Distribution Map},
  year         = {2022},
  howpublished = {\url{https://www.ine.es/dyngs/INEbase/en/operacion.htm?c=Estadistica_C&cid=1254736177088&menu=ultiDatos&idp=1254735976608}},
  note         = {Last accessed on October 1, 2025}
}

@misc{ine2025continuous,
  author       = {{INE}. Instituto Nacional de Estadística},
  title        = {Continuous Register Statistics},
  year         = {2025},
  howpublished = {\url{https://www.ine.es/dyngs/INEbase/en/operacion.htm?c=Estadistica_C&cid=1254736177012&menu=ultiDatos&idp=1254734710990}},
  note         = {Last accessed on October 1, 2025}
}

@misc{ministeriosanidad2023plan,
  author       = {{Ministerio de Sanidad, Gobierno de España}},
  title        = {Plan Nacional de actuaciones preventivas de los efectos del exceso de temperaturas sobre la salud 2023},
  year         = {2023},
  note         = {[in Spanish]},
  howpublished = {\url{https://www.sanidad.gob.es/ciudadanos/saludAmbLaboral/planAltasTemp/2023/docs/Plan_Excesos_Temperatura_2023.pdf}},
  note         = {Last accessed on August 28, 2024}
}

@article{tobias2023heat,
  title={Heat-attributable mortality in the summer of 2022 in Spain},
  author={Tob{\'\i}as, Aurelio and Roy{\'e}, Dominic and I{\~n}iguez, Carmen},
  journal={Epidemiology},
  volume={34},
  number={2},
  pages={e5-e6},
  year={2023},
  publisher={LWW},
  doi={10.1097/EDE.0000000000001583}
}

@article{watanabe2010asymptotic,
  title={Asymptotic equivalence of Bayes cross validation and widely applicable information criterion in singular learning theory.},
  author={Watanabe, Sumio and Opper, Manfred},
  journal={Journal of machine learning research},
  volume={11},
  number={12},
  year={2010},
  pages={3571-3594},
  doi={10.5555/1756006.1953045}
}

@misc{r-inla2025,
  author       = {{R INLA project}},
  title        = {{R INLA project}},
  year         = {2025},
  howpublished = {\url{http://www.r-inla.org/home}},
  note         = {Last accessed on October 1, 2025}
}

@article{rue2009approximate,
  title={Approximate Bayesian inference for latent Gaussian models by using integrated nested Laplace approximations},
  author={Rue, H{\aa}vard and Martino, Sara and Chopin, Nicolas},
  journal={Journal of the Royal Statistical Society Series B: Statistical Methodology},
  volume={71},
  number={2},
  pages={319--392},
  year={2009},
  publisher={Oxford University Press},
  doi={j.1467-9868.2008.00700.x}
}

@article{rue2017bayesian,
  title={Bayesian computing with INLA: a review},
  author={Rue, H{\aa}vard and Riebler, Andrea and S{\o}rbye, Sigrunn H and Illian, Janine B and Simpson, Daniel P and Lindgren, Finn K},
  journal={Annual Review of Statistics and Its Application},
  volume={4},
  number={1},
  pages={395--421},
  year={2017},
  publisher={Annual Reviews},
  doi={annurev-statistics-060116-054045}
}

@article{lindgren2011explicit,
  title={An explicit link between Gaussian fields and Gaussian Markov random fields: the stochastic partial differential equation approach},
  author={Lindgren, Finn and Rue, H{\aa}vard and Lindstr{\"o}m, Johan},
  journal={Journal of the Royal Statistical Society Series B: Statistical Methodology},
  volume={73},
  number={4},
  pages={423--498},
  year={2011},
  publisher={Oxford University Press},
  doi={j.1467-9868.2011.00777.x}
}

@article{van2023new,
  title={A new avenue for Bayesian inference with INLA},
  author={Van Niekerk, Janet and Krainski, Elias and Rustand, Denis and Rue, H{\aa}vard},
  journal={Computational Statistics \& Data Analysis},
  volume={181},
  pages={107692},
  year={2023},
  publisher={Elsevier},
  doi={10.1016/j.csda.2023.107692}
}

@article{simpson2017penalising,
  title={Penalising model component complexity: A principled, practical approach to constructing priors},
  author={Simpson, Daniel and Rue, H{\aa}vard and Riebler, Andrea and Martins, Thiago G and S{\o}rbye, Sigrunn H},
  journal={Statistical Science},
  volume={32},
  number={1},
  pages={1--46},
  year={2017},
  doi={10.1214/16-STS576}
}

@article{jimenez2010role,
  title={Role of {Saharan} dust in the relationship between particulate matter and short-term daily mortality among the elderly in {Madrid} ({Spain})},
  author={Jim{\'e}nez, Elena and Linares, Cristina and D{\'\i}az, Julio},
  journal={Science of the Total Environment},
  volume={408},
  number={23},
  pages={5729--5736},
  year={2010},
  doi={10.1016/j.scitotenv.2010.08.049}
}

@article{tobias2011short,
  title={Short-term effects of particulate matter on total mortality during {Saharan} dust outbreaks: a case-crossover analysis in {Madrid} ({Spain})},
  author={Tob{\'\i}as, Aurelio and P{\'e}rez, Laura and D{\'\i}az, Julio and Linares, Cristina and Pey, Jorge and Alastruey, Andr{\'e}s and Querol, Xavier},
  journal={Science of the Total Environment},
  volume={412--413},
  pages={386--389},
  year={2011},
  doi={10.1016/j.scitotenv.2011.10.027}
}
